\documentclass[doublecolumn]{pasj01}

\begin{document}
\SetRunningHead{T. Sasaki et al.}{X-Ray observations of a subhalo associated with the NGC~4839 group infalling toward the Coma cluster}

\title{X-Ray observations of a subhalo associated with the NGC~4839 group infalling toward the Coma cluster}



 \author{%
Toru \textsc{SASAKI}\altaffilmark{1}
 Kyoko \textsc{MATSUSHITA}\altaffilmark{1}
 Kosuke \textsc{SATO}\altaffilmark{1}
 and
Nobuhiro \textsc{OKABE}\altaffilmark{2,3}
}
\altaffiltext{1}{Department of Physics, Tokyo University of Science, 
1-3 Kagurazaka, Shinjuku-ku, Tokyo 162-8601 Japan}
\email{j1213703@gmail.com; matusita@rs.kagu.tus.ac.jp}
\altaffiltext{2}{Department of Physical Science, Hiroshima University,
1-3-1 Kagamiyama, Higashi-Hiroshima, Hiroshima 739-8526, Japan}
\altaffiltext{3}{Hiroshima Astrophysical Science Center, Hiroshima University, Higashi-Hiroshima,
Kagamiyama 1-3-1, 739-8526, Japan}





\maketitle

\begin{abstract}


We report {\it Suzaku} X-ray observations of the dark subhalo associated with the merging group of NGC 4839 in the Coma cluster. 
The X-ray image exhibits an elongated tail toward the southwest. 
The X-ray peak shifts approximately $1\arcmin$ away from the weak-lensing mass center toward the opposite direction of the Coma cluster center. 
We investigated  the temperature, normalization, pressure, and entropy distributions around the subhalo.
Excluding the X-ray tail, the temperature beyond the truncation radius is 8--10~keV, 
which is two times higher than that of the subhalo and the X-ray tail. 
The pressure is nearly uniform excluding southern part of the subhalo at two times of the truncation radius. 
We computed the gas mass within the truncation radius and the X-ray tail.
While the gas fraction within the truncation radius is about 5 times smaller than that of regular groups, 
the gas mass in the subhalo and the X-ray tail to weak-lensing mass ratio is consistent with that of regular groups.
Assuming the infall velocity,  $2000~\rm km~s^{-1}$, the ram pressure is 1.4 times greater than gravitational force per unit area. 
Assuming the Kelvin-Helmholtz instabilities, the total lost mass is approximately $3\times10^{11}~M\solar$.
If this gas had originally been within the truncation radius, the gas mass fraction of the subhalo would have been comparable with those of regular groups before infalling to the Coma cluster.

\end{abstract}

\section{Introduction}


Galaxy clusters have formed through mergers of clusters and/or accretions of smaller groups and galaxies.
X-ray observations have revealed features such as shocks, cold fronts, and gas stripping tails,  around accreting galaxy groups to clusters (e.g., \cite{Markevitch2000,Markevitch2002, Markevitch2007,Russell2010,Ghizzardi2010,Russell2014, Eckert2014, Ichinohe2015} ). 
Shock fronts are expected to occur in front of a merging subhalo and to heat the intracluster medium (ICM). 
Radio relics, which are diffuse synchrotron radio emissions,
are thought to be tracers of shock structures \citep{Ferrari2008} and
recent {\it XMM-Newton}, {\it Chandra} and  {\it Suzaku} observations have found jumps in  temperature across these radio relics (e.g., \cite{Finoguenov2010, Akamatsu2013a, Itahana2015, Akamatsu2015}).
\citet{Markevitch2000} found an edge \textcolor{blue}{in} surface brightness where the temperature jumped, and the gas pressure across the edge was continuous.
Such a feature is called a `cold front', and is caused by gas sloshing when a minor merger oscillates the cool dense core of a cluster (e.g., \cite{Ascasibar2006}). 
In addition, the accreting galaxies and groups experience the stripping of gas by ram pressure (e.g., \cite{Gunn1972}).
The gas stripped from the accreting objects looks like a tail structure in the X-ray band. 

Weak-lensing measurement is a direct, powerful technique to probe dark matter distribution of 
clusters without resting any assumptions of dynamical states.  
Since the lifetime of the dark matter subhalos is longer than that of gas subhalos \citep{Tormen2004}, 
weak-lensing enables us to measure the dark matter mass of the subhalos whose interaction triggers the gas features
discussed above.
Therefore, a joint weak-lensing and X--ray study is powerful to understand cluster mergers. 
To date, several joint studies of major mergers have revealed the interaction between ICM and the dark matter subhalo 
\citep{Clowe2006, Okabe2008, Okabe2011, Okabe2015, Medezinski2015}. 
However, joint studies on minor mergers are not yet sufficient. 
A recent weak-lensing study (\cite{Okabe2014}; see also \cite{Okabe2010}) for the Coma cluster detected 32 small subhalos 
whose mass was greater than $\sim 3\times10^{12}~h^{-1} M\solar$.  
The ratio between the subhalo mass and the virial mass reaches down to a few $10^{-3}$.  
Detections of such small subhalos was achievable by the large apparent size, 
because the cluster is very close to us.

There are four massive subhalos whose masses are greater than $\sim 1\times10^{13}~M\solar$, 
and labeled ``ID~1'', ``ID~2'', ``ID~9'', and ``ID~32'' in \citet{Okabe2014}. 
The massive subhalo signals excluding the ``ID~2'' in the Coma cluster, as well as stacked signals from all subhalos, 
were well represented with a truncated Navarro-Frenk-White (NFW) mass model \citep{Navarro1995} 
as expected from a tidal destruction model \citep{Okabe2014}.
These subhalos excluding the ``ID~9'' were observed with Suzaku \citep{Sasaki2015}.
While excess X-ray emission from the ``ID~1'' and ``ID~32'' subhalos was detected, 
the ``ID~2'' did not have excess X-ray emission. 
The gas mass to weak-lensing mass ratios of the ``ID~1'' and ``ID~32'' subhalos were 0.001, 
which is one to two orders of magnitude smaller than those of regular groups.
This supports that ram pressure with typical infall velocities 
can strip  significant amounts of gas from the infalling subhalos.

In this paper, we describe the thermodynamics around the third massive subhalo "ID~9"  whose mass and truncation radius ($r_t$) are $(1.58\pm0.26)\times10^{13}~M\solar$ and 3.43\arcmin$_{-0.47}^{+0.28}$, respectively, as detected  by weak-lensing analysis \citep{Okabe2014}. 
The subhalo ``ID~9" is associated with the NGC~4839 group, which
 is the most famous merging substructure in the Coma cluster.
It is located at a projected distance of approximately 50$\arcmin$ ($\sim r_{500}$
\footnote{$r_{500}$ is the radius within which the mean density of the cluster is 500 times the critical density of the universe.}).
The NGC 4839 group has an X-ray tail toward the southwest or opposite to the direction of the Coma cluster center \citep{Briel1992}.
With {\it XMM-Newton}, \citet{Neumann2001} reported complex temperature structures around the NGC~4839 group, 
with indications of a bow shock and of ram pressure stripping. 
At the southwest of the NGC~4839 group, a radio relic was discovered around the virial radius of the Coma cluster \citep{Ballarati1981}.  
The X-ray observations found that there is temperature discontinuity at the relic, which corresponds to a shock with the {\cal M}ach number $\sim$ 2 \citep{Ogrean2013, Akamatsu2013b}.



We summarize the observation logs and data reductions in section \ref{sec:obs}.
Section \ref{sec:ana} shows an X-ray image around subhalo ``ID~9" with the mass distribution derived from weak-lensing analysis, and results of spectral analysis. 
We discuss thermodynamics around the massive subhalo in section \ref{sec:dis}.

In this paper, we use $\Omega_{m,0}=0.27, \Omega_{\Lambda}=0.73$, 
and $H_{0}=$70~km~s$^{-1}$~Mpc$^{-1}$.
The redshift of the Coma cluster is $z$ = 0.0231 \citep{Struble1999}.
At this redshift, $1\arcmin$ corresponds to 28.9~kpc.
The solar abundance table is given by \citet{Lodders2003}. 
Unless noted otherwise, the errors are in the 68\% confidence region for the single parameter of interest.

\section{Observations and data reduction}
\label{sec:obs}

\begin{table*}[t]
  \caption{{\it Suzaku} observation logs around the NGC~4839 group.}
  \label{tb:suzakuobslog}
  \begin{center}
    \begin{tabular}{lllll}
      \hline
Field name & Sequence &  Date-Obs.\footnotemark[$*$] & (R.A., decl.)\footnotemark[$\dagger$]  &  Exposure\footnotemark[$\ddagger$]  \\
& Number & & (J2000.0) & (ksec) \\ \hline
ID9~East   & 808020010 & 2013-Jun-09T13:12:43 &\timeform{12h58m01.7s}, \timeform{27D22'08.0"} & 16.2 \\
ID9~North  & 806047010 & 2011-Jun-23T12:58:52 &\timeform{12h57m44.5s}, \timeform{27D43'35.8"} & 6.6 \\
ID9~West   & 802047010 & 2007-Dec-02T10:08:56 &\timeform{12h57m01.4s}, \timeform{27D34'17.0"} & 23.9 \\
ID9~South  & 802048010 & 2007-Dec-04T04:47:49 &\timeform{12h56m06.0s}, \timeform{27D25'30.7"} & 29.3 \\
ID~2           & 808021010 & 2013-Jun-10T00:39:15 & \timeform{12h55m55.3s}, \timeform{27D45'17.6s} & 23.7 \\
ID~1           & 808022010 & 2013-Jun-10T14:12:00 & \timeform{12h55m28.0s}, \timeform{27D31'00.1s} & 18.4 \\  
      \hline
  \multicolumn{5}{@{}l@{}}{\hbox to 0pt{\parbox{180mm}{\footnotesize
      \par\noindent
      \footnotemark[$*$] Start date of the observation written in the event FITS files as DATE-OBS keyword.      
      \par\noindent
      \footnotemark[$\dagger$] The nominal position of the observation written in the event FITS files as RA\_NOM and DEC\_NOM  keywords. 
      \par\noindent
      \footnotemark[$\ddagger$] After data screening. 
       }    \hss}}
    \end{tabular}
  \end{center}
\end{table*}

We analyzed {\it Suzaku} data of six pointing observations
around the NGC~4839 group (labeled as the ``ID 9" subhalo in \cite{Okabe2014}).
The observation logs are summarized in table \ref{tb:suzakuobslog}.
The X-ray Imaging Spectrometers (XIS) data were only used in this study. 
The XIS was operated in the normal clocking mode (8~s exposure per frame) with $3\times3$ or $5\times5$ editing mode.
The data reduction criteria were the same as those of \citet{Sasaki2015}.
In addition, we excluded flickering pixels from our analysis 
\footnote{http://www.astro.isas.jaxa.jp/suzaku/analysis/xis/nxb\_new/}.
The reduction and analysis were performed with HEASOFT version 6.16.
We applied the latest calibration database version 2015-March-12. 


\section{Analysis and results}
\label{sec:ana}

\subsection{X-ray image of the subhalo}
\label{sec:image}

\begin{figure*}[htpd]
\begin{center}
\includegraphics[width=0.7\textwidth,angle=0,clip]{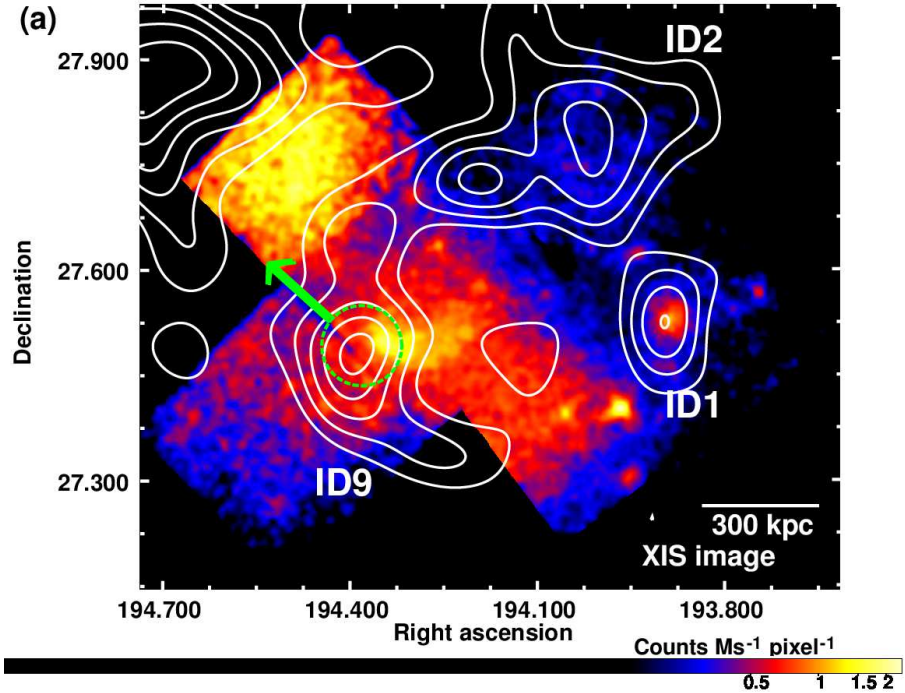}
\caption{
{\it Suzaku} XIS mosaic image around subhalo ``ID~9" in the 0.5--5.0 keV energy band\@.
While the exposure time and instrumental background were corrected, the effect of the vignetting were not corrected. 
The image was smoothed by a Gaussian of $\sigma=$16 pixels $\approx 17\arcsec$.  
The numbers below the color bars have units of counts~Ms$^{-1}$~pixel$^{-1}$.
The white contour shows the mass distribution in a linear scale derived from weak-lensing analysis \citep{Okabe2014}.
The dashed circle indicates the $r_t$ of the subhalo ``ID~9".
The arrow indicates the direction toward the Coma cluster center. 
}
\label{fig:image}
\end{center}
\end{figure*}

\begin{figure*}[htpd]
\begin{center}
\includegraphics[width=0.5\textwidth,angle=0,clip]{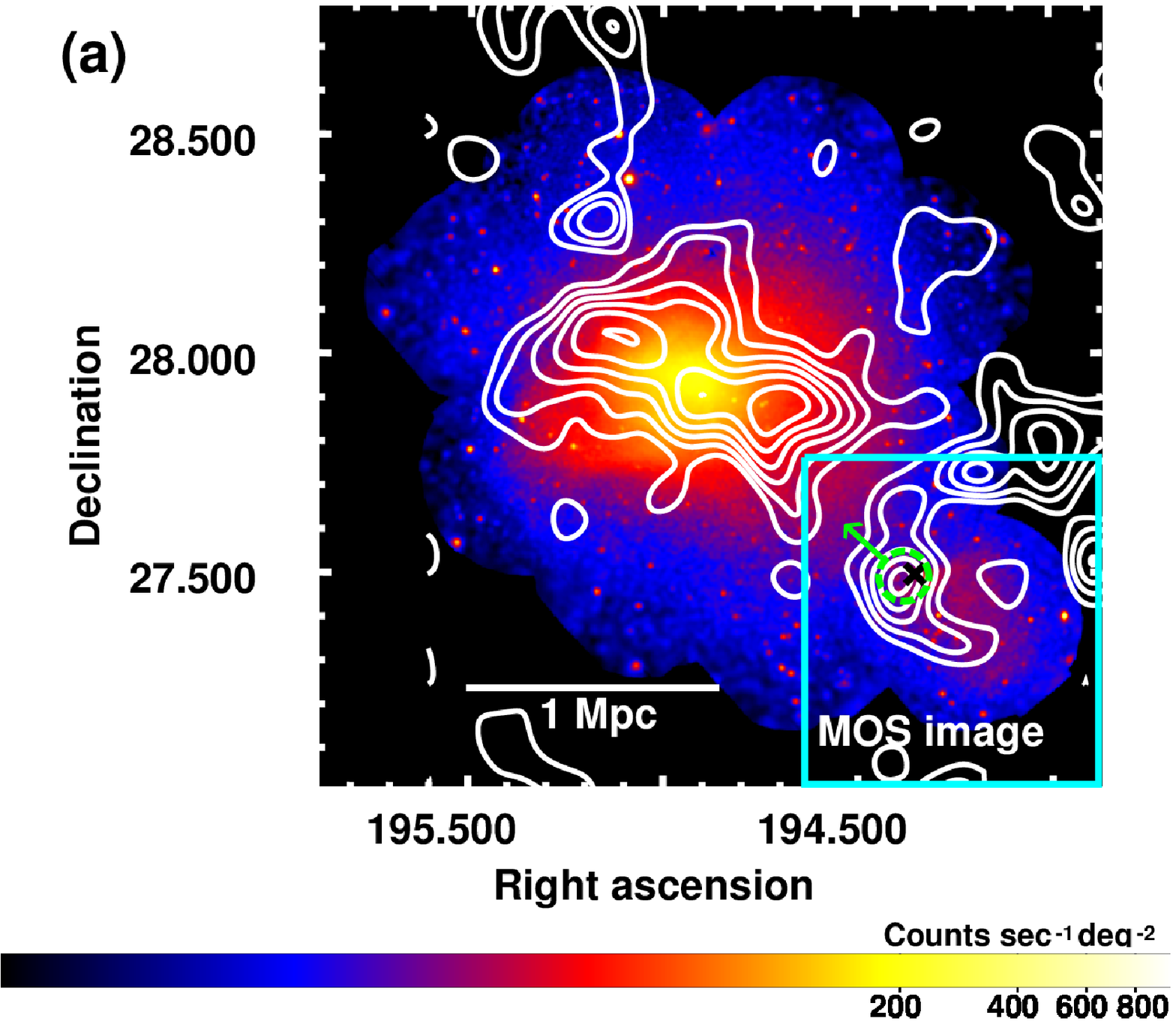}
\includegraphics[width=0.45\textwidth,angle=0,clip]{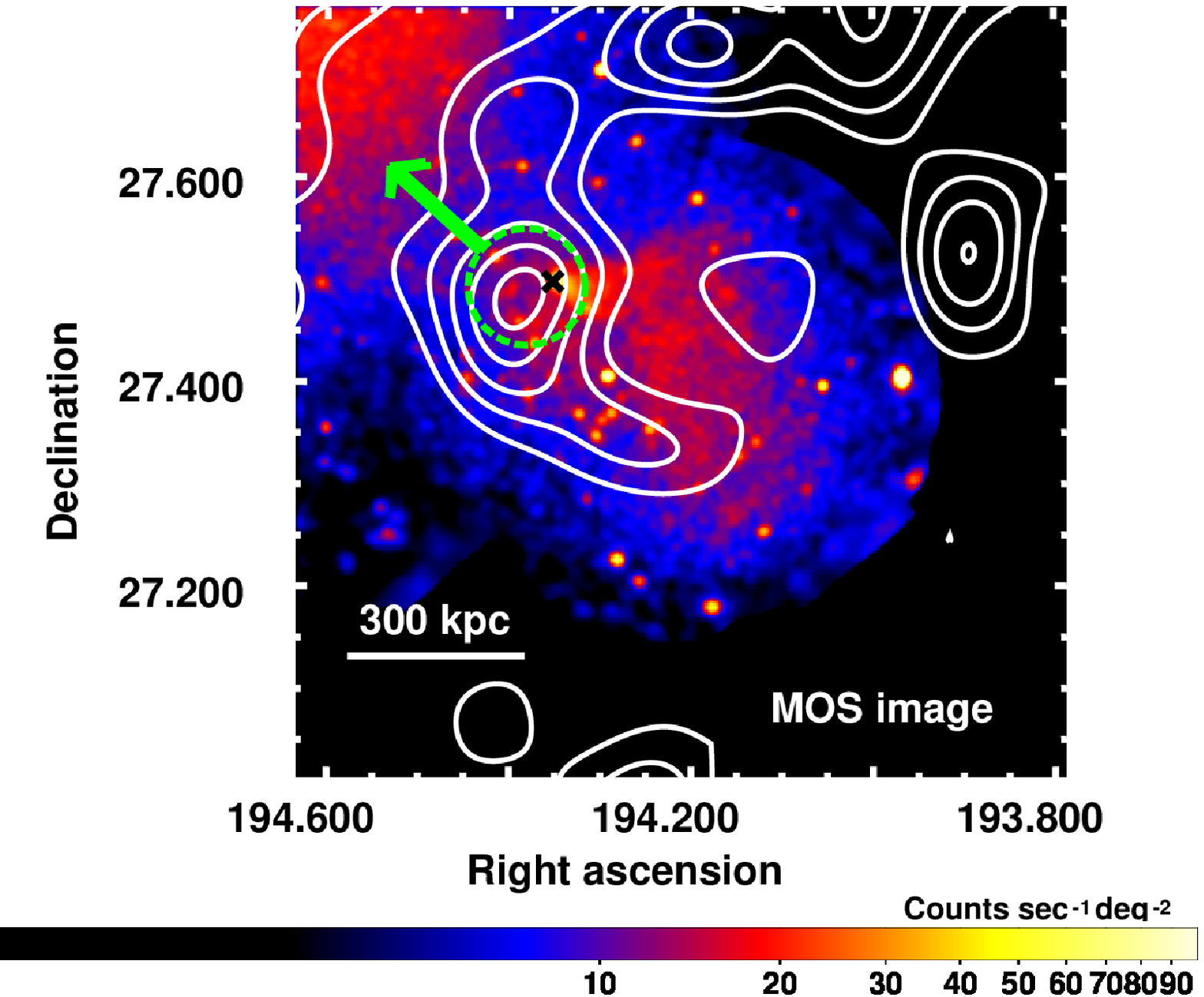}
\caption{
(a) {\it XMM-Newton} public image of the Coma cluster \footnote{http://heasarc.gsfc.nasa.gov/docs/xmm/gallery/esas-gallery/xmm\_gal\_science\_coma.html}. 
The energy band is 0.4--1.3~keV.
The image was smoothed by a Gaussian of $\approx 12\arcsec$.  
The cross indicates the position of the NGC~4839 obtained from NED. 
The notations of contour, circle, and arrow are the same as the figure \ref{fig:image}.
(b) The same figure as figure (a), but around subhalo ``ID~9" highlighted by cyan box in figure (a).
}
\label{fig:imageXMM}
\end{center}
\end{figure*}

Figure \ref{fig:image} (a) shows the XIS image in the 0.5-5.0~keV energy range around the subhalo ``ID~9".
Here, the differences in exposure time and instrumental background  were corrected by the  ``xisexpmapgen" and ``xisnxbgen" Ftools tasks, while the vignetting effect was not. 
We also obtained {\it XMM-Newton} public image in 0.4-1.3~keV from ``{\it XMM-Newton Science data images}\footnote{http://heasarc.gsfc.nasa.gov/docs/xmm/xmm\_gal\_science.html}".
Figure \ref{fig:imageXMM} (a) is the {\it XMM-Newton} image of the Coma cluster with mass contour\citep{Okabe2014}.  
The subhalo ``ID~9" is highlited in figure \ref{fig:imageXMM} (b).

The X-ray peak derived from the {\it XMM-Newton} image, which is spatially coincident with the NGC~4839 galaxy, is shifted approximately 1$\arcmin$ ($0.3~r_t$) away from the weak-lensing mass center toward the opposite direction of the Coma cluster center. 
The position uncertainty of weak-lensing mass center is $\sim 20\arcsec$ because of a finite sample of background galaxies. Since the X-ray peak is determined with an accuracy of a few arcsecond, the two peaks are offset. 
The {\it XMM-Newton} image shows that at the head of the NGC~4839 group, 
the surface brightness has an edge like structure. 
The bright X-ray tail is elongated toward the southwest direction or the outskirts of the Coma cluster,
 out to approximately $6~r_t$, $\sim$ 600~kpc from the X-ray peak.
In addition to the tail,  {\it Suzaku} image shows an extended excess X-ray emission around the weak-lensing mass contours from the northeast to the southwest direction. 
The apparent size of the excess is about several times of $r_t$.

\subsection{Spectral modeling for the foreground, background and ICM components}
\label{sec:bgd}

\begin{figure*}[tpd]
\begin{center}
\includegraphics[width=0.45\textwidth,angle=0,clip]{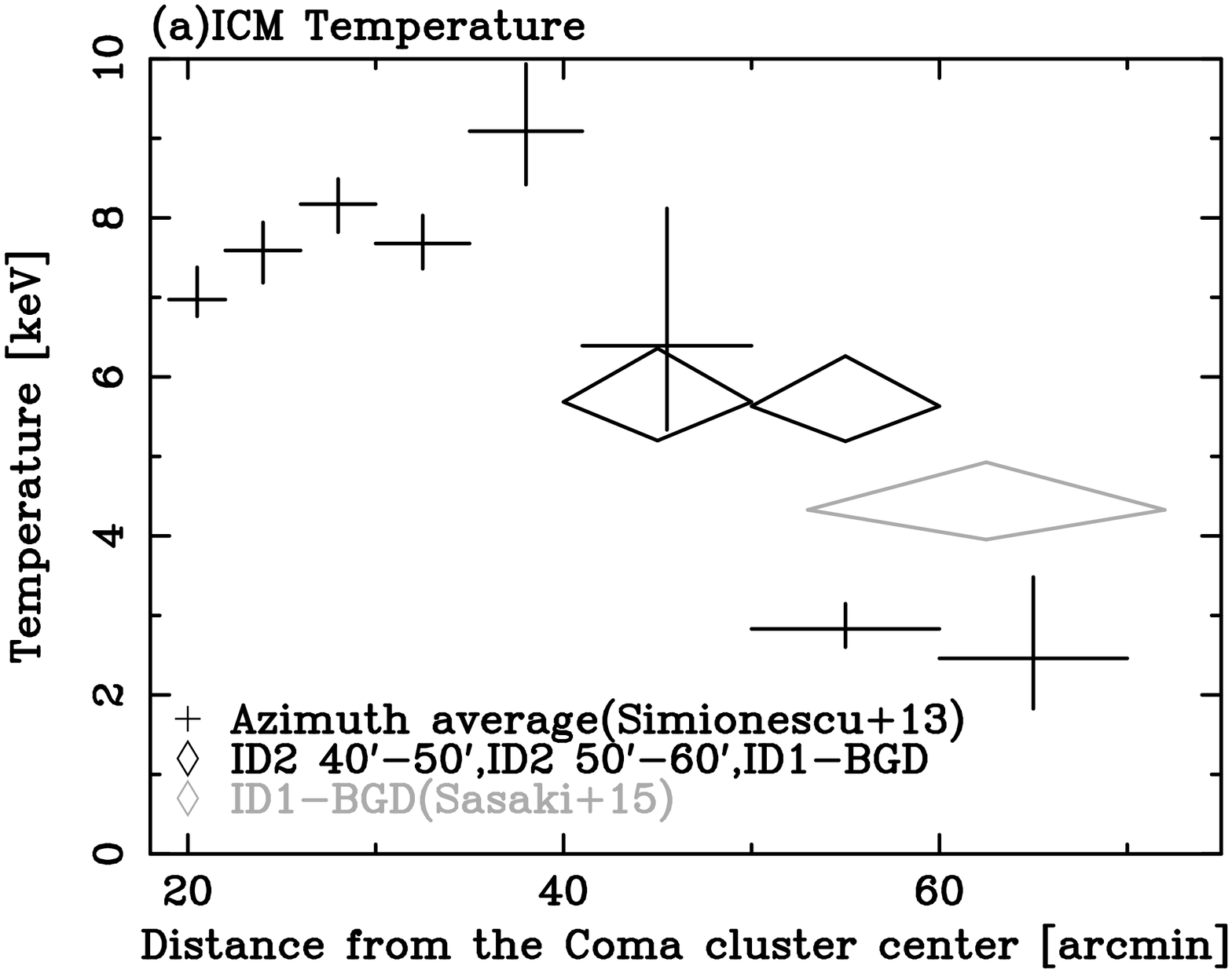}
\includegraphics[width=0.45\textwidth,angle=0,clip]{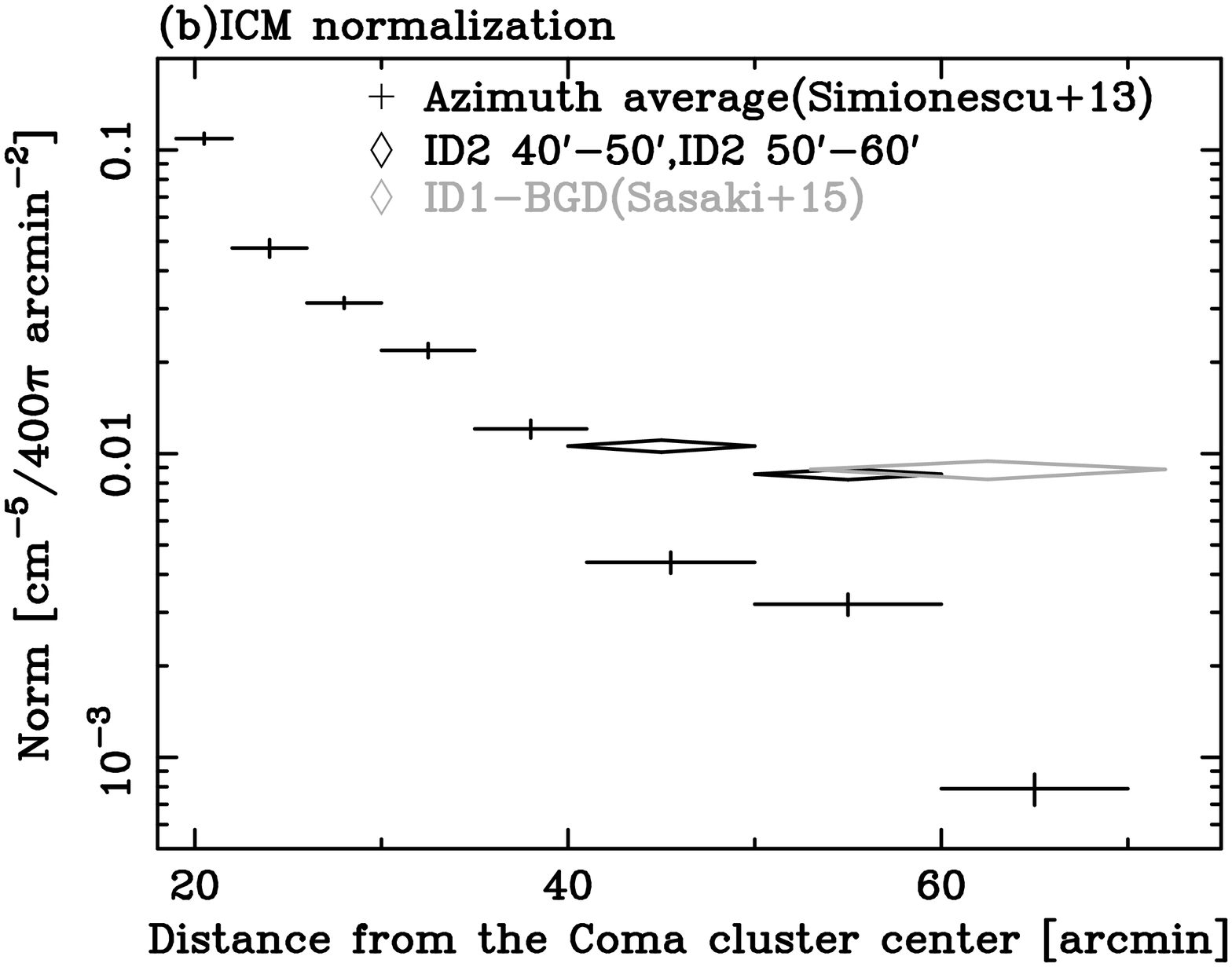}
\end{center}
\caption{
The radial profiles of  (a) temperature and (b) normalization of the ICM. 
The normalization is defined as $Norm = \int n_{\rm e} n_{\rm H} dV \,/~\,[4\pi\,(1+z)^2 D_{\rm A}^{~2}] \times 10^{-14}$ cm$^{-5}$~/ $(20^2\pi)$~arcmin$^{-2}$. 
The crosses show the azimuthal average profiles \citep{Simionescu2013} excluding the southwest direction which includes the subhalo, "ID~9".
The black diamonds are those derived from the regions of  ``ID2 within 50\arcmin", ``ID2 beyond 50\arcmin".
Those derived from the ``ID1-BGD" region are plotted with light-gray diamonds \citep{Sasaki2015}.
}
\label{fig:ICMprofile}
\end{figure*}

To investigate the excess emission associated with the subhalo, ``ID~9", 
we included the ICM emission in the spectral model.
The ICM emission was modeled by a single thermal component with 
 APEC plasma code \citep{Smith2001} for the ICM emission ($apec_{\rm ICM}$)
modified by the photoelectric absorption model, $phabs$, for the Galactic absorption ($phabs_{\rm GAL}$).
We adopted the azimuthally-averaged radial profiles of the ICM
temperature and normalization derived  by \citet{Simionescu2013},
which are plotted in Figure \ref{fig:ICMprofile}.
Here, the southwest direction, which  includes the subhalo ``ID~9", was excluded. 
In this paper, the normalization is defined as following formula;  
$Norm = \int n_{\rm e} n_{\rm H} dV \,/~\,[4\pi\,(1+z)^2 D_{\rm A}^{~2}]  \times 10^{-14}$ cm$^{-5}$~/ $(20^2\pi)$~arcmin$^{-2}$ (see \cite{Ishisaki2007} for detail). 
In the following spectral fits, 
we restricted the ICM parameters within the error ranges 
of the azimuthally-averaged profiles at the same 
distance from the cluster center (hereafter, {\it Default}-case). 
The abundance of the ICM was fixed at 0.3 solar \citep{Simionescu2013}.
We also fitted the spectra by changing the abundance of the ICM to be 0.2 solar,
 and obtained almost the same temperatures and normalizations.

We assumed the X-ray foreground and background emissions to be composed of 
the local hot bubble (LHB), milky way halo (MWH), and cosmic X-ray background (CXB). 
The LHB and MWH are modeled by unabsorbed and  absorbed APEC model, 
$apec_{\rm LHB}$ and $phabs_{\rm GAL}\times apec_{\rm MWH}$, respectively. 
The CXB was assumed as a absorbed power-law whose photon index was fixed at 1.4 (e.g., \cite{Kushino2002}). 
The parameters of $apec_{\rm LHB}$, $apec_{\rm MWH}$, and $powerlaw_{\rm CXB}$ were restricted within 
the error range estimated from 110$'$-130$'$ regions
of the Coma cluster by \citet{Sasaki2015}.

\subsection{Spectral fittings of the excess emission}
\label{sec:excess}

\begin{table}[t]
  \caption{Regional notations}
  \label{tb:suzakuregion}
  \begin{center}
    \begin{tabular}{lll}
      \hline
Area name & Angle & radius   \\
& (degree) & ($r_t$) \\ \hline
Center & 0-360 & 0-1\\
NWW &  20-48    & 1-2, 2-3, 3-4 \\
NW    &  48-94    & 1-2, 2-3, 3-4  \\
NE     &  94-140  & 1-2, 2-3, 3-4 \\
E        & 140-228 & 1-2, 2-3, 3-4 \\
S        & 228-305 & 1-2, 2-3, 3-4 \\
SW     & 305-340 & 1-2, 2-3, 3-4 \\
Tail     & 340-380, 315-355\footnotemark[$*$] & 1-2, 2-3, 3-4, 4-6, 6-8 \\
      \hline
  \multicolumn{3}{@{}l@{}}{\hbox to 0pt{\parbox{180mm}{\footnotesize
      \par\noindent
      \footnotemark[$*$] For $4-6~r_t$ and $6-8~r_t$ regions.
       }    \hss}}
    \end{tabular}
  \end{center}
\end{table}

\begin{figure*}
\begin{center}
\includegraphics[width=0.7\textwidth,angle=0,clip]{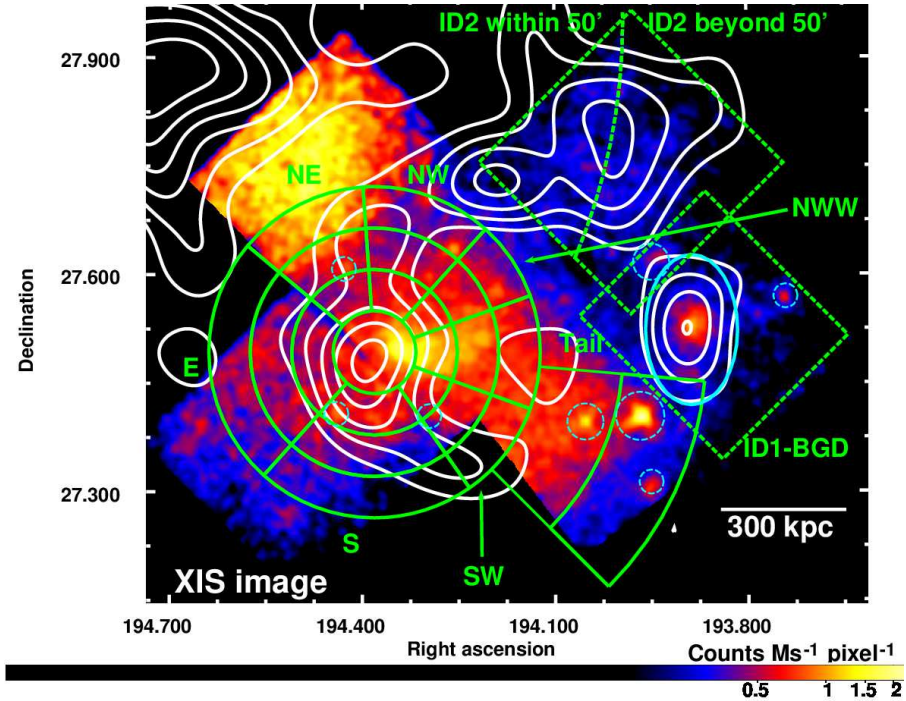}
\caption{
The same figure as figure \ref{fig:image}, but overlaid with the regions from which we extracted the spectra. 
The green dashed regions are the areas for which background was estimated. 
Note that the radius of the innermost circle is equal to the truncation radius of the subhalo. 
The cyan dashed circles indicate the point sources excluded from the analysis. 
The cyan solid ellipse shows the subhalo ``ID~1'', which was excluded from the spectral analysis for the "ID1-BGD" region by \citet{Sasaki2015}. 
}
\label{fig:regions}
\end{center}
\end{figure*}

We extracted the spectra within a circular region whose radius is the same as $r_t$, 3.43\arcmin, 
centered on the weak-lensing signal peak, (R.A., Decl.) = ($\timeform{12h57m31.44s}, \timeform{27D29"34.80'}$). 
As shown in figure \ref{fig:regions}, beyond $r_t$, we used annular regions whose radii were 
$1-2~r_t$, $2-3~r_t$, and $3-4~r_t$ and divided into 7 sectors labeled as ``NW", ``NE", ``E", ``S", ``SW", ``Tail", and ``NWW".
In addition, we extracted the spectra from annular regions whose radii were $4-6~r_t$ and $6-8~r_t$ for ``Tail'' sector.  
The angles of the each sector and the radii of the annular regions are summarized in table \ref{tb:suzakuregion}.
Using the ``wavdetect'' tool in CIAO package, we searched for point sources in the {\it Suzaku} XIS image.
Circular regions with 1$\arcmin$ radius around the detected point sources were excluded from the spectral analysis.
The exclusion radii were  1.5$\arcmin$ and  2$\arcmin$ for the point source flux 
brighter than $2\times10^{-13} {\rm erg~s^{-1}~cm^{-2}}$ and
$4\times10^{-13} {\rm erg~s^{-1}~cm^{-2}}$ in 2-10~keV energy range, respectively.
The calibration corners are also removed from the spectral analysis.

We added an absorbed thermal model, $phabs_{\rm GAL}\times apec_{\rm subhalo}$, 
to the spectral model used for the background regions, 
and fitted the spectra around the subhalo.
As a result, the spectra were fitted by following formula; 
$constant \times ( apec_{\rm LHB} + phabs_{\rm GAL} \times ( apec_{\rm MWH} 
+ apec_{\rm ICM} +  apec_{\rm subhalo} $ $ + powerlaw_{\rm CXB} ) ).$
The abundance of  $apec_{\rm subhalo}$ was fixed at 0.3 solar.
All spectral fitting was performed in 0.7-8.0~keV(BI) and 0.8-8.0~keV(FI) energy range.
As shown in figure \ref{fig:spec2T} and figure \ref{fig:spec2TTail}-\ref{fig:spec2TE} in Appendix,
the spectra were well-represented by our spectral-modeling.

\begin{figure*}
\begin{center}
\includegraphics[width=0.30\textwidth,angle=0,clip]{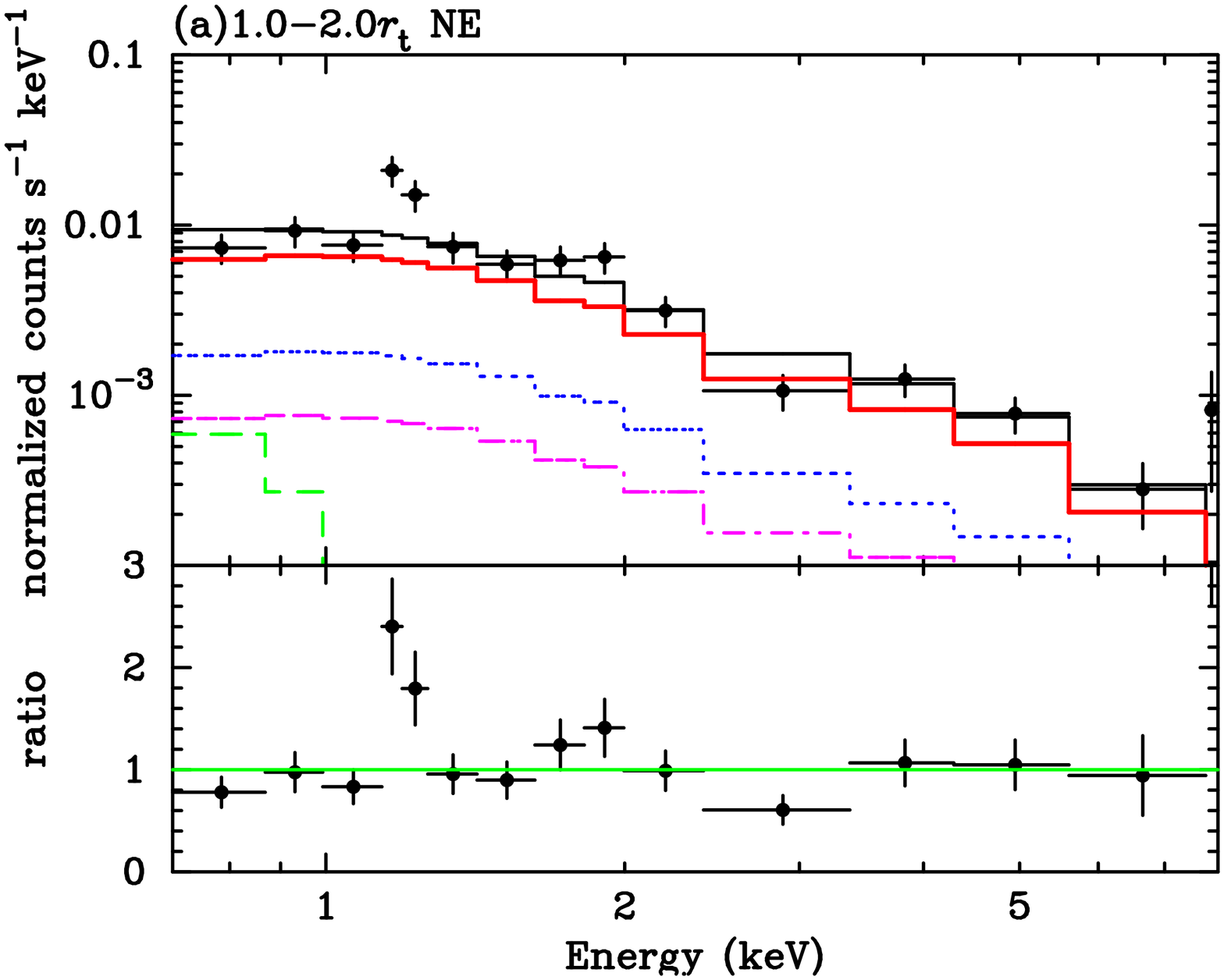}
\includegraphics[width=0.30\textwidth,angle=0,clip]{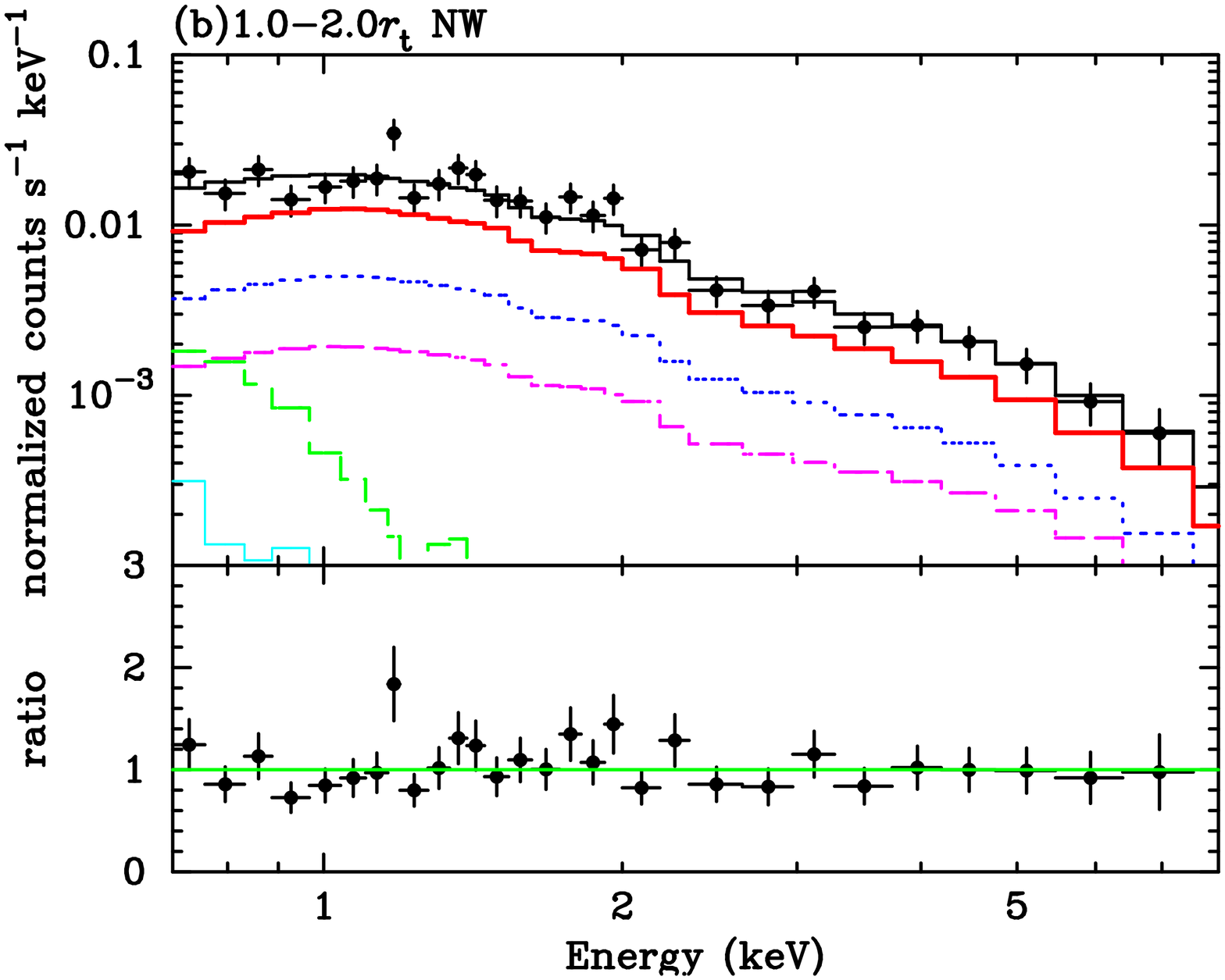}
\includegraphics[width=0.30\textwidth,angle=0,clip]{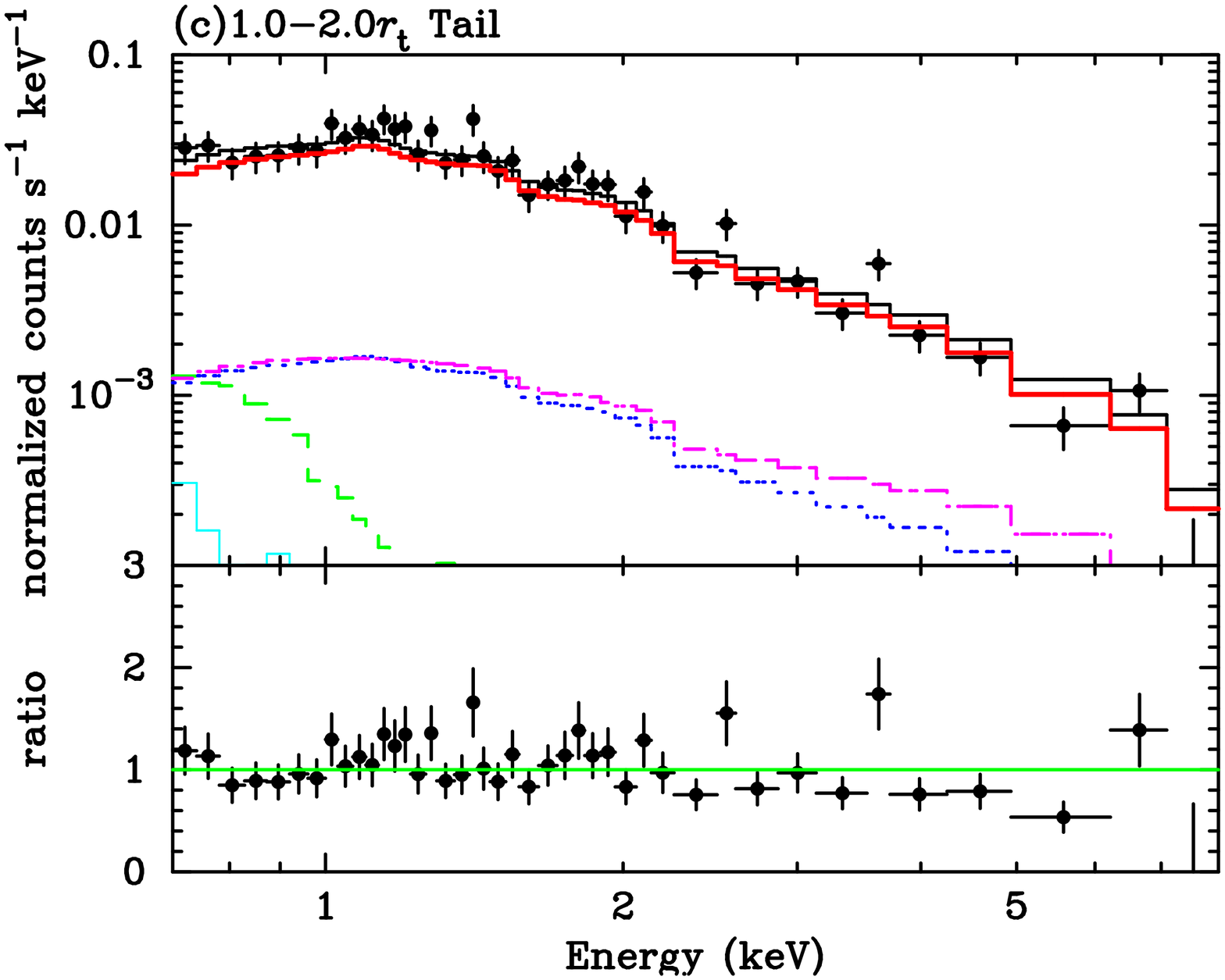}
\includegraphics[width=0.30\textwidth,angle=0,clip]{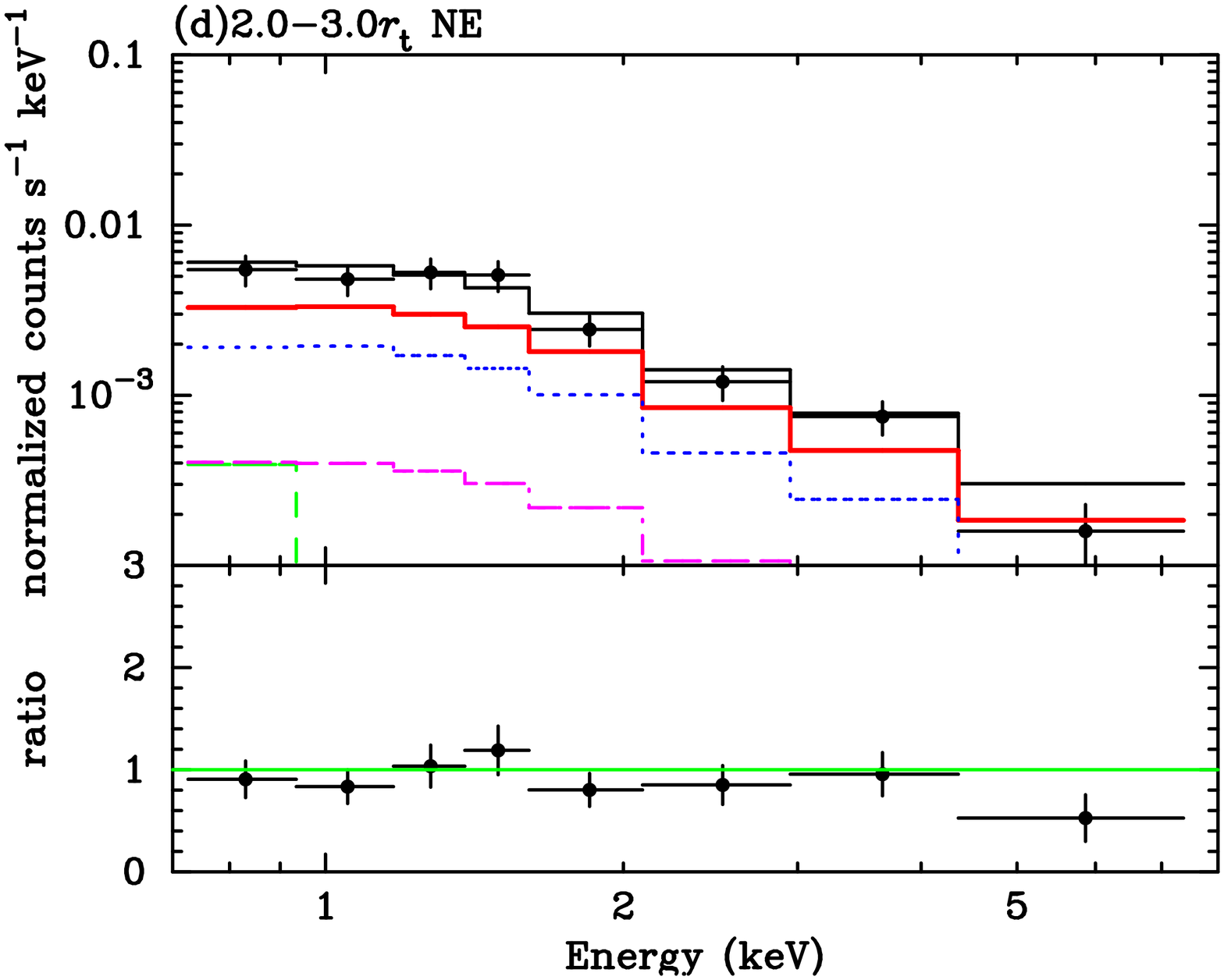}
\includegraphics[width=0.30\textwidth,angle=0,clip]{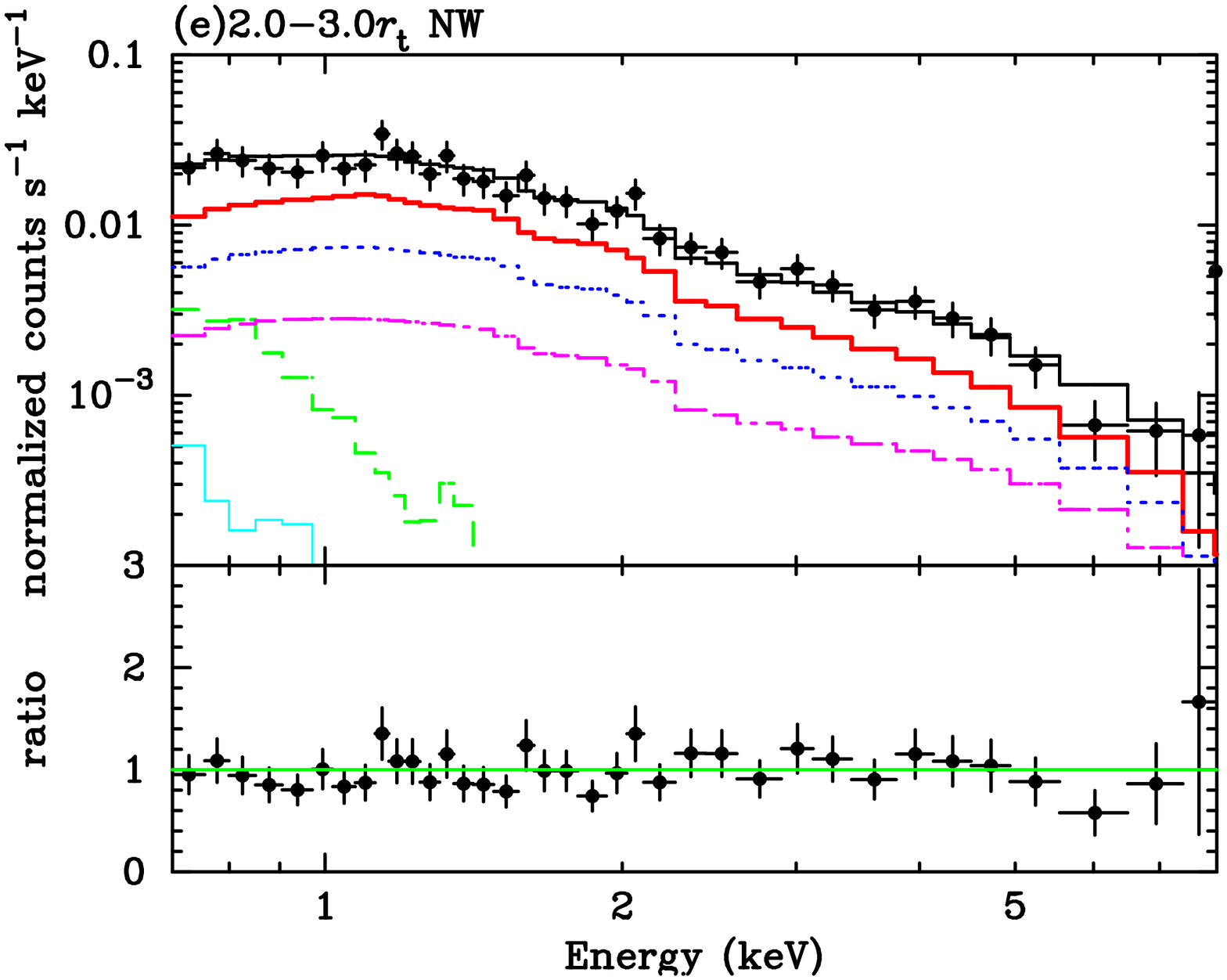}
\includegraphics[width=0.30\textwidth,angle=0,clip]{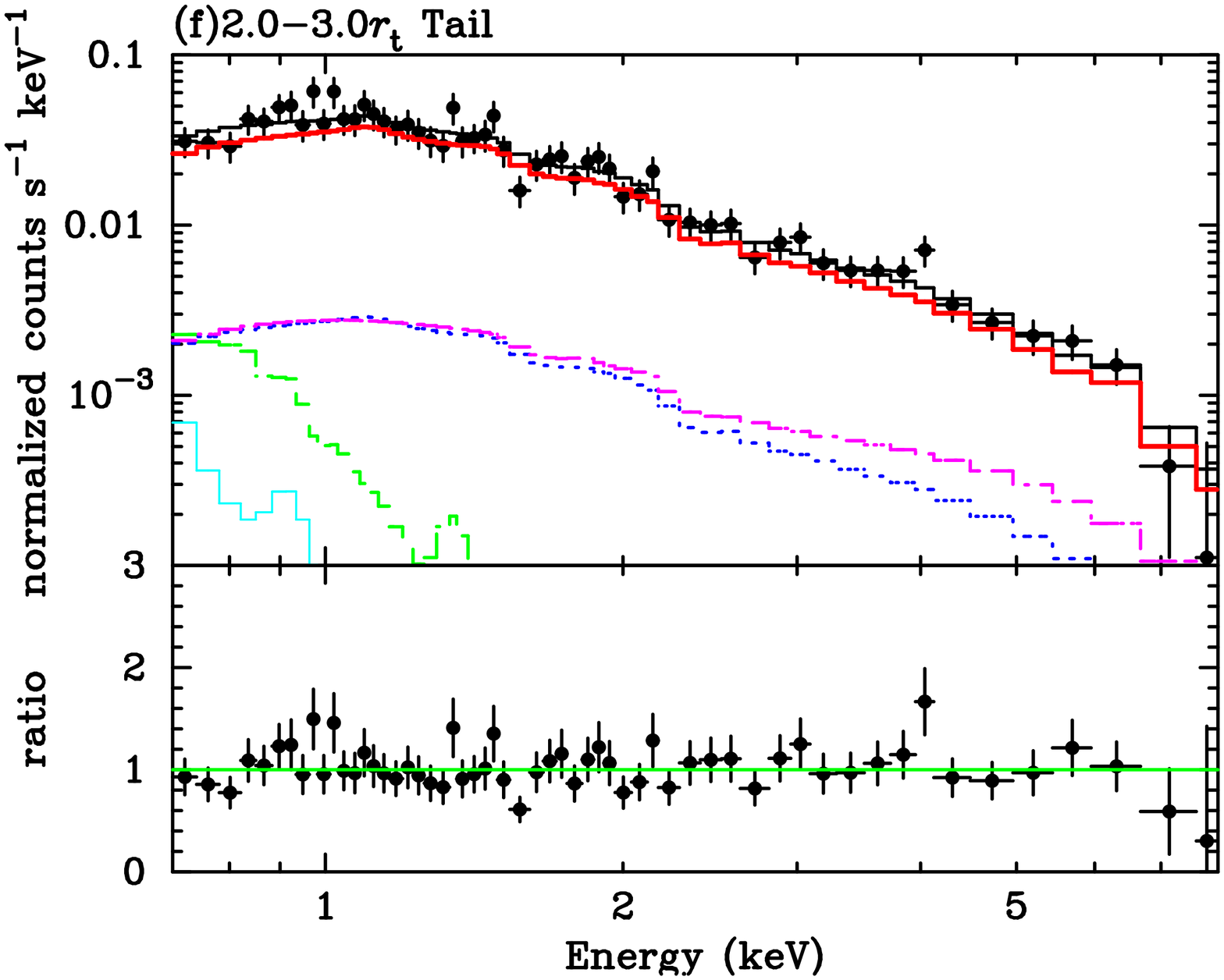}
\caption{
(Top panels) The panels from left to right panels  are the ``NE", ``NW", and ``Tail" sectors with $1-2~r_t$ annular region. 
The spectra were rebinned here for display purposes only.
Upper panels show the NXB subtracted XIS~1 spectra (black crosses).  
The subhalo component is plotted as a (red) bold line.
The ICM, CXB, LHB, and MWH components are indicated by 
(blue) dotted, (magenta) dash-dotted, (cyan) thin, and (green) dashed lines, respectively.
The lower panels show the data-to-model ratios.
(Bottom panels) The same as the top panels, but the radius is $2-3~r_t$.
}
\label{fig:spec2T}
\end{center}
\end{figure*}

\subsection{Temperature and normalization profiles of the subhalo component}
\label{sec:TempandNorm}

\begin{figure*}[tpd]
\begin{center}
\includegraphics[width=0.4\textwidth,angle=0,clip]{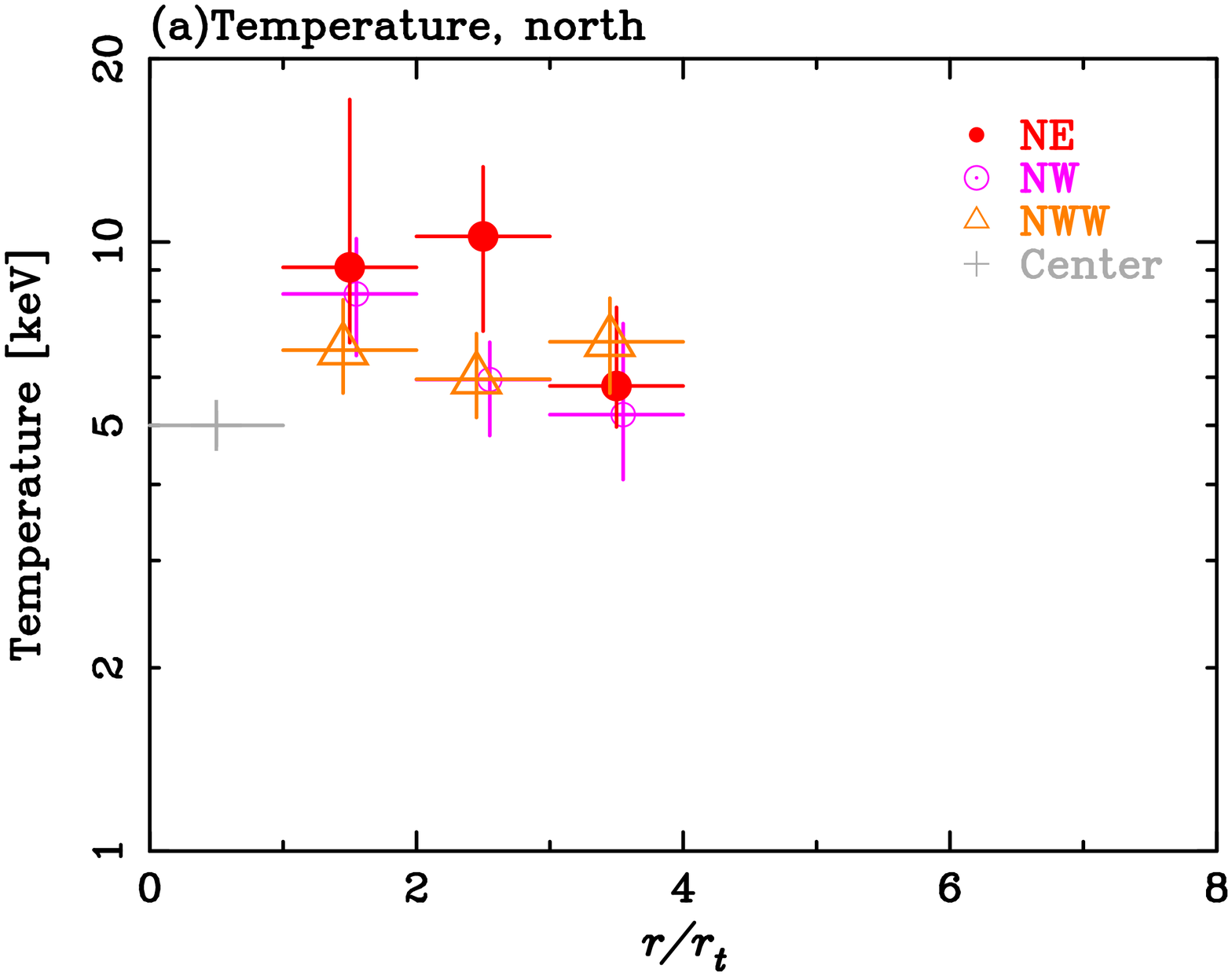}
\includegraphics[width=0.4\textwidth,angle=0,clip]{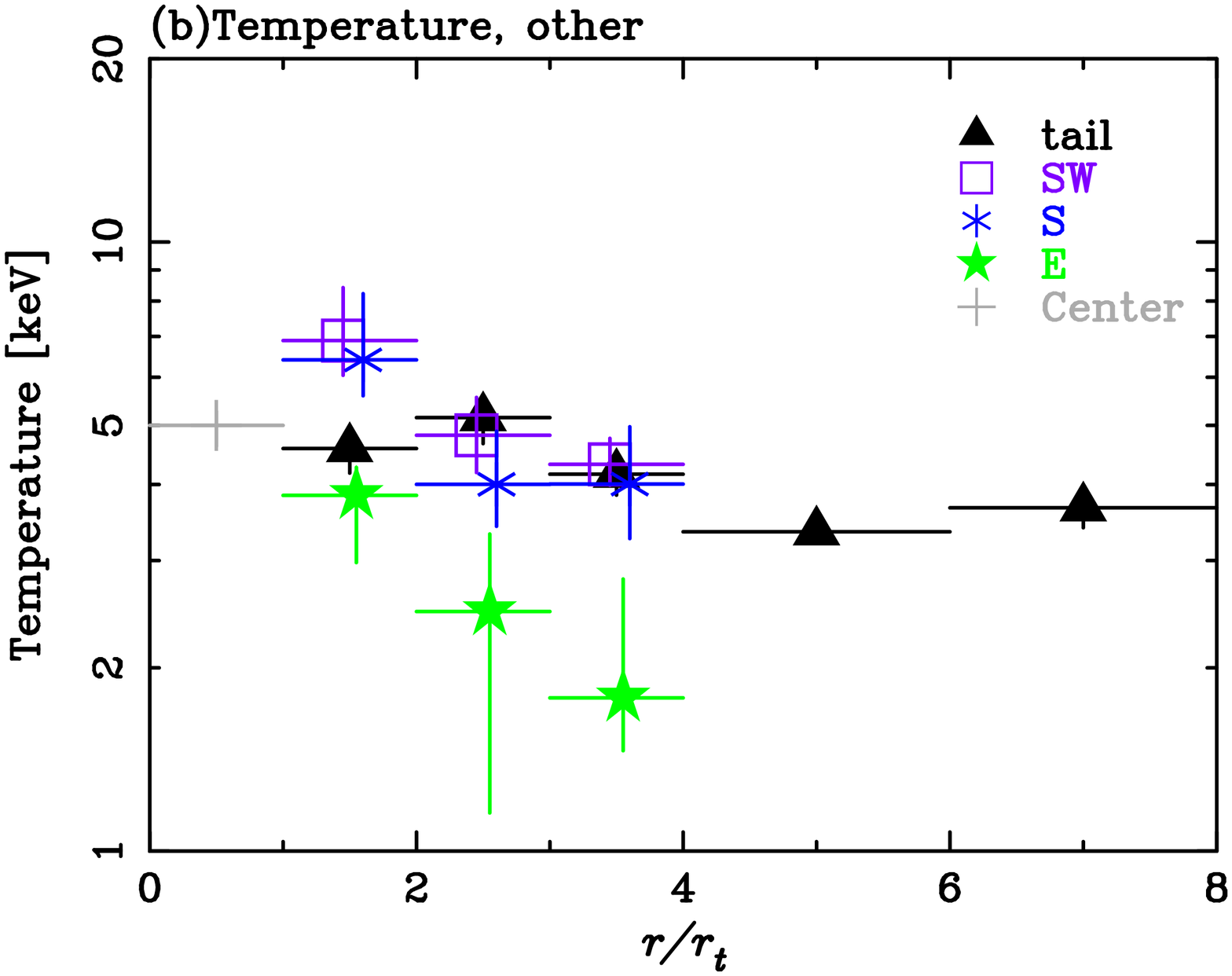}
\includegraphics[width=0.4\textwidth,angle=0,clip]{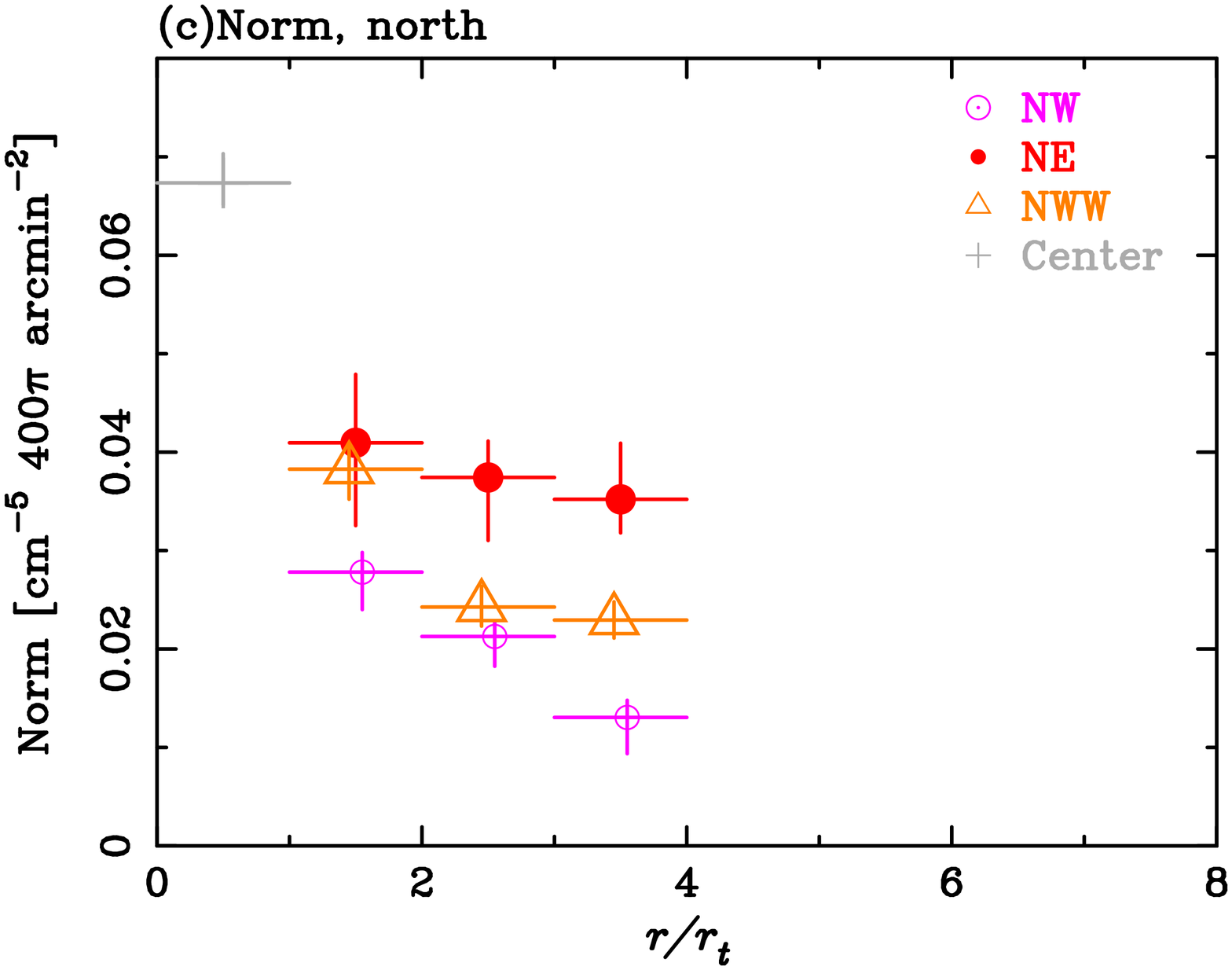}
\includegraphics[width=0.4\textwidth,angle=0,clip]{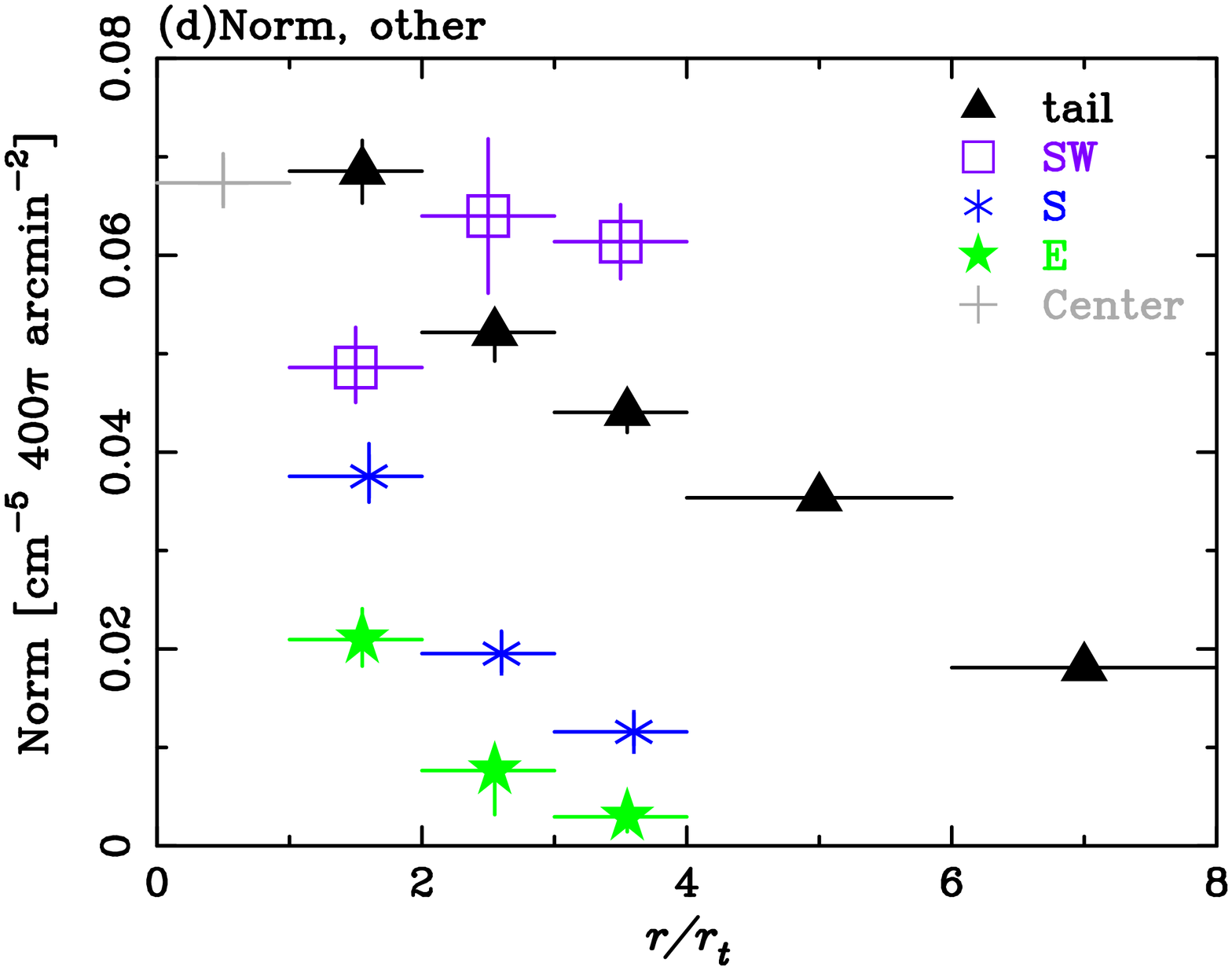}
\caption{
(Top panels) Radial temperature profiles derived from subhalo-model fitting.
The left and right panels are the northern part of the subhalo (labeled ``NE", ``NW", and ``NWW") 
and the other direction, respectively.
The marks and color notations are shown in the figures.
The radius is normalized by $r_t$.
The vertical lines indicate the error range of the fitting result of the ``ID9-BGD" region.
(Bottom panels) The same as the top panels, but showing normalization profiles.
}
\label{fig:result2T}
\end{center}
\end{figure*}

\begin{figure*}[htpd]
\begin{center}
\includegraphics[width=0.43\textwidth,angle=0,clip]{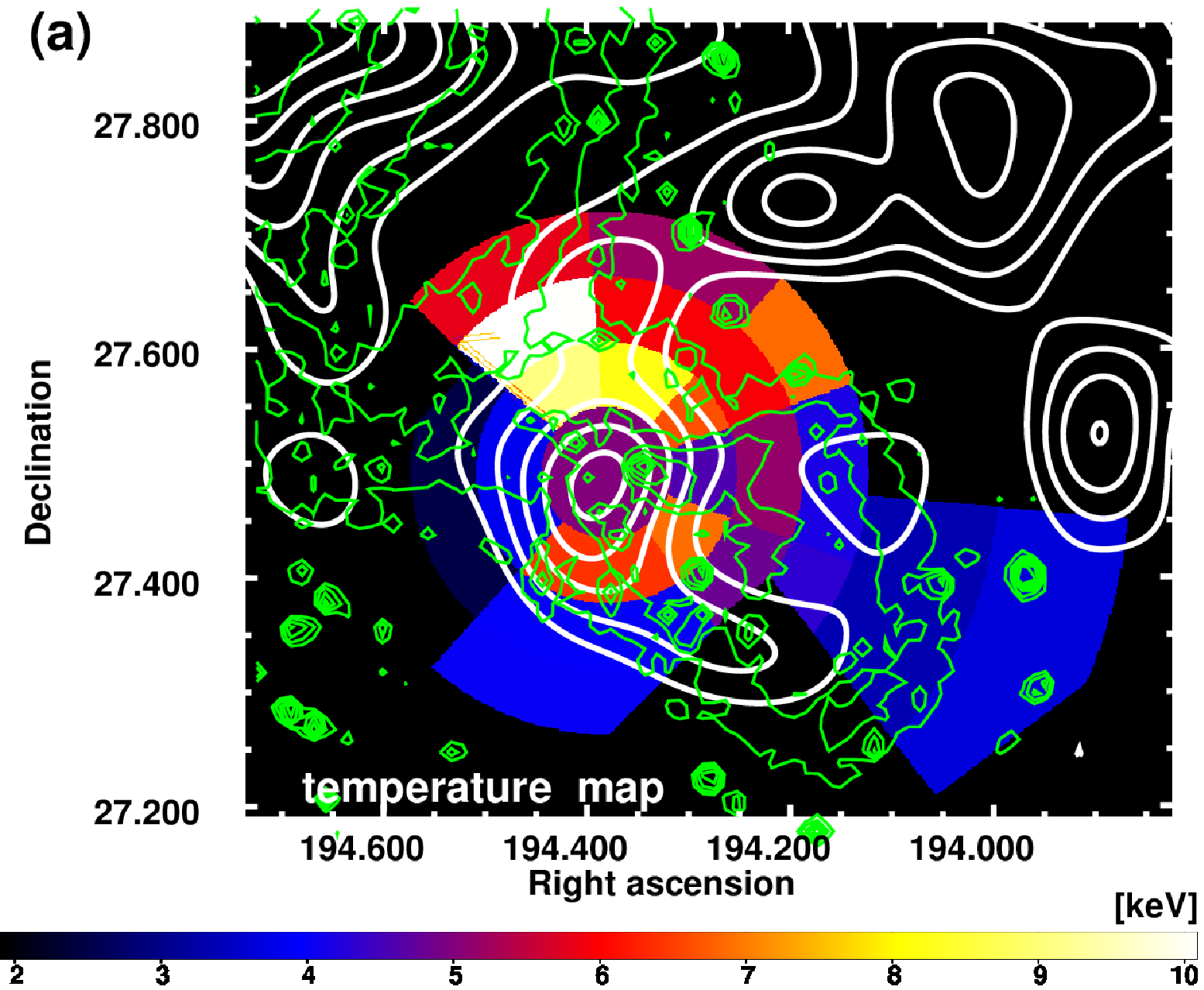}
\includegraphics[width=0.43\textwidth,angle=0,clip]{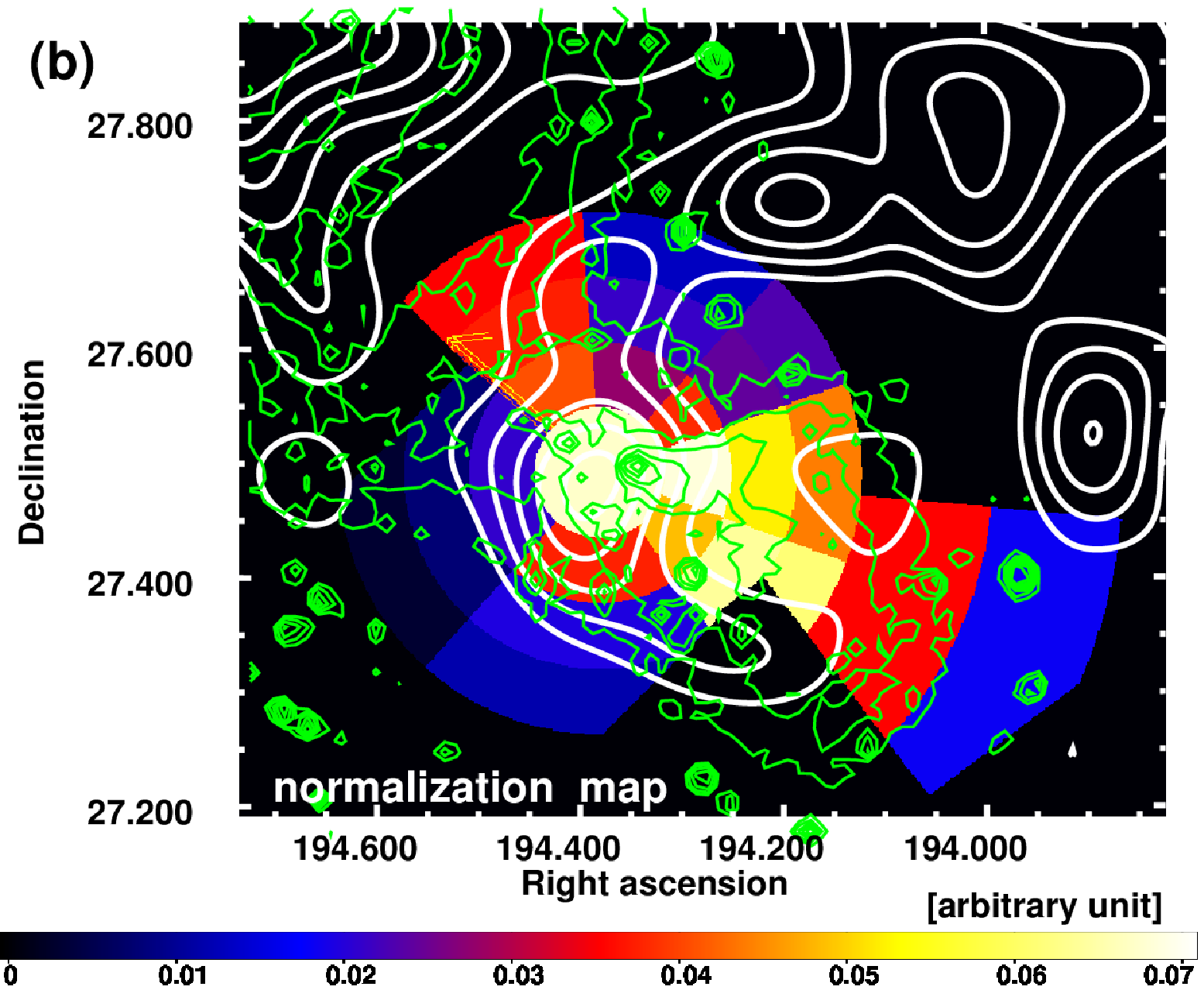}
\caption{
(a)Temperature map and (b) normalization map of the subhalo component.
The unit of normalization is arbitrary. 
The thin and bold contours are surface brightnesses obtained from the XMM-Newton public image 
in figure \ref{fig:imageXMM} (a) and mass contour from weak-lensing analysis \citep{Okabe2014}, respectively. 
}
\label{fig:2TMap}
\end{center}
\end{figure*}

The radial profiles of the temperatures along the individual directions and 
temperature map 
are shown in the top panel of figure \ref{fig:result2T} and figure \ref{fig:2TMap} (a), respectively. 
The temperature within $r_t$ is 5~keV, 
While the temperatures at $1-2~r_t$ in the ``NE", ``NW", ``SW", and ``S"  sectors increase up to 8--10~keV,
these in the other sectors are 4--7~keV. 
The temperature profile of the ``Tail" sector is flat at $\sim$5~keV up to $4~r_t$, 
and about $\sim$3~keV beyond this radius. 
In the ``E" sector, the temperature is likely to be decreasing with the radius, albeit with large error bars. 
\citet{Neumann2001} studied the temperature structure around the NGC~4839 group, 
and obtained $8.0^{+5.0}_{-2.4}$~keV (90\% confidence level) in  0.5--1.5 $r_t$ of the ``SW" sector.
\citet{Akamatsu2013b} also reported the temperature profiles of southwest and north areas of the NGC~4839 group. 
We note that southwestern and northern areas in earlier works in \citet{Akamatsu2013b} coincides with the "Tail" and "SW" sectors and the "NE", "NW" and "NWW" sectors respectively. 
Therefore, our spatial resolution of  temperature measurements is finer than those of \citet{Akamatsu2013b}. 
They reported the temperature profiles of southwest areas within 10$\arcmin$ ($\sim 3~r_t$) from the NGC~4839 group.
Our profiles in ``Tail'' region are consistent with that of southwest area in \citet{Akamatsu2013b}.
Using the same data with our work, 
the best-fit temperature of a north area derived \citet{Akamatsu2013b} was $\sim$7~keV, 
and consistent with our results. 

The radial profiles and map  of normalizations are shown 
in the bottom panels of figure \ref{fig:result2T} and figure \ref{fig:2TMap} (b), respectively. 
We find that the normalizations in the ``NE" sector are roughly constant.  
Conversely, the normalizations of the sectors other than "SW" decreased with radius.
In particular, the gradient of the ``S" and ``E" sectors were steeper than that of the other sectors.
The normalization in the ``Tail" sector smoothly decreases with the radius out to $6~r_t$, 
and the normalization of $6-8~r_t$ is two times lower than that of $4-6~r_t$.

\begin{figure*}[htpd]
\begin{center}
\includegraphics[width=0.3\textwidth,angle=0,clip]{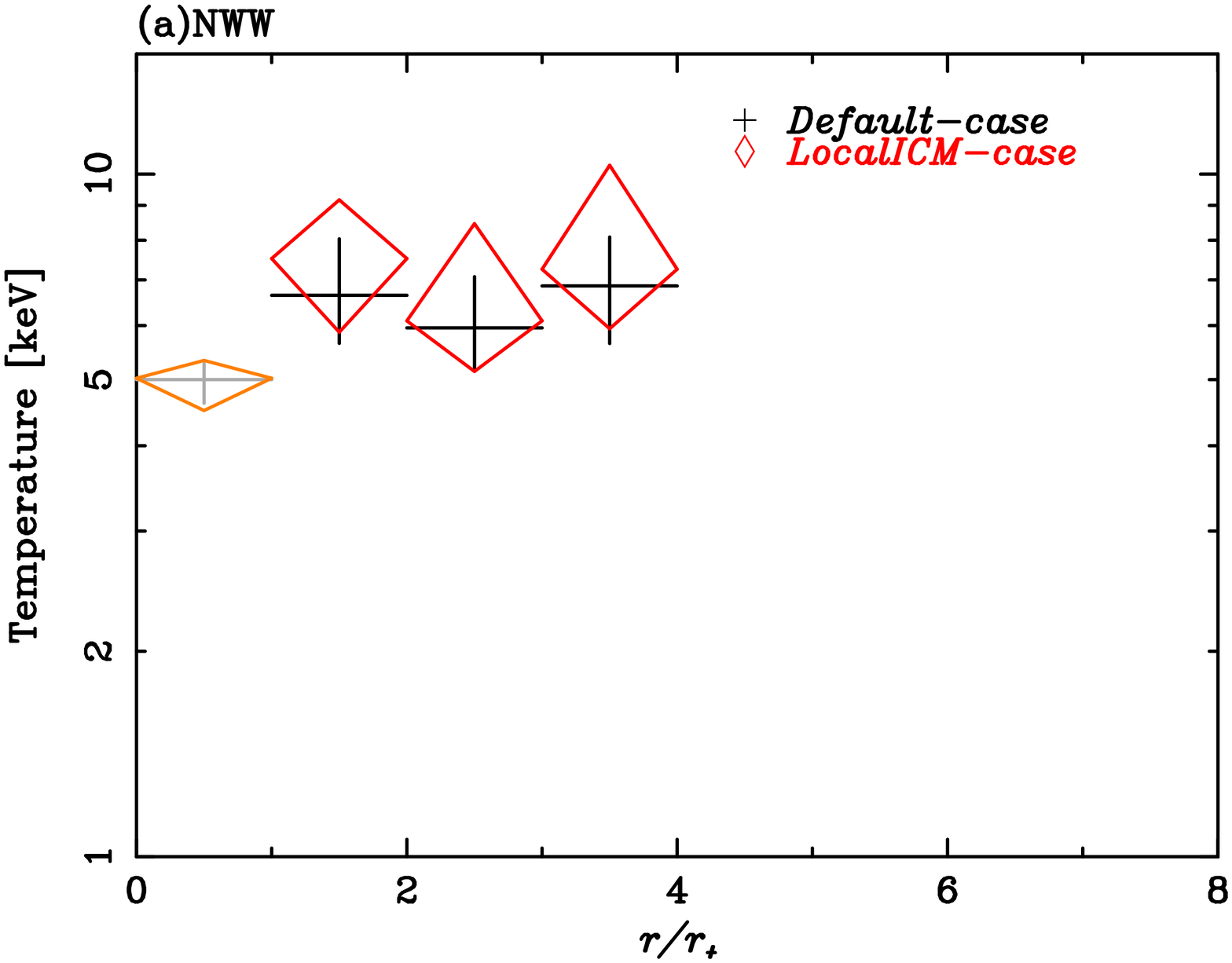}
\includegraphics[width=0.3\textwidth,angle=0,clip]{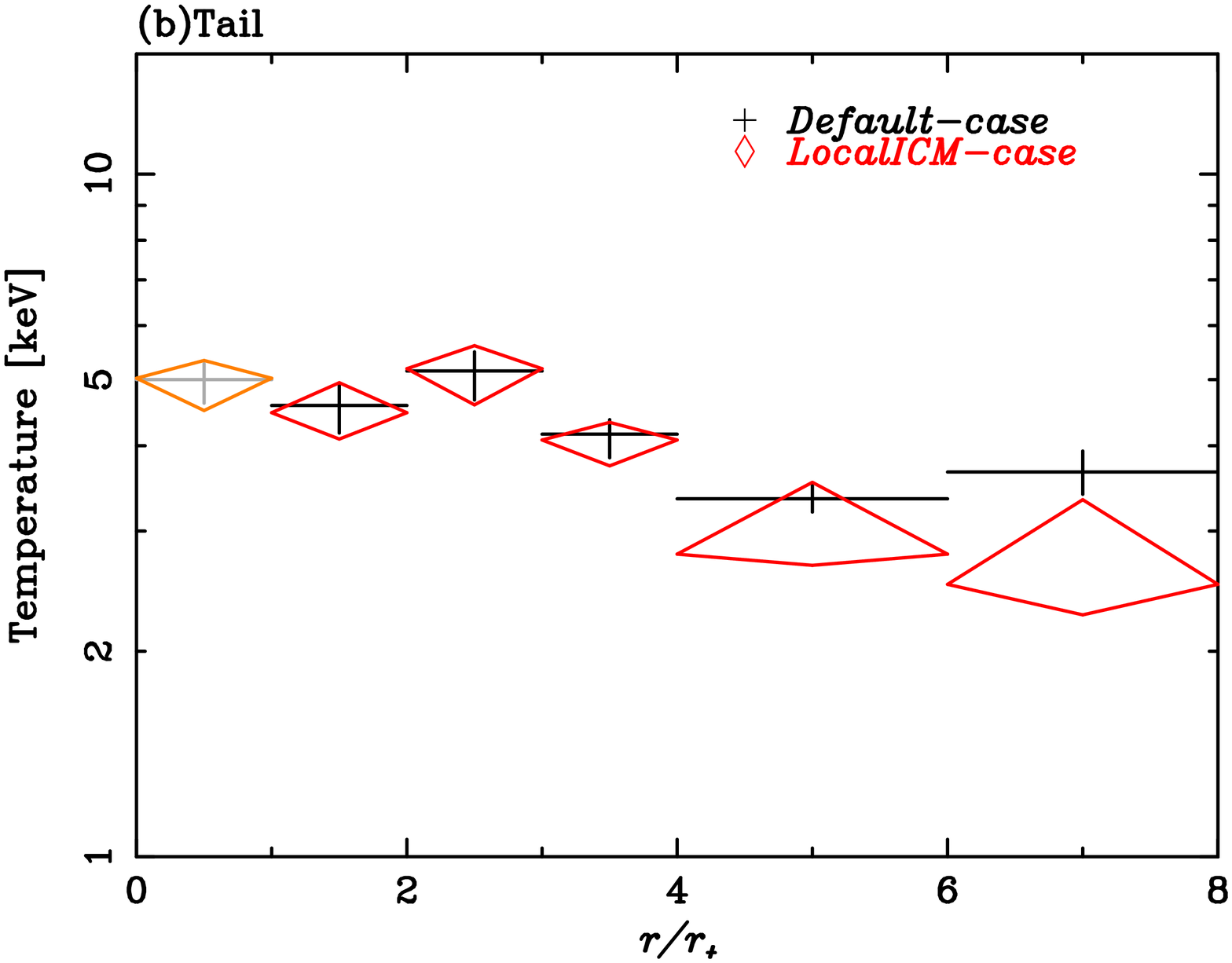}
\includegraphics[width=0.3\textwidth,angle=0,clip]{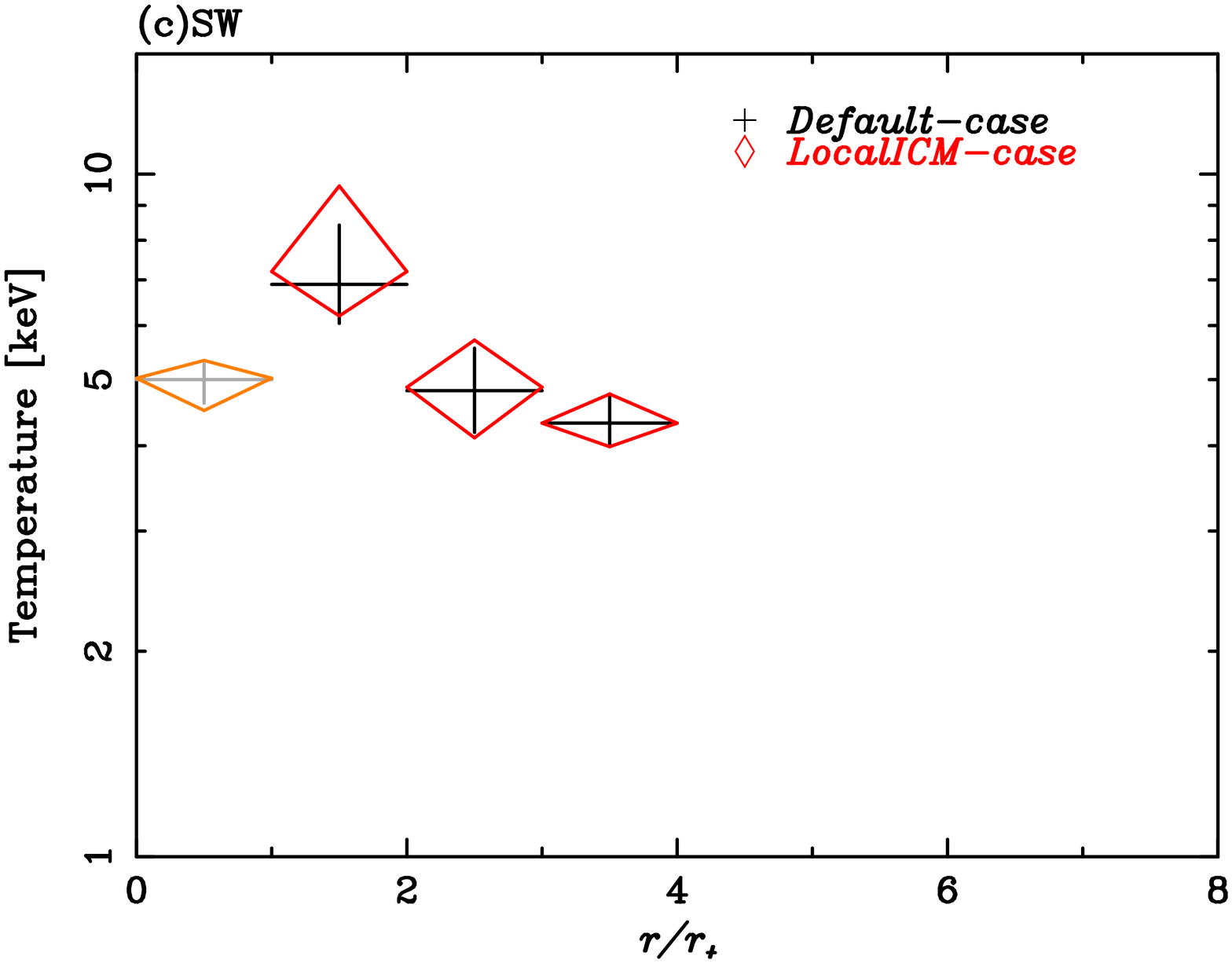}
\includegraphics[width=0.3\textwidth,angle=0,clip]{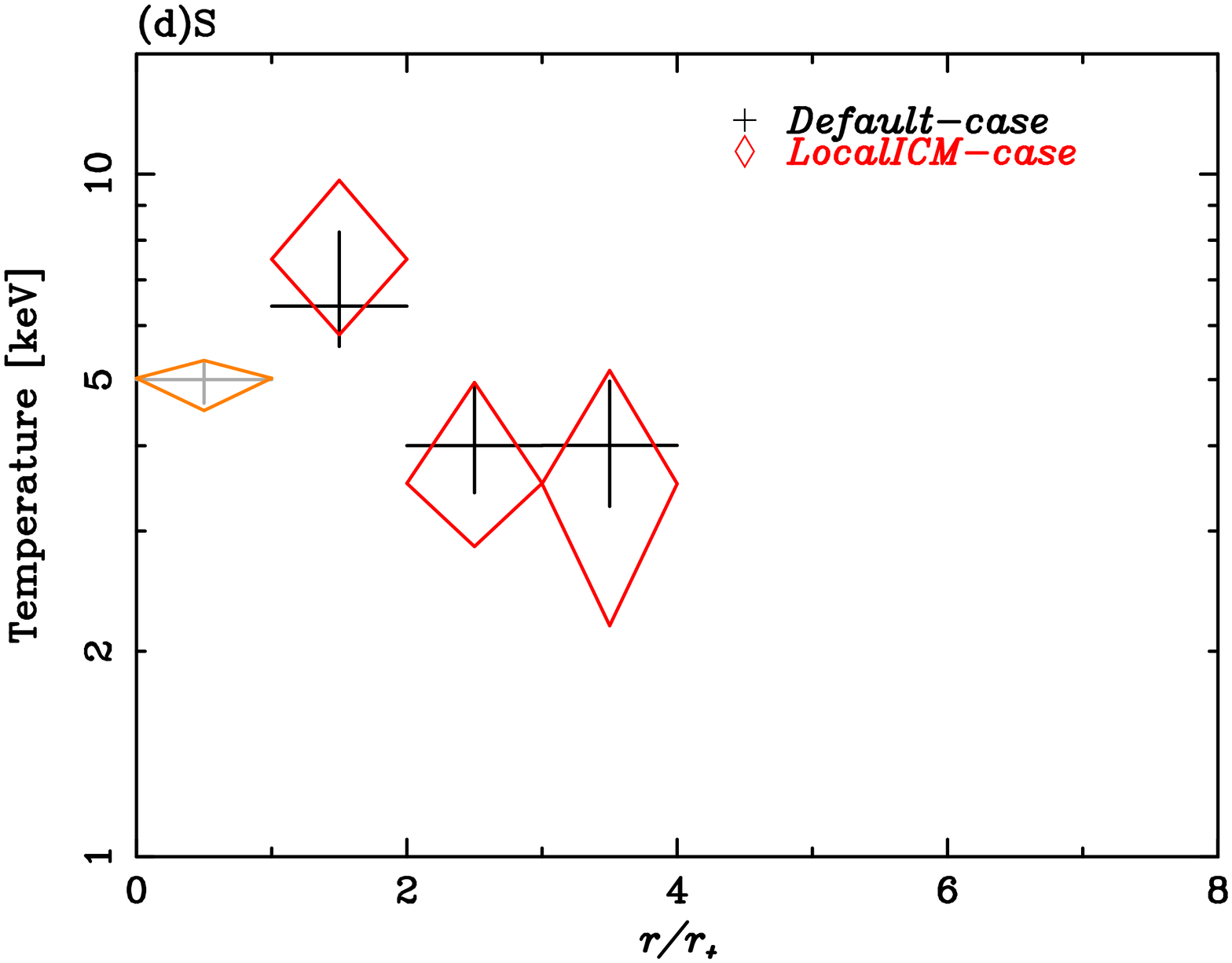}
\caption{
Radial temperature profiles of the subhalo component along the each direction. 
The radius is normalized by $r_t$.
The crosses are the results derived from the ICM parameters estimated from \citet{Simionescu2013} ({\it Default}-case).
The diamonds indicate the results with ICM parameters with {\it LocalICM}-case.
For details, see text. 
}
\label{fig:temp2Tsys}
\end{center}
\end{figure*}

\begin{figure*}[htpd]
\begin{center}
\includegraphics[width=0.3\textwidth,angle=0,clip]{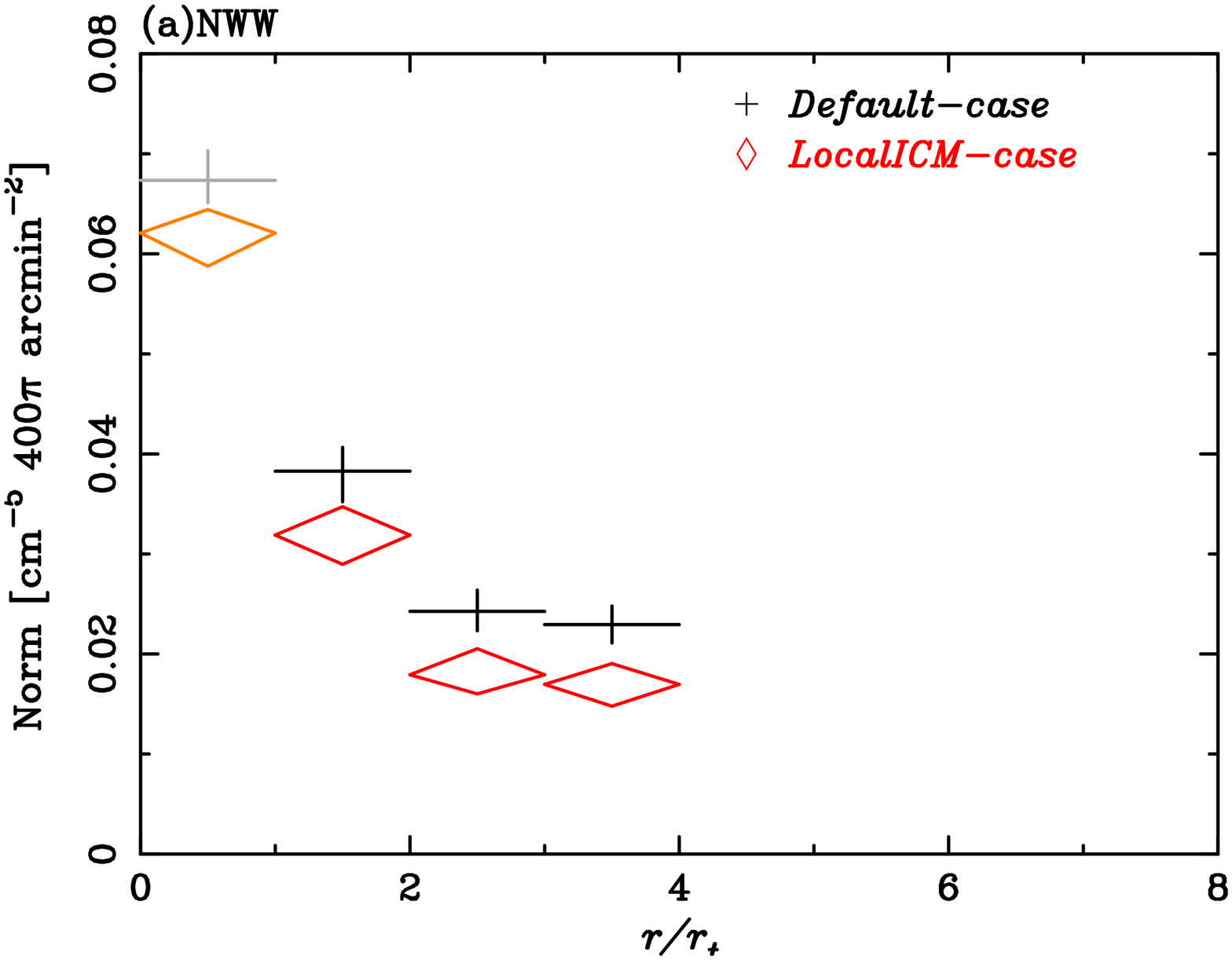}
\includegraphics[width=0.3\textwidth,angle=0,clip]{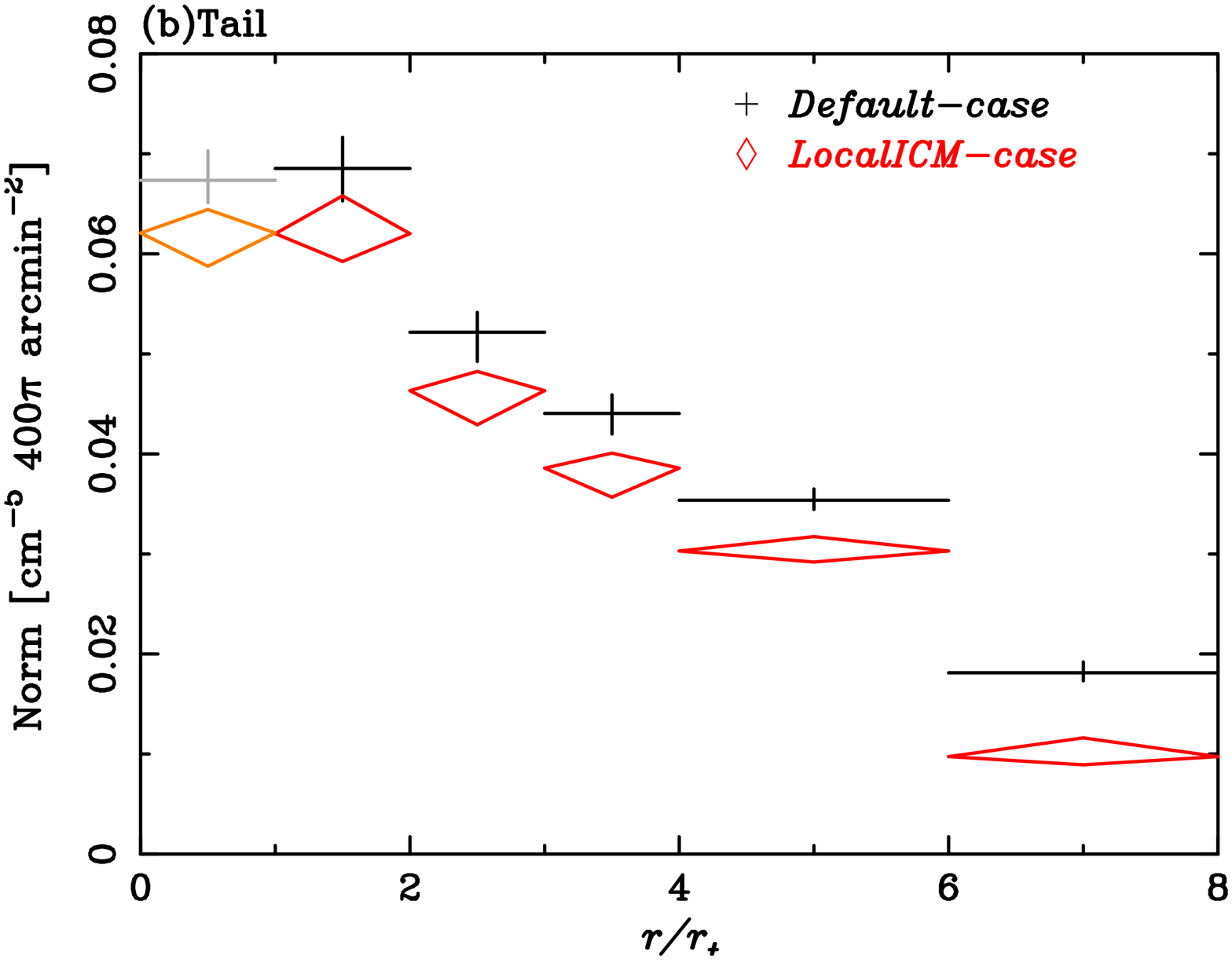}
\includegraphics[width=0.3\textwidth,angle=0,clip]{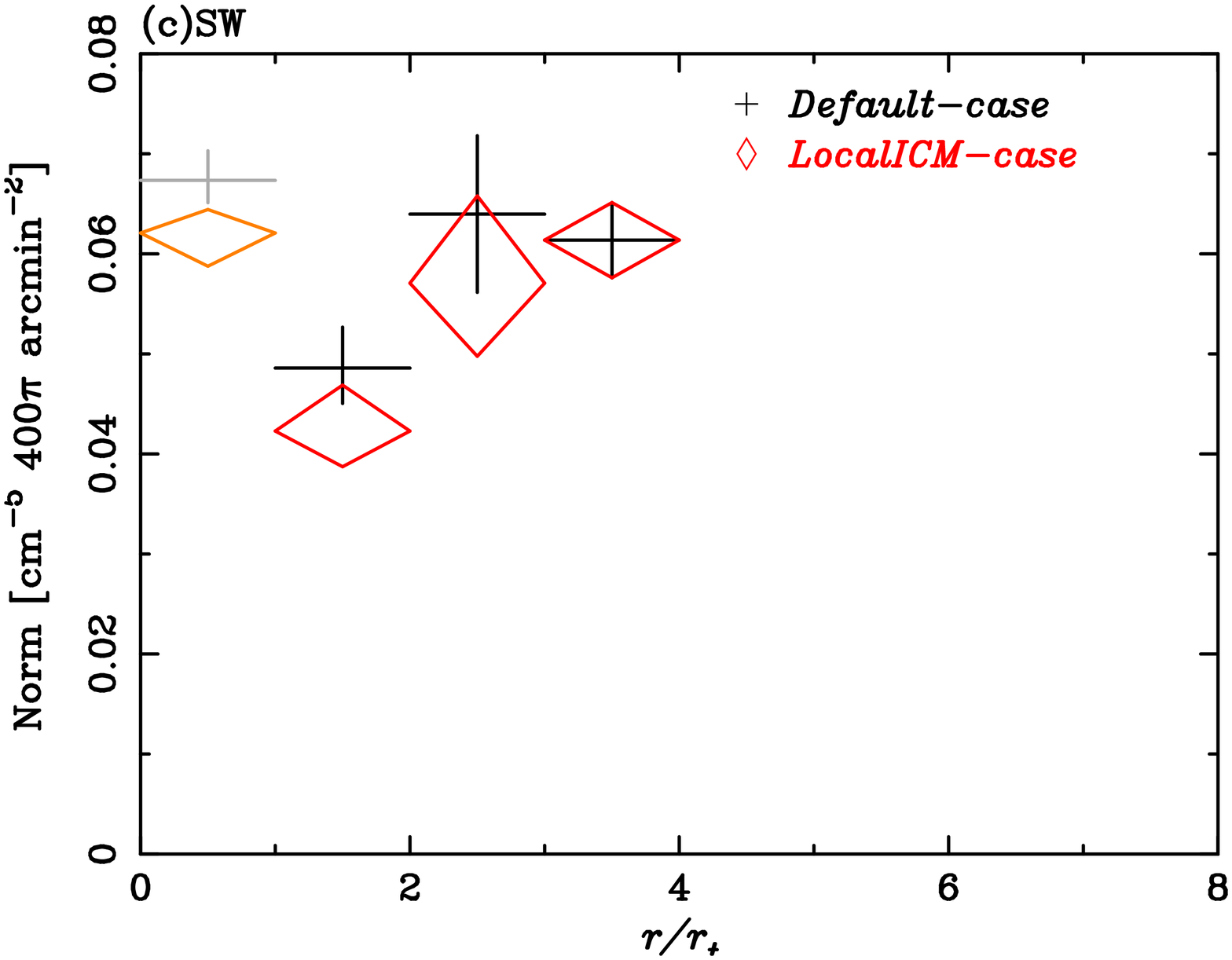}
\includegraphics[width=0.3\textwidth,angle=0,clip]{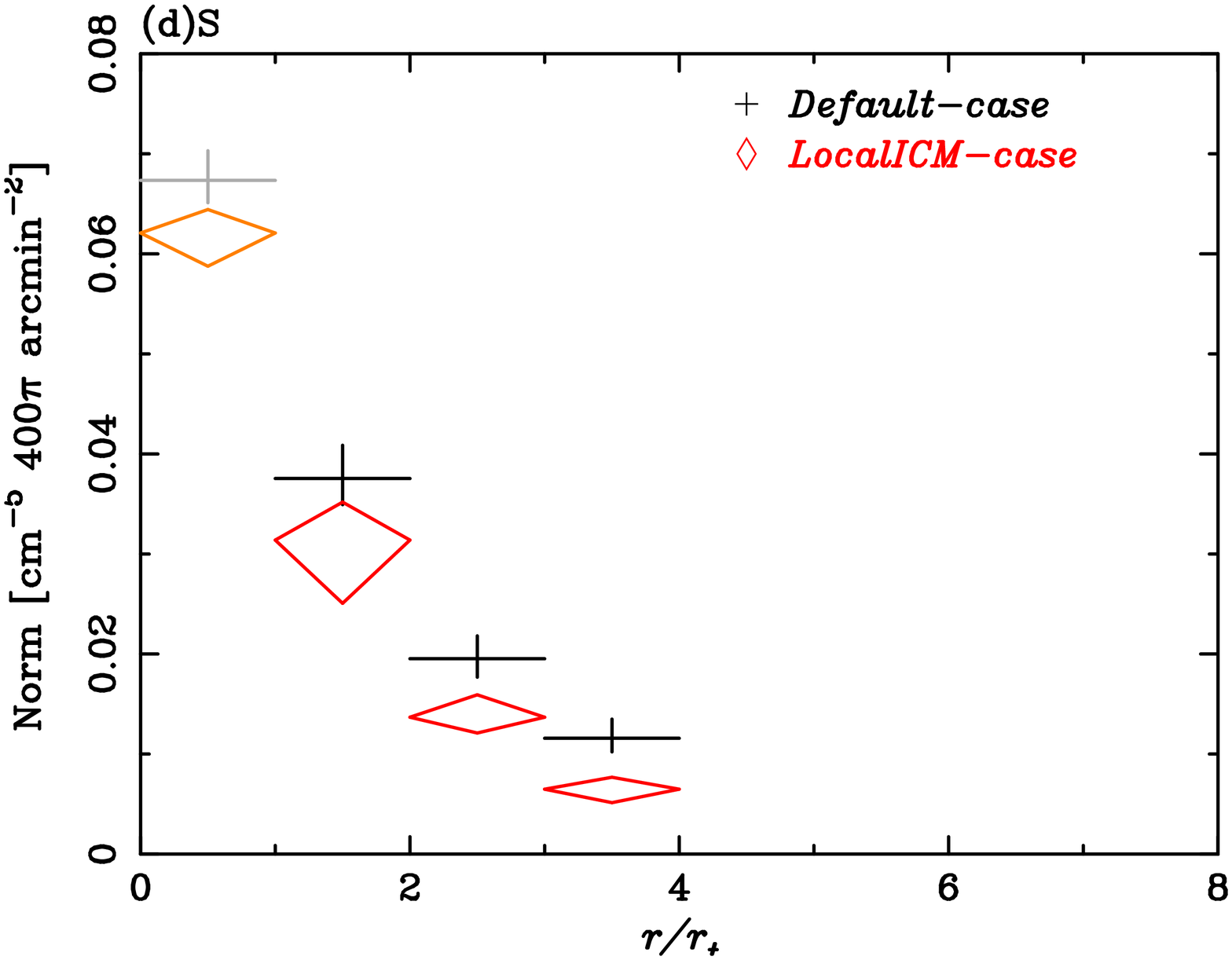}
\caption{
The  same as figure \ref{fig:temp2Tsys}, but for normalization profiles.
}
\label{fig:norm2Tsys}
\end{center}
\end{figure*}

\subsection{Systematic uncertainties by the ICM subtraction}

The radial profiles of the temperature and normalization
of the  ICM in the Coma cluster show  significant azimuthal variations
\citep{Simionescu2013}.
As plotted in Figure \ref{fig:ICMprofile}, the ICM around 
the subhalo "ID~1" derived by \citet{Sasaki2015} (the ID1-BGD region 
plotted in Figure \ref{fig:regions}),
has significantly higher temperature and normalization
than the azimuthally averaged values by \citet{Simionescu2013}.
Therefore, to study the local enhancement of the subhalo, "ID~9"
compared to the surrounding ICM component, 
we extracted the spectra from the observation of subhalo ``ID~2", 
which shows no significant excess emission as reported by \citet{Sasaki2015}.
As shown  in figure \ref{fig:regions}, the field-of-view of the ``ID~2" observation 
was divided into two regions; within and beyond 50$\arcmin$ 
from the center of the Coma cluster, (R.A., Decl.) = ($\timeform{12h59m44.81s}, \timeform{27D56'49.92"}$).
These regions were labeled as ``ID2 within 50\arcmin" and ``ID2 beyond 50\arcmin".
We fitted the spectra extracted from these two regions using 
the same background and foreground parameters described in section \ref{sec:bgd}.
Here, the ICM abundance was fixed at 0.3 solar.
We also fitted the spectra by changing the abundance of the ICM to be 0.2 solar,
 and obtained almost the same temperatures and normalizations.
The resultant ICM parameters are summarized in table \ref{tb:bgd} and shown in figure \ref{fig:ICMprofile}.
The derived temperatures and normalizations from 
"ID2" regions smoothly connect the azimuthally-averaged profiles
at 40$\arcmin$ from the cluster center.
However,  beyond this radius, the temperatures and normalizations 
from the  "ID2" and "ID1-BGD"  regions are nearly flat with radius,
and significantly higher than the azimuthally averaged profiles.

We fitted the subhalo "ID~9" spectra which were extracted from  the regions  
beyond 40$\arcmin$ from the cluster center in the same way as in 
section \ref{sec:excess}, but using the ICM parameters derived from 
the two "ID~2" and "ID~1" observations (hereafter {\it LocalICM}-case). 
Figure \ref{fig:temp2Tsys} and \ref{fig:norm2Tsys} show
the derived temperatures and normalizations, respectively,
 for the sectors "NWW", "Tail", "SWW", and "S" and the "Center" region, 
 which are located beyond 40$\arcmin$ from the cluster center.
The temperatures did not change within statistic error ranges excluding 6--8$~r_t$ in the ``Tail" sector.
The normalizations for the {\it LocalICM}-case are slightly smaller than 
these with {\it Default}-case in figure \ref{fig:norm2Tsys}. 
The difference is caused by the increase the ICM normalizations of {\it LocalICM}-case, 
which is several times higher than that of {\it Default}-case as shown in figure \ref{fig:ICMprofile} (b). 

\begin{table}
  \begin{center}
  \caption{Fitting results of the ICM component at the {\it Suzaku} BGD regions.}
  \label{tb:bgd}
    \begin{tabular}{llll}
      \hline
  Field name & Distance\footnotemark[$*$] [$\arcmin$] &$kT$ [keV] & Normalization\footnotemark[$\dagger$] \\ \hline 
   ID2 within 50\arcmin    & $40\arcmin-50\arcmin$ & $5.45^{+0.61}_{-0.46}$ & $10.6^{+0.47}_{-0.69}$ \\
   ID2 beyond 50\arcmin & $50\arcmin-60\arcmin$ & $5.43^{+0.67}_{-0.48}$ &  $8.31^{+0.58}_{-0.34}$ \\
      \hline
 \multicolumn{3}{@{}l@{}}{\hbox to 0pt{\parbox{85mm}{\footnotesize
      \par\noindent
      \footnotemark[$*$] Distance from the Coma cluster center. \\
      \footnotemark[$\dagger$] The normalization is defined as $Norm = \int n_{\rm e} n_{\rm H} dV / [4\pi\,(1+z)^2 D_{\rm A}^{~2}] \times 10^{-17}$ cm$^{-5}~ /(20^2 \pi)$~arcmin$^{-2}$. 
       }    \hss}}
    \end{tabular}
  \end{center}
\end{table}

\section{Discussion}
\label{sec:dis}

We analyzed the {\it Suzaku} data from the six pointing observations 
around a massive subhalo "ID~9"  detected by the weak-lensing survey \citep{Okabe2014}, 
which is associated with the NGC~4839 group.
We first discuss the pseudo-pressure and pseudo-entropy profiles around the subhalo in section \ref{sec:press}.
Section \ref{sec:shock} discusses the possible shock heating at the high-temperature region. 
We compare the X-ray properties with those of regular groups in section \ref{sec:kt-m}.
Section \ref{sec:ram} discusses the effect of ram pressure stripping.

\subsection{Pressure and entropy distributions around the subhalo}
\label{sec:press}

\begin{figure}[htpd]
\begin{center}
\includegraphics[width=0.45\textwidth,angle=0,clip]{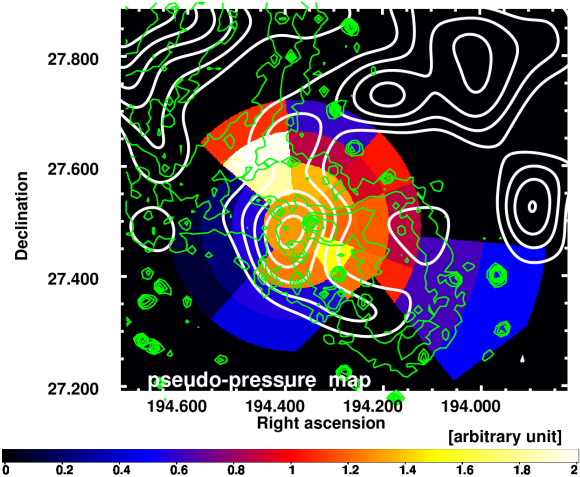}
\caption{
Pseudo-pressure map of the subhalo component. 
The units of the pressure map are arbitrary. 
The thin and bold contours are surface brightness from the {\it XMM-Newton}
and mass contour from weak-lensing analysis \citep{Okabe2014}, respectively. 
}
\label{fig:PMap}
\end{center}
\end{figure}

\begin{figure*}[htpd]
\begin{center}
\includegraphics[width=0.4\textwidth,angle=0,clip]{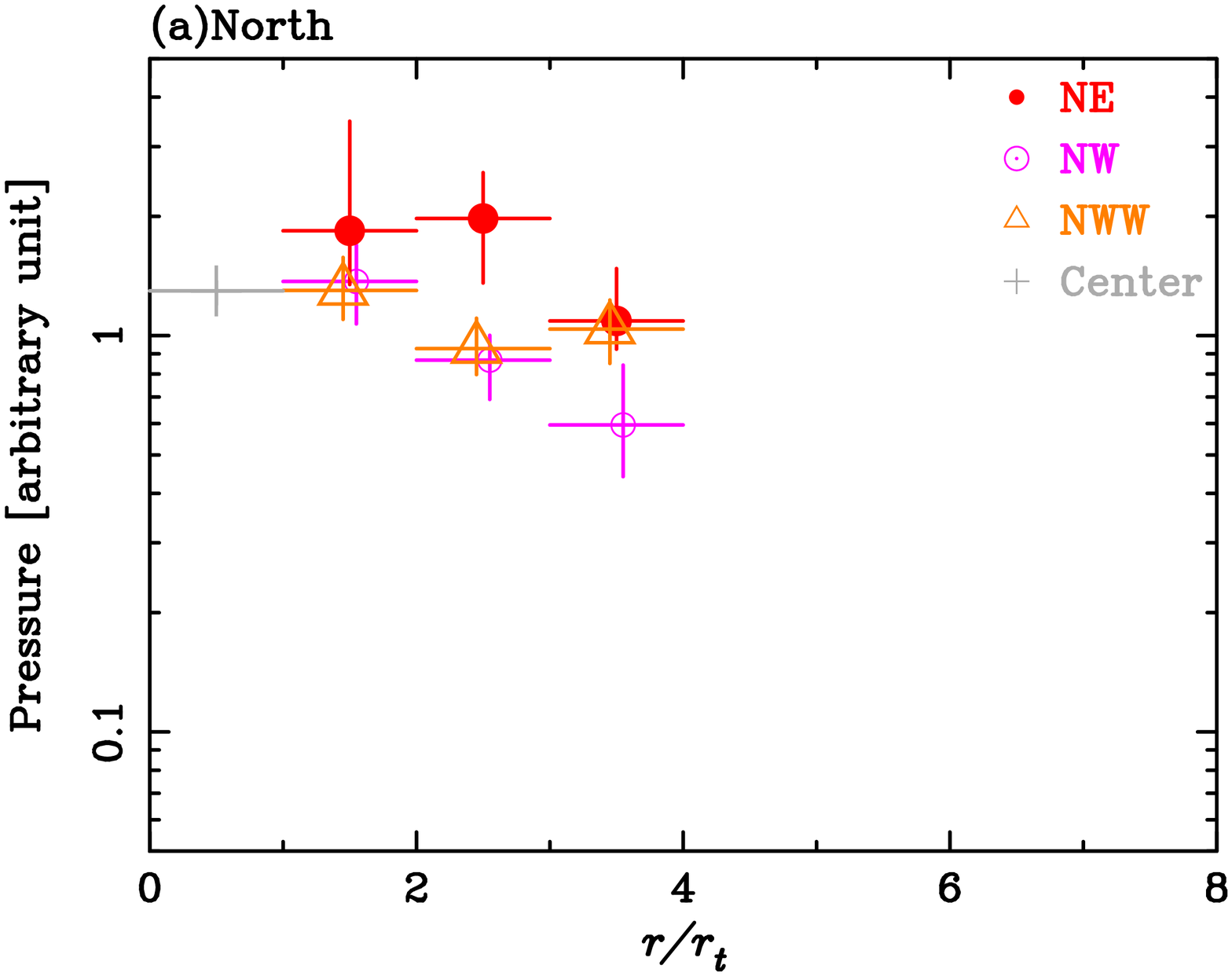}
\includegraphics[width=0.4\textwidth,angle=0,clip]{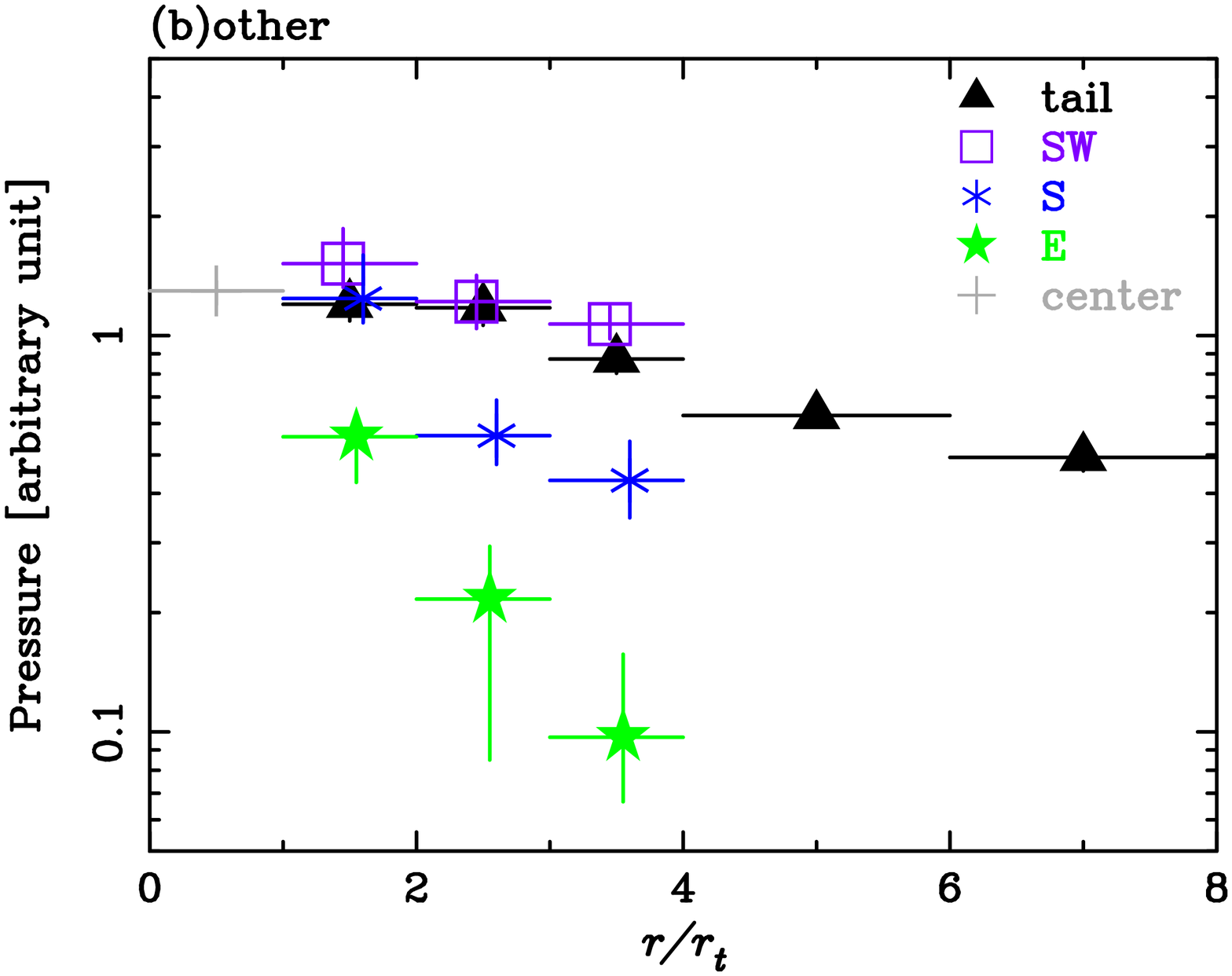}
\caption{
Radial pseudo-pressure profiles of the subhalo component. 
Panel (a) and (b) are the northern parts and the other directions, respectively. 
The radius is normalized by $r_t$.
}
\label{fig:press}
\end{center}
\end{figure*}

To investigate the thermodynamical structure, we study the pseudo-pressure and pseudo-entropy distributions around the subhalo. 
Since the normalization derived from spectral fitting is proportional to the square of the electron density
\footnote{The definition of the normalization is summarized in Xspec manual page;
https://heasarc.gsfc.nasa.gov/xanadu/xspec/manual/XSmodelApec.html}, 
we obtained the density profiles using the square root of the normalizations.
Therefore, thermal pressure, defined as $P=kT~n_e$, is converted to $P\propto kT~norm^{0.5}$.
The resultant pseudo-pressure map and profiles of each sector are shown in figures \ref{fig:PMap} and \ref{fig:press}, respectively.
The pseudo-pressure is nearly continuous around the subhalo ($< 3~r_t$).  
Beyond $r_t$, the pseudo-pressure distributions of the ``NE"  and ``NWW" sectors are nearly flat.
In contrast, at $2~r_t$ of the ``S" and ``E" sectors, the pseudo-pressure is discontinuous; 
pressures of $2-3~r_t$ in the ``S" and ``E" sectors are approximately half of those of $1-2~r_t$, respectively.
In the ``E" sector, however, the temperature at $2~r_t$ is not discontinuous due to large error range.  
The pressure profile of the ``Tail" sector smoothly decreases with radius up to $8~r_t$.

\begin{figure}[htpd]
\begin{center}
\includegraphics[width=0.45\textwidth,angle=0,clip]{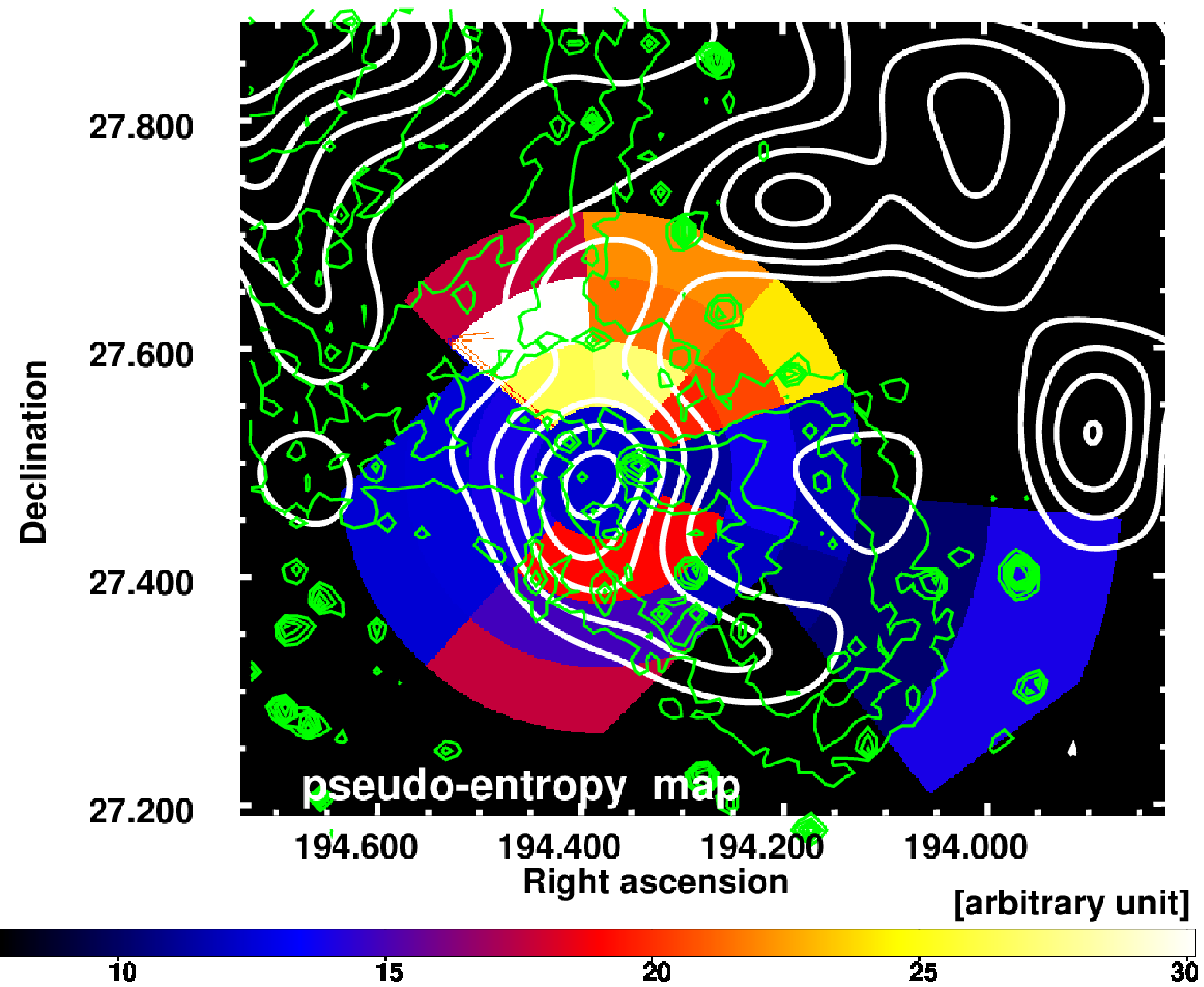}
\caption{
Same as figure \ref{fig:PMap}, but for pseudo-entropy profiles. 
}
\label{fig:EntMap}
\end{center}
\end{figure}
\begin{figure*}[htpd]
\begin{center}
\includegraphics[width=0.4\textwidth,angle=0,clip]{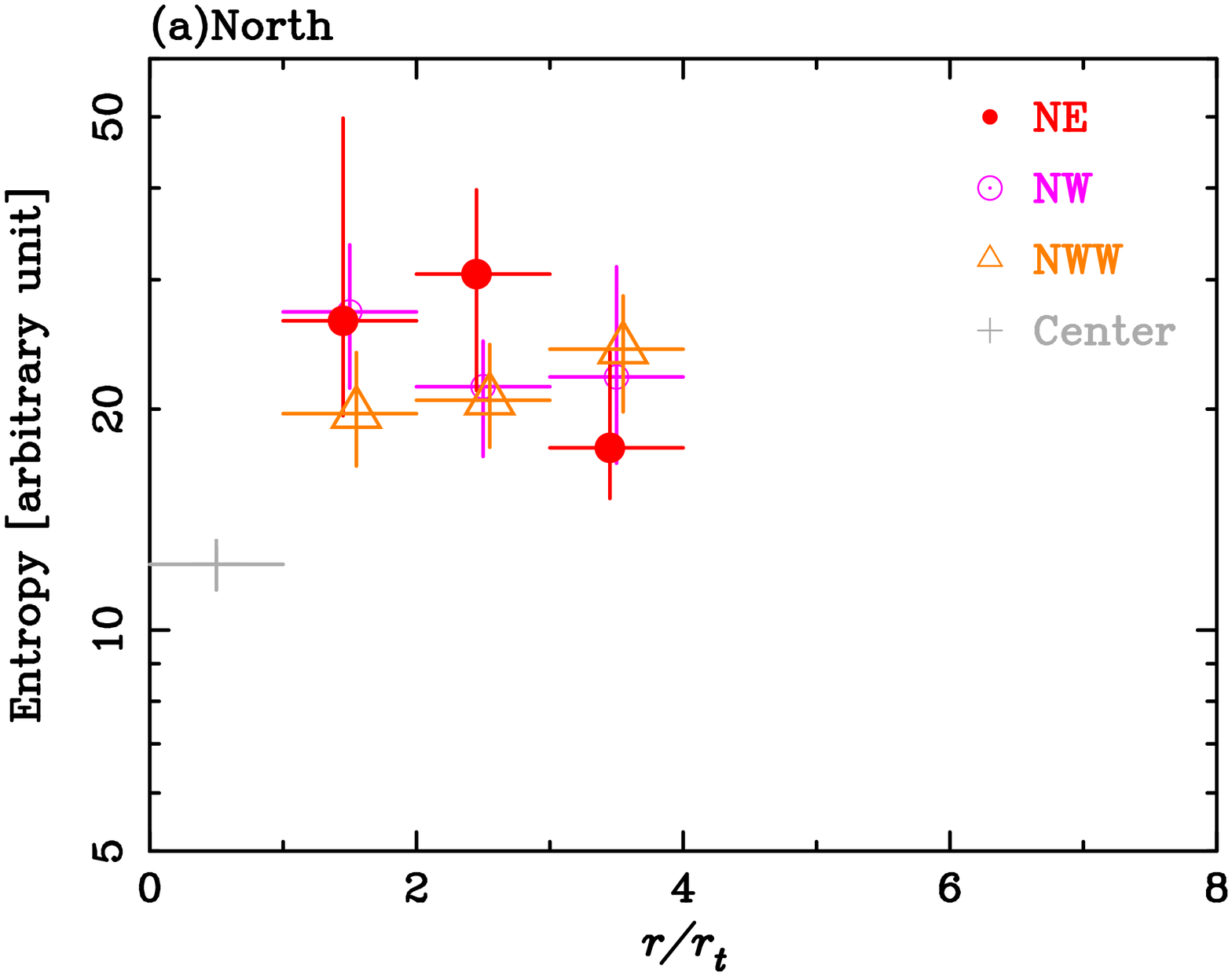}
\includegraphics[width=0.4\textwidth,angle=0,clip]{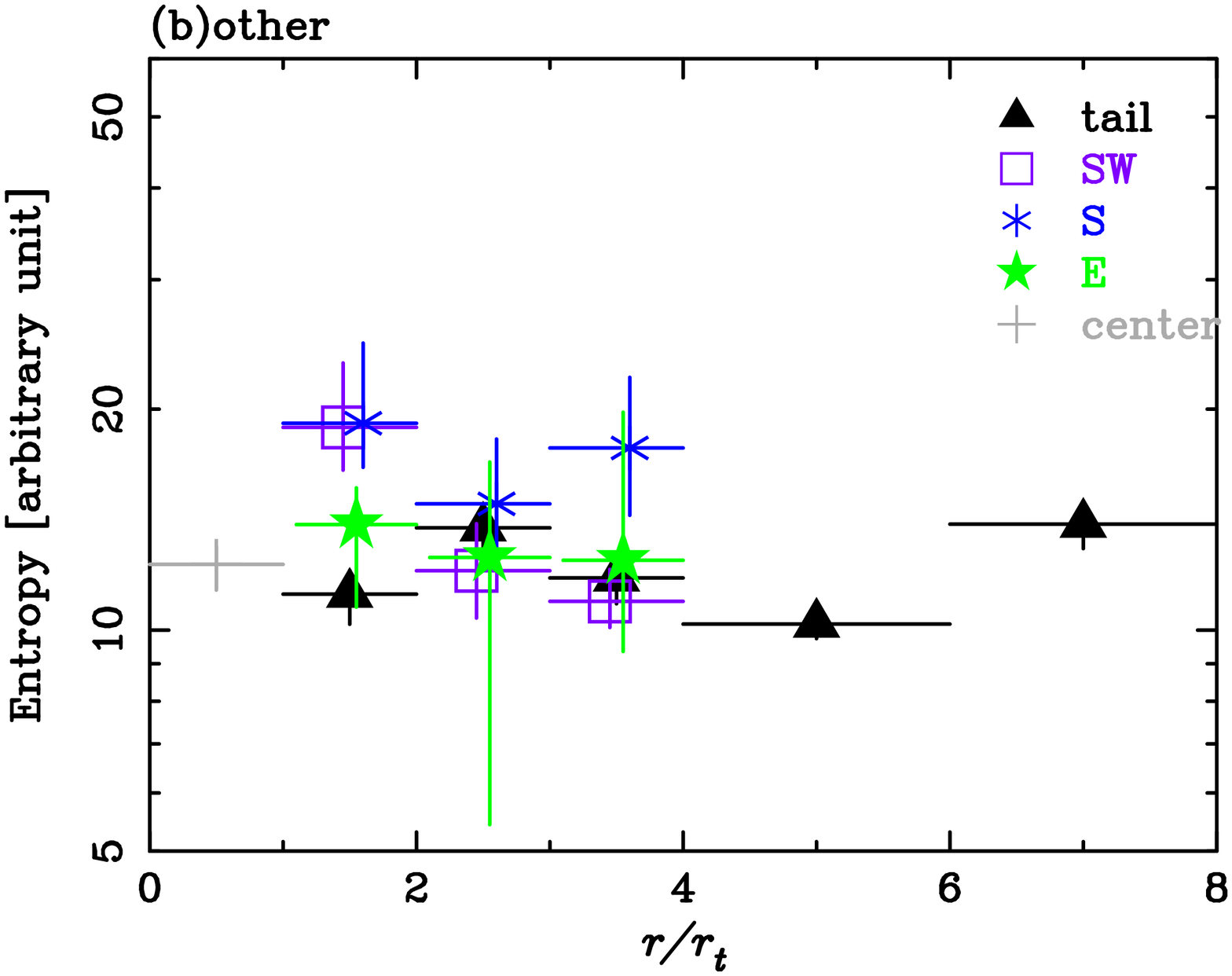}
\caption{
Same as figure \ref{fig:press}, but for pseudo-entropy profiles. 
}
\label{fig:Entropy}
\end{center}
\end{figure*}

The entropy, defined as  $K=kT~n_e^{-2/3}$, is a useful parameter for 
investigating the thermodynamics of galaxy clusters. 
We computed the pseudo-entropy defined as $K\sim kT~norm^{-1/3}$. 
The resultant pseudo-entropy map and profiles are shown in figures \ref{fig:EntMap} and \ref{fig:Entropy}, respectively.
The entropy in the ``Tail'' regions are almost constant and equal to that in the ``Center'' region.
The entropy in the ``NE", ``NW", and ``NWW" sectors are relatively higher than that in the ``Center''. 
Beyond $2~r_t$, the pseudo-entropy in the ``S" and ``SW" sector 
are comparable with that in the ``Tail" sector.
Removal of the low entropy gas due to ram-pressure creates a low entropy tail.
In contrast, the ``NE" and  ``NW" sectors have 2-3 times higher entropies than that in the ``Center".
Entropy in the ``NWW" sectors also is about two times higher than that in the ``Center". 

\subsection{Possible shock heating around the subhalo}
\label{sec:shock}

In order to study the possible shock heating, we calculated the Mach number $\cal M$
 under the Rankine-Hugoniot condition using the following formula; 
\begin{equation} 
\frac{T_2}{T_1} = \frac{5{\cal M}^4 + 14{\cal M}^2 - 3}{16{\cal M}^2}.
\end{equation}
Here, the subscripts 1 and 2 indicate pre-shock and post-shock values, respectively. 
We have assumed a specific heat ratio of $\gamma=5/3$.

\citet{Akamatsu2013b} found temperature jump at the radio relic caused by the accretion of the NGC~4839 group.
The Much number of the shock at the radio relic is about 2.
At the $2~r_t$ in the ``S" sector, we found the temperature jump from $\sim$7~keV to $\sim$4 keV. 
Adopting the temperatures of $1-2~r_t$ and $2-3~r_t$ in the ``S" sector, the resultant Mach number is $1.60^{+0.64}_{-0.48}$.
At the $3~r_t$ in the ``NE" sector, there is also possible temperature jump although the statistic error is too large to make sure. 
Adopting the temperature of $2-3~r_t$ and $3-4~r_t$, the Mach number is $1.74^{+0.68}_{-0.82}$. 
Due to high-temperature and complex emissions of the excess structure,  
we could not detect the shock heated gas significantly.
Although our results have large error bars, there is no discrepancy with \citet{Akamatsu2013b}.
In order to detect the shock heated gas clearly, 
hard X-ray spectrometers like {\it NuSTAR} should observe around the NGC~4839 groups.

\subsection{Comparison of X-ray properties with other regular groups and clusters}
\label{sec:kt-m}

To compare the X-ray properties with those of the regular groups and clusters,
we first calculated the mean density within $r_t$.
The resultant value is approximately 30000 times greater than the critical density of the Universe.
With Chandra's results \citep{Vikhlinin2006}, we computed the $kT-M$ and $M_{gas}-M$ relations at the same overdensity as that of the subhalo. 

\begin{figure*}[htpd]
\begin{center}
\includegraphics[width=0.5\textwidth,angle=0,clip]{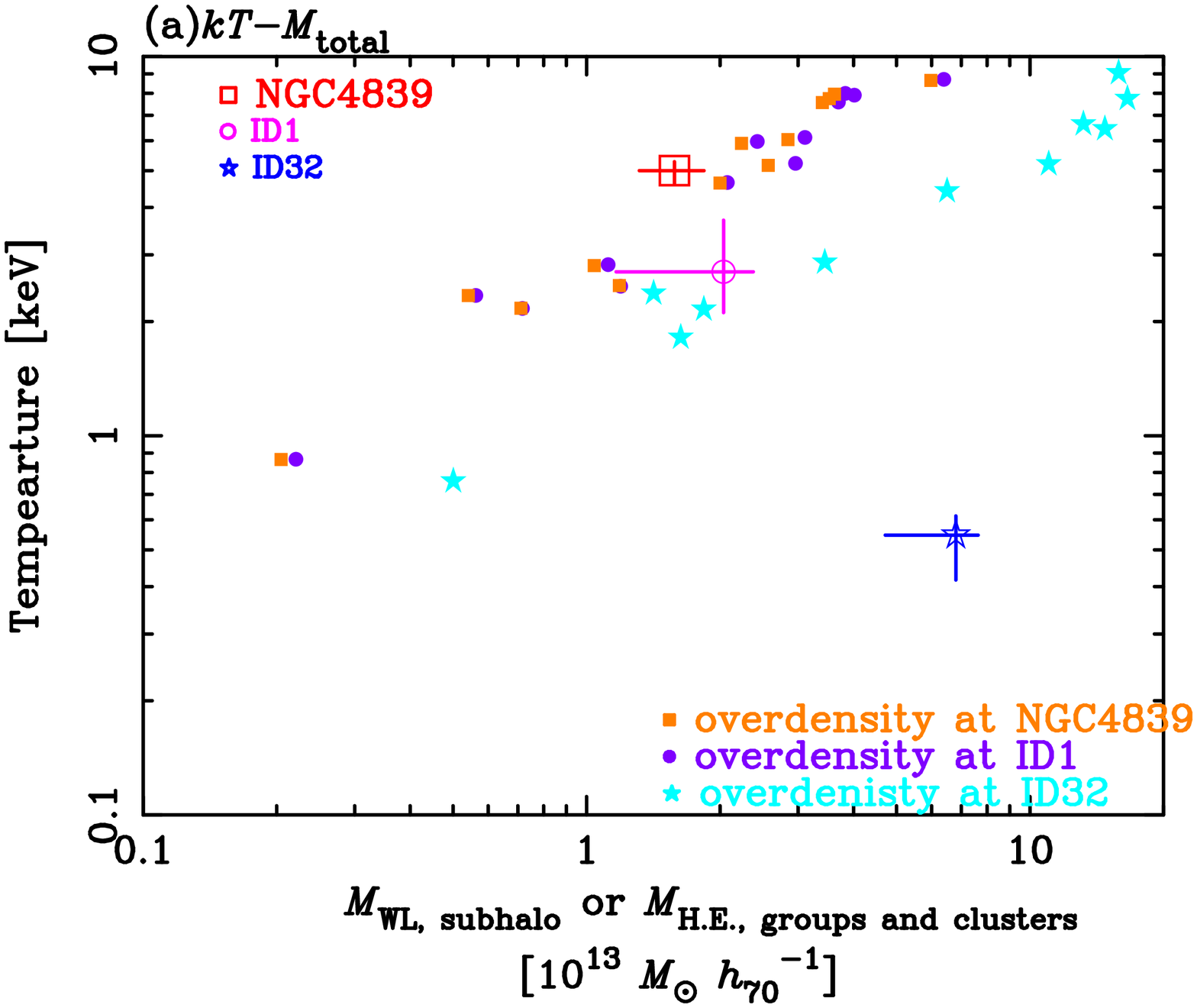}
\includegraphics[width=0.38\textwidth,angle=0,clip]{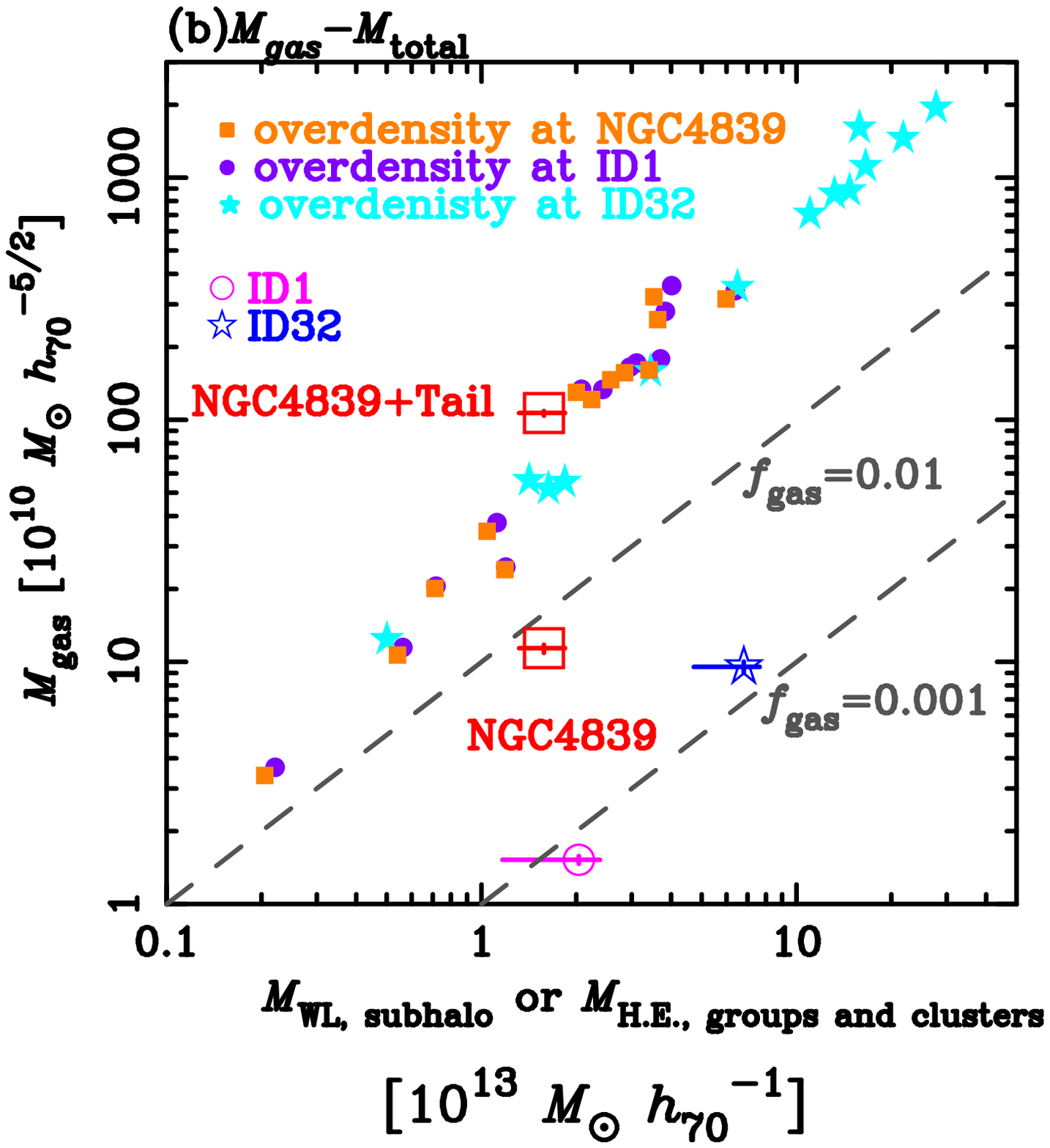}
\caption{
(a)The temperature as a function of the gravitational mass. 
While the gravitational masses of the subhalos are the weak-lensing masses \citep{Okabe2014}, those of the groups are the hydrostatic masses computed using the temperature and density profiles derived by \citet{Vikhlinin2006}.
The open box indicates the subhalo associated with the NGC4839 group (this work).
For comparison, the other subhalos labeled ``ID1", and ``ID32" by \citet{Sasaki2015} are plotted with circles and stars, respectively.
The filled boxes,  circles, and stars show the temperature against hydrostatic mass for galaxy groups and clusters at the radius within which the mean densities of groups and clusters are equivalent to the overdensity of each subhalo within $r_t$.
For details, see text. 
(b)The same as (a), but with gas mass against gravitational mass. 
(A color version of this figure is available in the online journal.)
}
\label{fig:relation}
\end{center}
\end{figure*}

In figure \ref{fig:relation} (a), the $kT-M$ relation of the subhalo associated with the NGC~4839 group is  compared with that of the other subhalos and regular groups. 
In contrast to the other two subhalos reported by \citet{Sasaki2015},
if we adopt the temperature within $r_t$, the $kT-M$ relation of the subhalo is consistent  with that of the regular groups.


To estimate the gas mass of the excess emission of the  subhalo, "ID~9",  
we estimated the electron density for each spectral fitting region.
Within the truncation radius, we assumed a spherical region with uniform electron density.
The geometries of the other regions were approximated as truncated cones 
whose radii are the same as those of individual regions.
Integrating the derived electron densities, we computed the gas mass of the subhalo.
The gas mass within $r_t$, $M_{\rm gas,Center}$, is approximately $1.1\times10^{11}~M\solar$.
The estimated total gas mass up to $8~r_t$ was $M_{\rm gas,Center}+M_{\rm gas,Tail}(<8~r_t)=(10.7\pm0.3)\times10^{11}~M\solar$.
To study the uncertainties of the geometry, we also computed the electron density 
and gas mass assuming with spherical shells.
The gas masses of each sector with spherical shell are few tens of percents larger than these assuming cones.

The resultant $M_{gas}$--$M$ relations for the ``Center" within $r_t$ , $M_{\rm gas,Center}$--$M_{WL}$, 
is shown in figure \ref{fig:relation} (b).
The gas fraction within $r_t$, or the ratio of $M_{\rm gas,Center}(<r_t)$ to the weak-lensing mass by \citet{Okabe2014}, 
is about 5 times lower than those of the regular galaxy groups and clusters.
For comparison, we also plotted the same relation for the other two subhalos derived from \citet{Sasaki2015}.
The gas mass fractions of these two subhalos are much lower than those of 
regular galaxy groups, and most of the gas may have already been stripped away.
Therefore, the most of the gas associated with the group has now been stripped.
Assuming that the stripped gas was now in the ``Tail" sector, 
we computed again the gas fraction in the ``Tail" sector by changing the radius. 
As a result, the fraction out to $8~r_t$ in the ``Tail" sector, $M_{\rm gas,Center}(<r_t)+M_{\rm gas,Tail}(<8~r_t)$ 
to the weak-lensing mass ratio, is comparable within the dispersion of that of regular groups.

\subsection{Ram pressure stripping}
\label{sec:ram}

The thermodynamic properties around the subhalo associated with the NGC~4839 group indicate the gas stripping.
When the ram pressure is greater than the gravitational storing force per area, 
the gas associated with the subhalo is effectively stripped. 
To investigate the ratio between ram pressure and gravitational force per unit area ratio, 
we adopted the same equation described in \citet{Sasaki2015}: 
\begin{eqnarray}
Ratio &=& \frac{P_{ram}}{F_{grav} / Area} \nonumber \\
&=& 0.8 \left( \frac{r}{100~\rm kpc}\right)^4
\left(\frac{n_{e,\rm ICM}}{ 10^{-4}~\rm cm^{-3}}\right)
\left(\frac{M_{\rm subhalo}}{10^{13}~M\solar}\right)^{-1}  \nonumber  \\ 
&& \left(\frac{M_{\rm gas, subhalo}}{10^{11}~M\solar}\right)^{-1}
\left(\frac{v}{ 2000~\rm km~s^{-1} }\right)^{2} 
\end{eqnarray}
Here, the $n_{\rm e,ICM}$, $v_{\rm gal}$, $r$,  $M_{\rm subhalo}$, and $M_{\rm gas, subhalo}$ 
are the electron density of the ICM, velocity of the subhalo, radius, 
the mass of subhalo, and gas mass of the subhalo, respectively.
The normalization of the ICM component of the ``ID2 within 50$'$" region is approximately 4 times greater 
than those at a similar distance from the cluster center in other directions \citep{Simionescu2013}.
Therefore, we assumed that the $n_{\rm e,ICM}$ around the subhalo was $2\times10^{-4}~\rm cm^{-3}$, 
and two times greater than that of the other arms.
The $M_{\rm gas}$ and $M_{\rm subhalo}$ were adopted as the gas mass and weak-lensing masses within $r_t$. 
The infall velocity of $\sim 2000\rm~km~s^{-1}$ was computed from the NFW profile for the main halo derived from \citet{Okabe2014}. 
Adopting 5~keV and a Mach number $1.73^{+0.25}_{-0.16}$ at $2~r_t$ in the ``S" sector, 
the velocity calculated from the Mach number, $2000\pm300~\rm km~s^{-1}$, was comparable with the infall velocity.
At $r_t$, the ratio was $1.4 \times (n_{\rm e,ICM} / 2\times10^{-4}~\rm cm^{-3})$, 
and the ram pressure could strip the gas associated with the subhalo.
This result is well consistent with the X-ray tail morphology.

The gas mass in the ``Tail" region was estimated to be approximately 
$1.0\times10^{12}~M\solar$ as descrived in section \ref{sec:kt-m}, 
and we assumed the gas was originally located within $r_t$ before being stripped. 
Since the length of the X-ray tail is approximately $8~r_t$, the gas at the end of tail had been stripped 
approximately $400\times (v/2000~\rm km~s^{-1})$~Myr ago assuming a constant infall velocity.
This value implies that mass-loss rate is approximately $2500~\times (v/2000~\rm km~s^{-1})^{-1}~M\solar{\rm yr}^{-1}$. 
This value is comparable with those of merging groups in clusters (e.g., \cite{Eckert2014}).

\citet{Nulsen1982} estimated the mass-loss rate caused by turbulent stripped via Kelvin-Helmholtz instabilities; 
$\dot{M}_{KH} \approx \pi r^2 \rho_{ICM} v$.
Here, $r$ and $\rho_{ICM}$ are the radius of the subhalo and gas density of the ICM, respectively.
Since the mass-loss rate depend on the density profile of the ICM, 
we fitted the azimuthal average electron density profiles
excluding the direction of the subhalo, "ID~9",  
by \citet{Simionescu2013} with a $\beta$-model.
We adopted the best-fit $\beta$-model profile out to the virial radius for calculation. 
Assuming that the groups was infalling from the virial radius of the Coma cluster 
and a constant infall velocity, $v=2000~{\rm km~s}^{-1}$, 
the timescale of the mass-loss was approximately 700 Myr. 
By integrating the mass-loss rate over the time scale, 
we obtained the total lost mass, $3\times10^{11}$~M\solar.
If this gas had originally been within $r_t$, the gas mass fraction of the subhalo 
would be comparable with that of regular groups before infalling to the Coma cluster. 
The gas mass in the ``Tail" region is about three times greater than 
the lost mass via Kelvin-Helmholtz instabilities. 
This suggests that the ram pressure had been removed the gas more effective than Kelvin-Helmholtz instabilities.

\section{Summary and conclusions}

We observed the third massive subhalo ``ID~9"  detected from 
weak-lensing analysis \citep{Okabe2014} with {\it Suzaku}. 
By comparing the subhalo mass distribution with thermodynamics obtained by X-ray observation, 
we first investigated cluster evolution through the accretion of objects.

While the X-ray peak is shifted approximately $1\arcmin$ away from the mass center,  
the NGC~4839 galaxy coincided with the X-ray peak. 
The X-ray tail was elongated toward the southwest direction or the outskirts of the Coma cluster, 
and the length was approximately $6~r_t$, $\sim$ 600~kpc.
The temperatures of the central core and X-ray tail regions were approximately 5~keV. 
At a distance $r_t$ toward the north, corresponding to the head of the subhalo, 
we found a temperature jump, from 5 keV to 8--10 keV.
While the temperature of the southern part was also 7~keV at 1-2~$r_t$, 
temperature profile beyond 2~$r_t$ decreased to 4 keV. 
At $2~r_t$ in the southern part, we estimated a Mach number to be 
${\cal M}=1.73^{+0.25}_{-0.16}$ under the Rankine-Hugoniot condition. 
The gas mass fraction of the subhalo within $r_t$ was approximately 0.7\%.
It is approximately 5 times smaller than that of regular groups and poor clusters. 
With the infall velocity estimated from the weak-lensing mass of the Coma cluster \citep{Okabe2010}, 
the ram pressure was comparable with the gravitational force per unit area. 
Assuming the Kelvin-Helmholtz instabilities, the removed gas mass 
was estimated to be approximately $3\times10^{11}~M\solar$.
Assuming that the removed gas had been within $r_t$ before infalling to the Coma cluster, 
the gas mass fraction of the subhalo had been comparable with those of regular groups.


\bigskip

We thank the anonymous referee for careful reading the manuscript and providing valuable comments. 
We gratefully acknowledge all members of the {\it Suzaku} operation, calibration, and software development teams.
TS is supported by JSPS Research Fellows. 
We acknowledge the support of  Grants-in-Aid for Scientific Research from the MEXT, No. 25400235(K. M.), 25800112 (K. S.), and 26800097(N. O.).
This work was supported by the Funds for the Development of Human
Resources in Science and Technology under MEXT, Japan, and 
 Core Research for Energetic Universe in Hiroshima University (the MEXT
 program for promoting the enhancement of research universities, Japan).

\appendix

\section{Spectra of each region}

\begin{figure*}
\begin{center}
\includegraphics[width=0.30\textwidth,angle=0,clip]{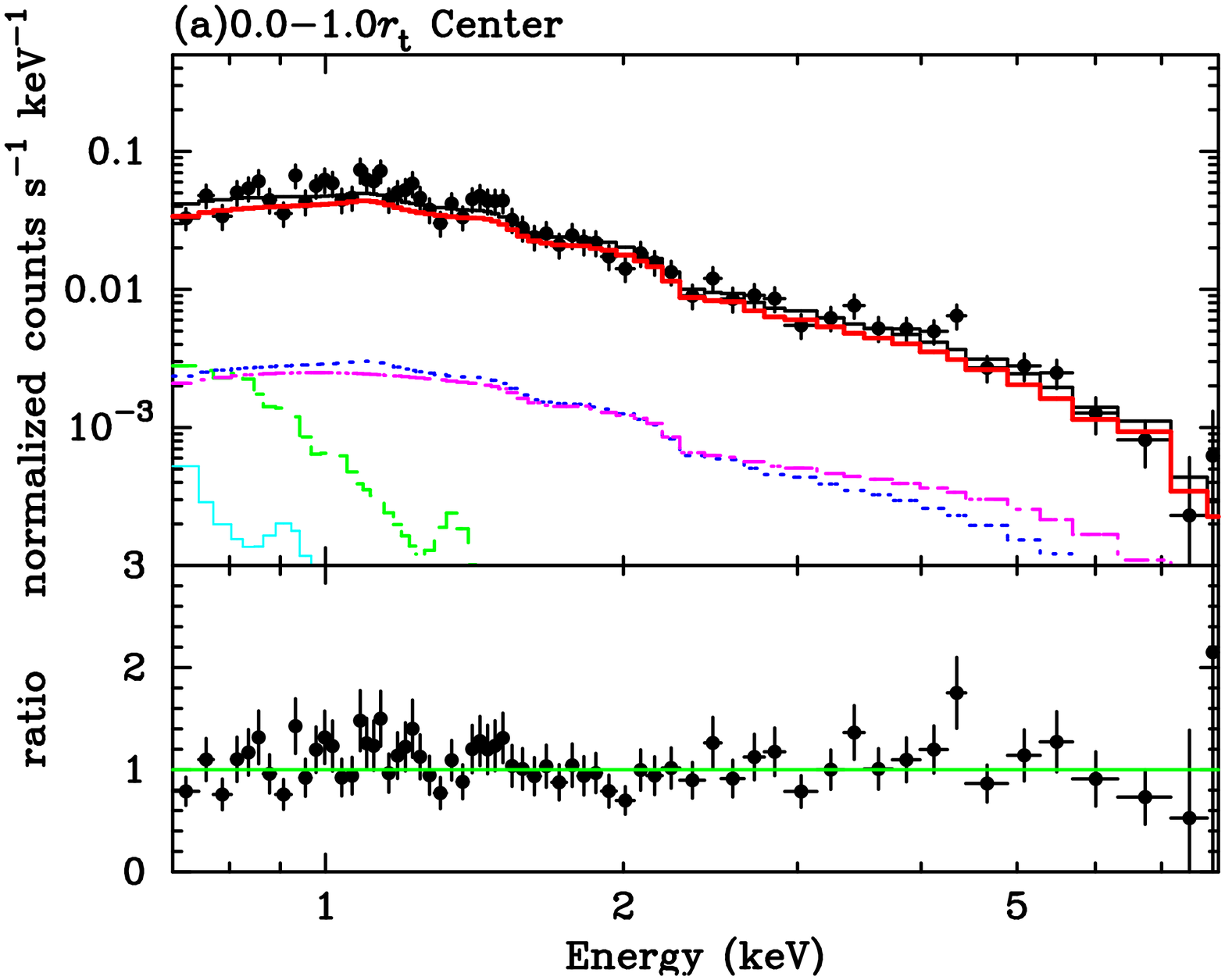}
\includegraphics[width=0.30\textwidth,angle=0,clip]{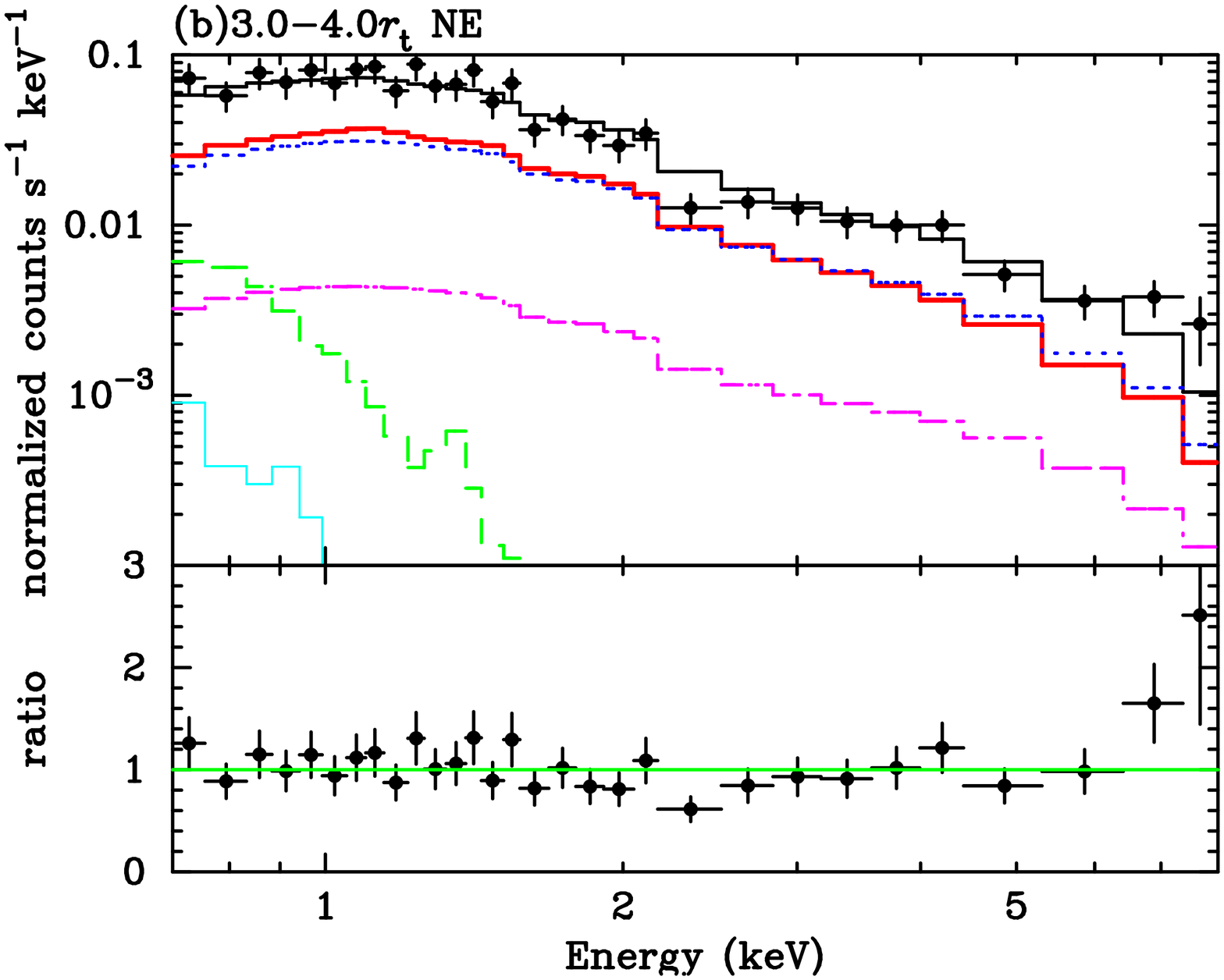}
\includegraphics[width=0.30\textwidth,angle=0,clip]{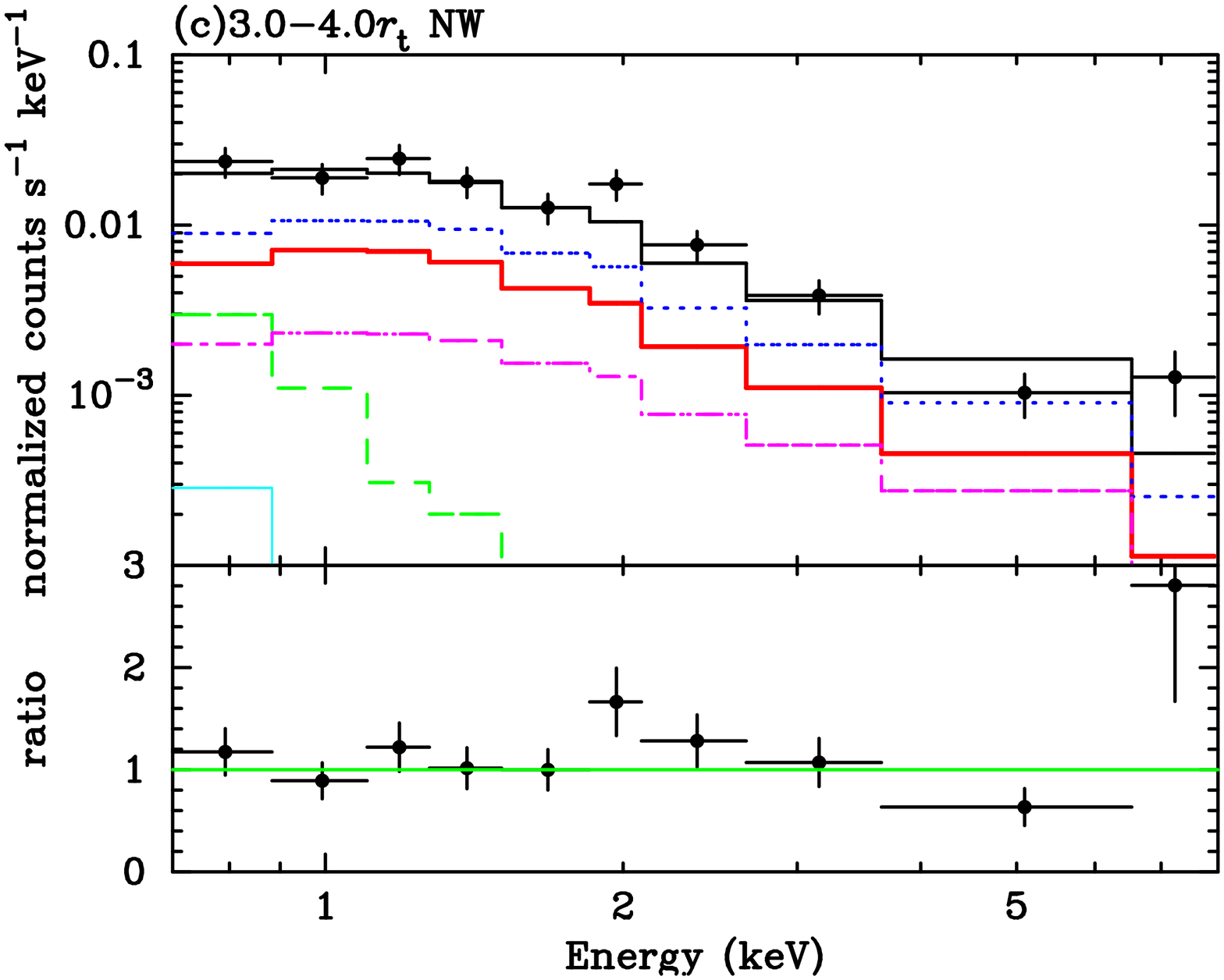}
\caption{
The spectra in (a) ``Center",  (b)3-4~$r_t$ in the ``NE" sector, and (c)3-4~$r_t$ in the ``NW" sector. 
The spectra were rebinned here for display purposes only.
Upper panels show the NXB subtracted XIS~1 spectra (black crosses).  
The subhalo component is plotted as a (red) bold line.
The ICM, CXB, LHB, and MWH components are indicated by 
(blue) dotted, (magenta) dash-dotted, (cyan) thin, and (green) dashed lines, respectively.
The lower panels show the data-to-model ratios.
}
\label{fig:spec2TTail}
\end{center}
\end{figure*}

\begin{figure*}
\begin{center}
\includegraphics[width=0.30\textwidth,angle=0,clip]{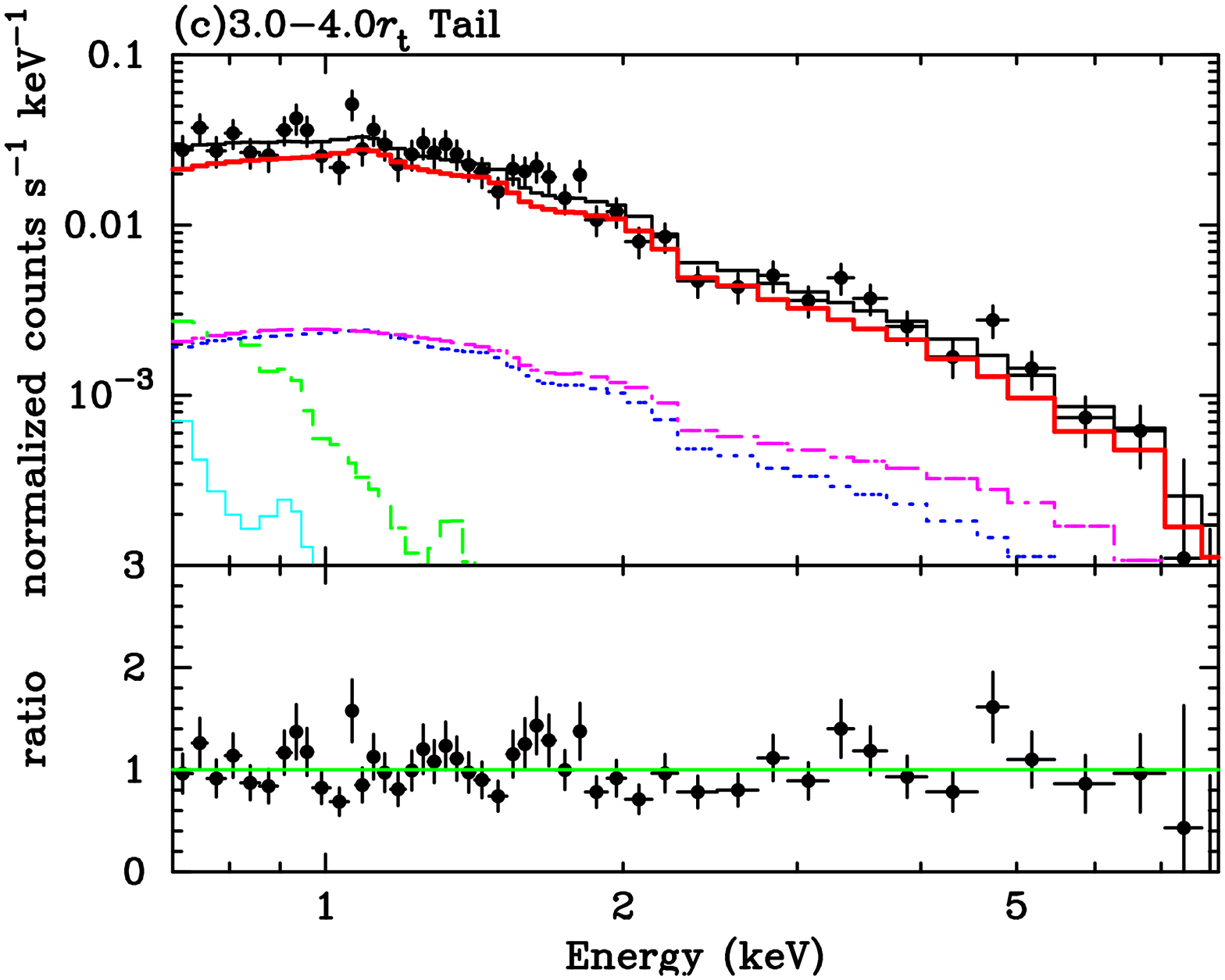}
\includegraphics[width=0.30\textwidth,angle=0,clip]{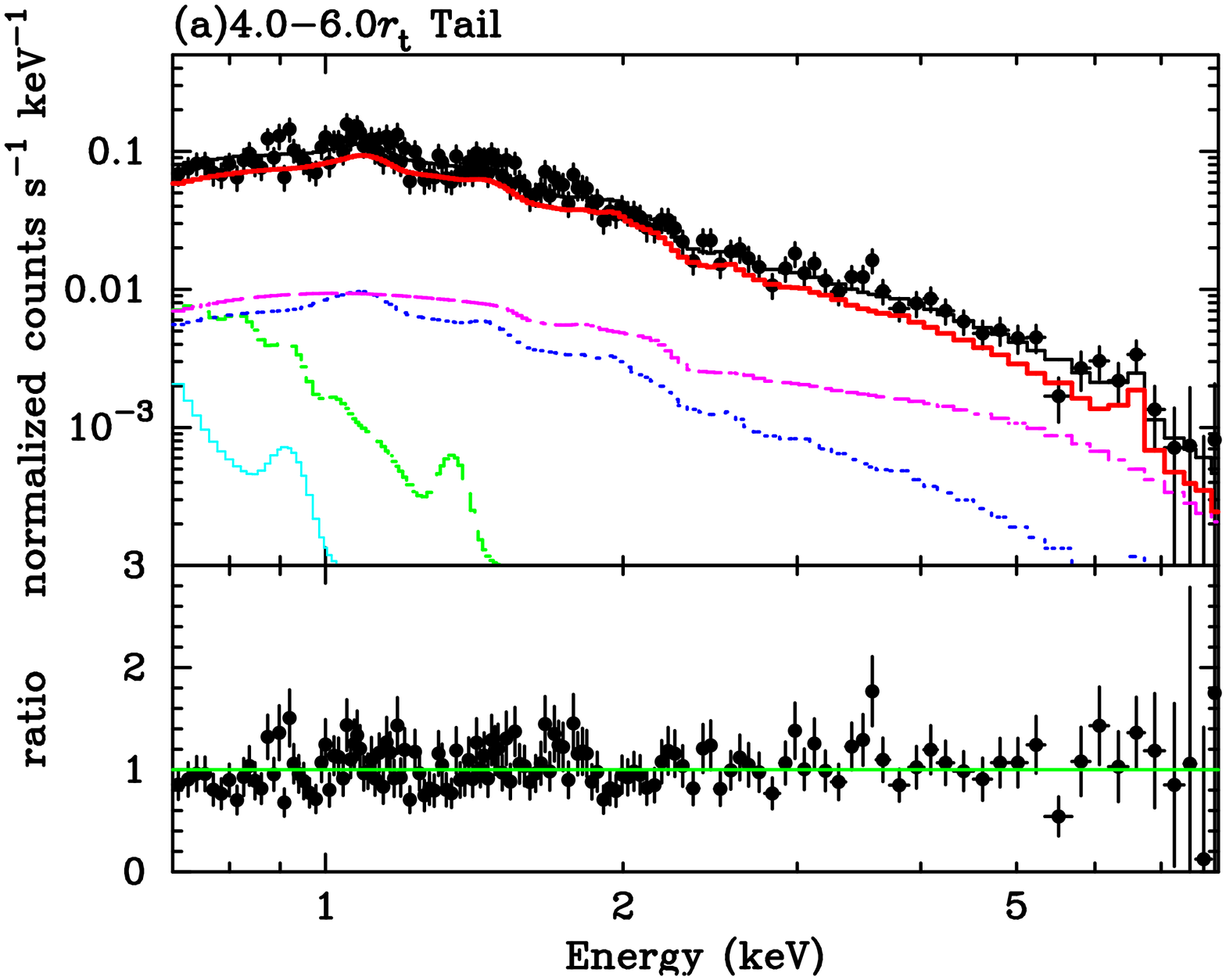}
\includegraphics[width=0.30\textwidth,angle=0,clip]{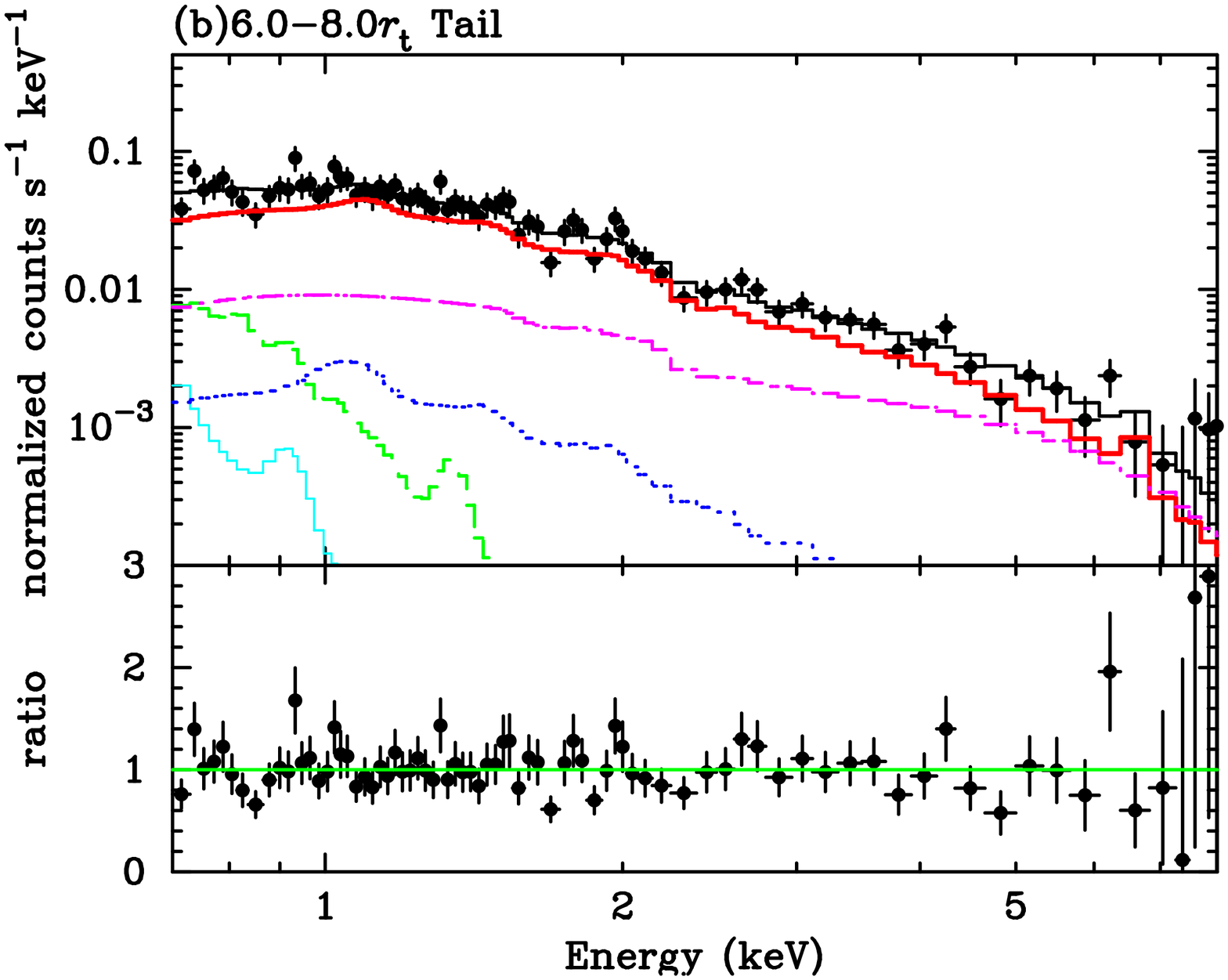}
\caption{
The spectra of (a)3-4~$r_t$, (b)4-6~$r_t$, and (c)6-8~$r_t$  in the ``Tail" sector. 
The spectra of 1-2~$r_t$ and 2-3~$r_t$ in this sector are shown in figure \ref{fig:spec2T}.
The color notations are the same as figure \ref{fig:spec2TTail}.   
}
\label{fig:spec2TNE}
\end{center}
\end{figure*}

\begin{figure*}
\begin{center}
\includegraphics[width=0.30\textwidth,angle=0,clip]{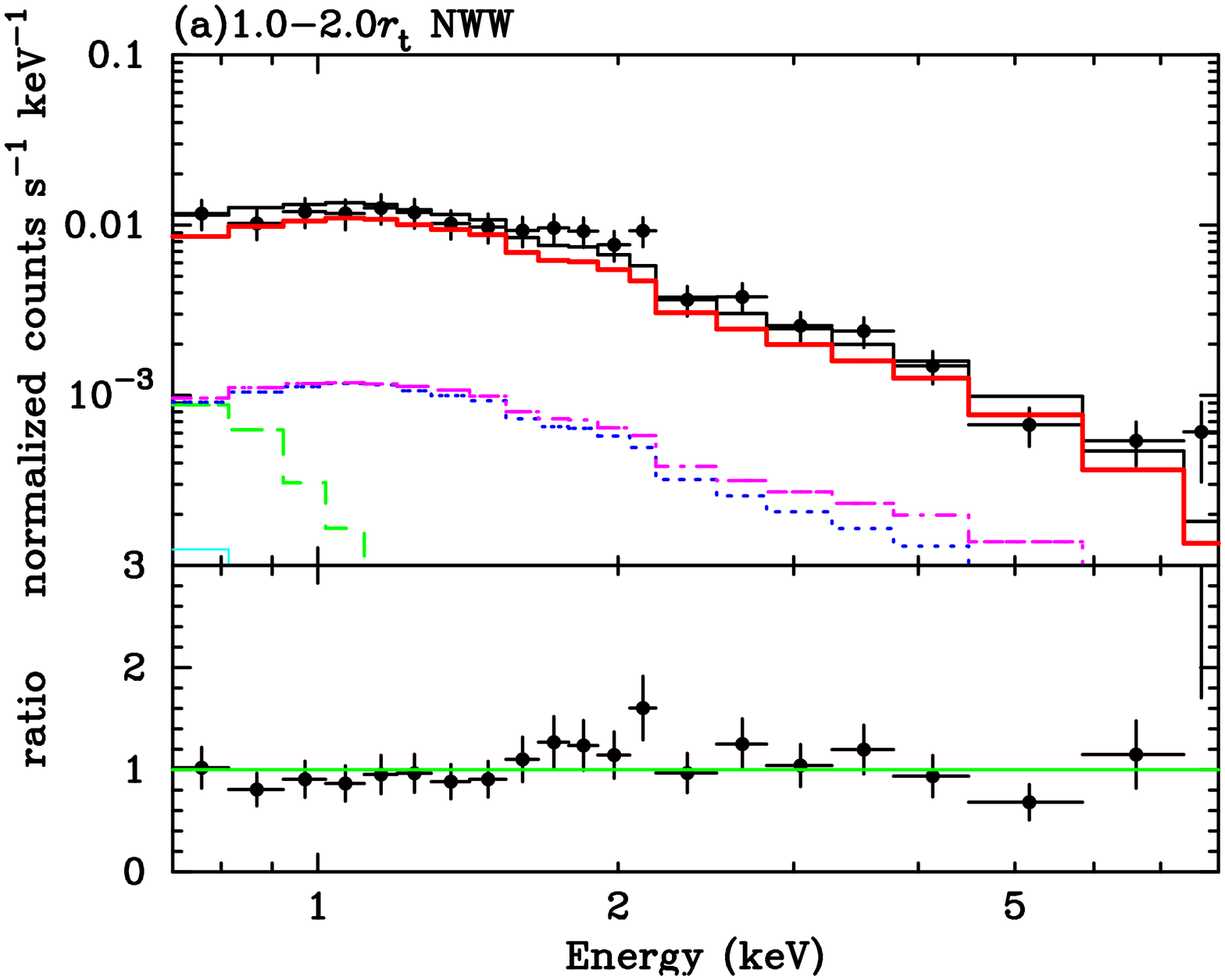}
\includegraphics[width=0.30\textwidth,angle=0,clip]{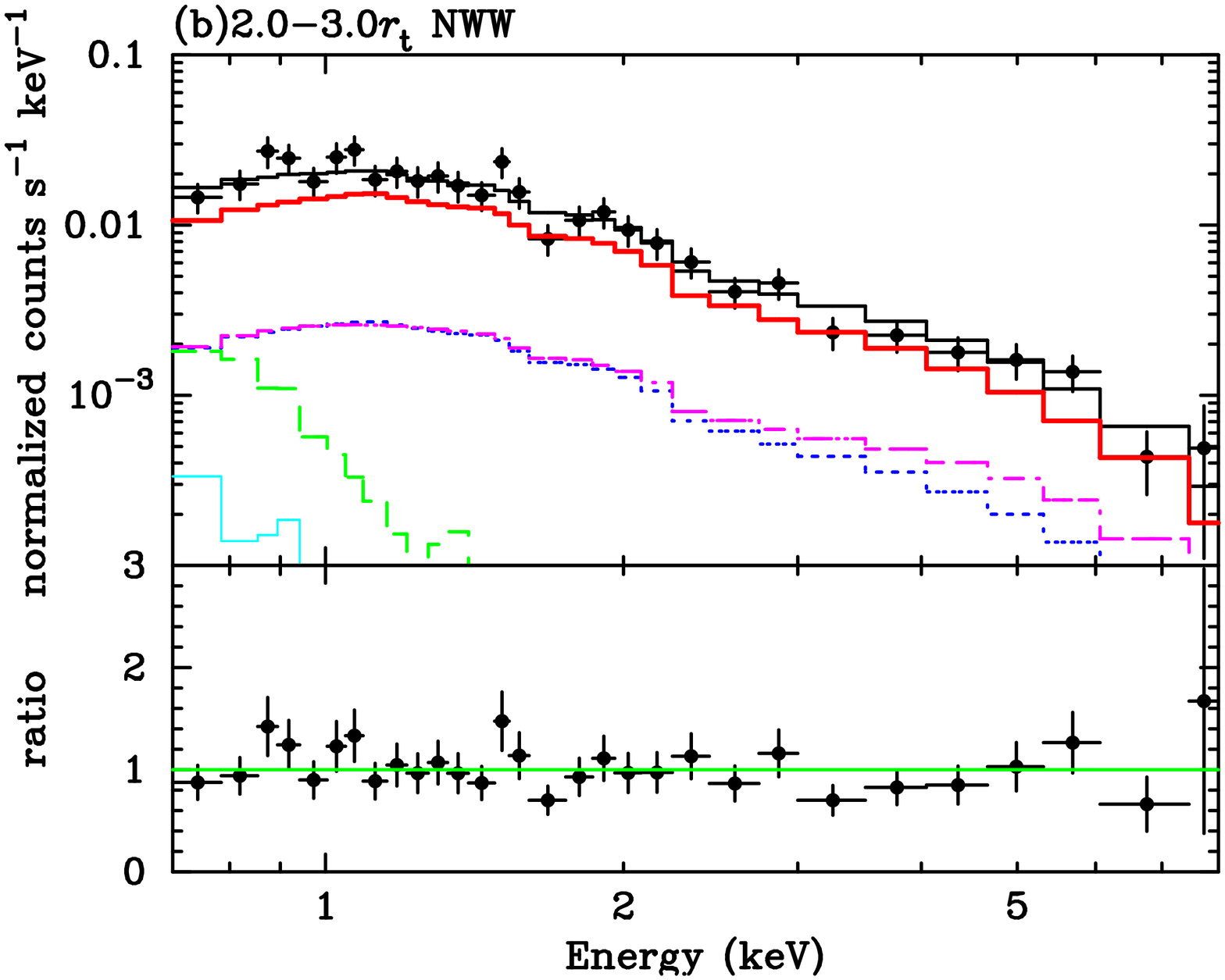}
\includegraphics[width=0.30\textwidth,angle=0,clip]{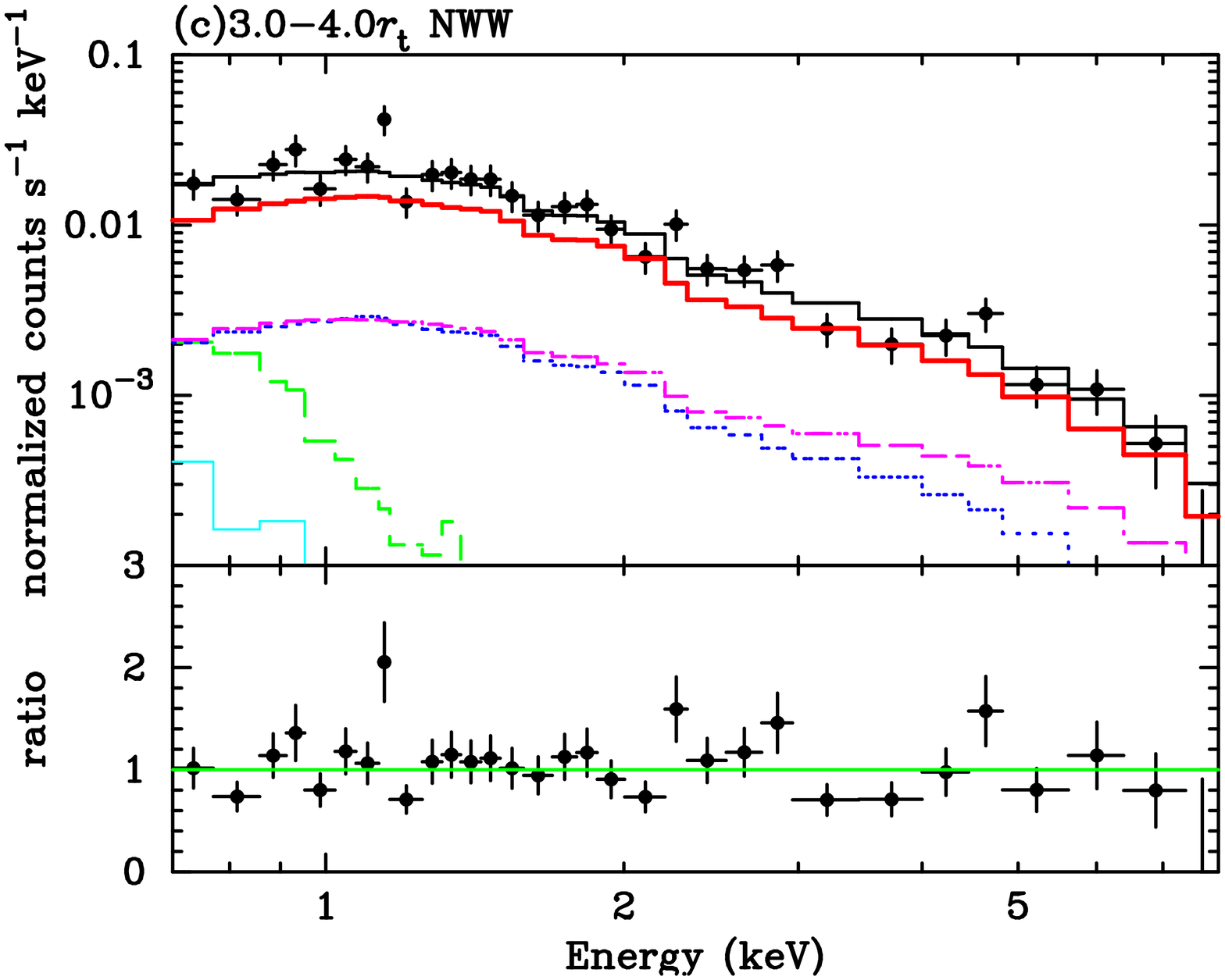}
\caption{
The spectra of (a)1-2~$r_t$, (b)2-3~$r_t$, and (c)3-4~$r_t$  in the ``NWW" sector. 
The color notations are the same as figure \ref{fig:spec2TTail}.   
}
\label{fig:spec2TNWW}
\end{center}
\end{figure*}

\begin{figure*}
\begin{center}
\includegraphics[width=0.30\textwidth,angle=0,clip]{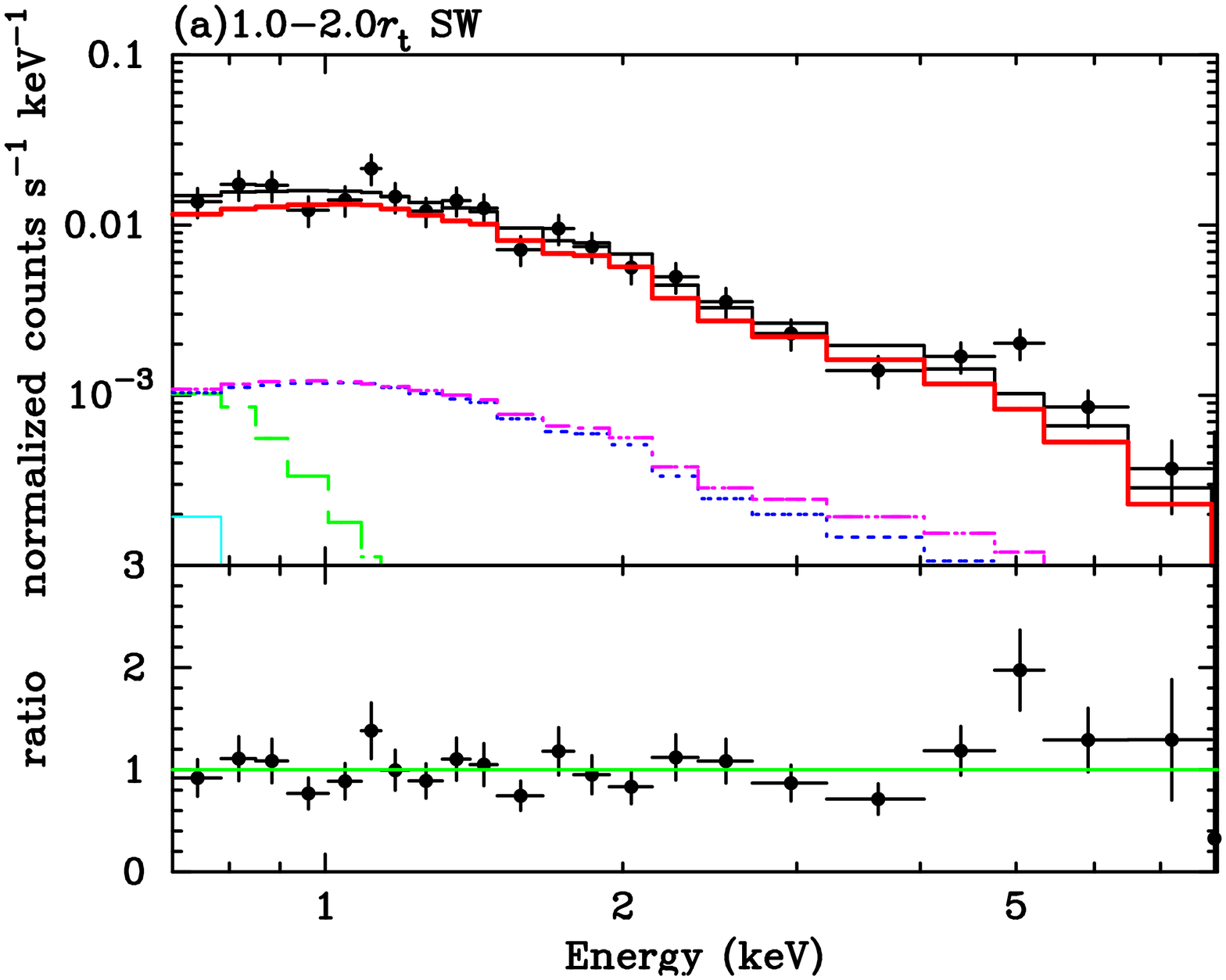}
\includegraphics[width=0.30\textwidth,angle=0,clip]{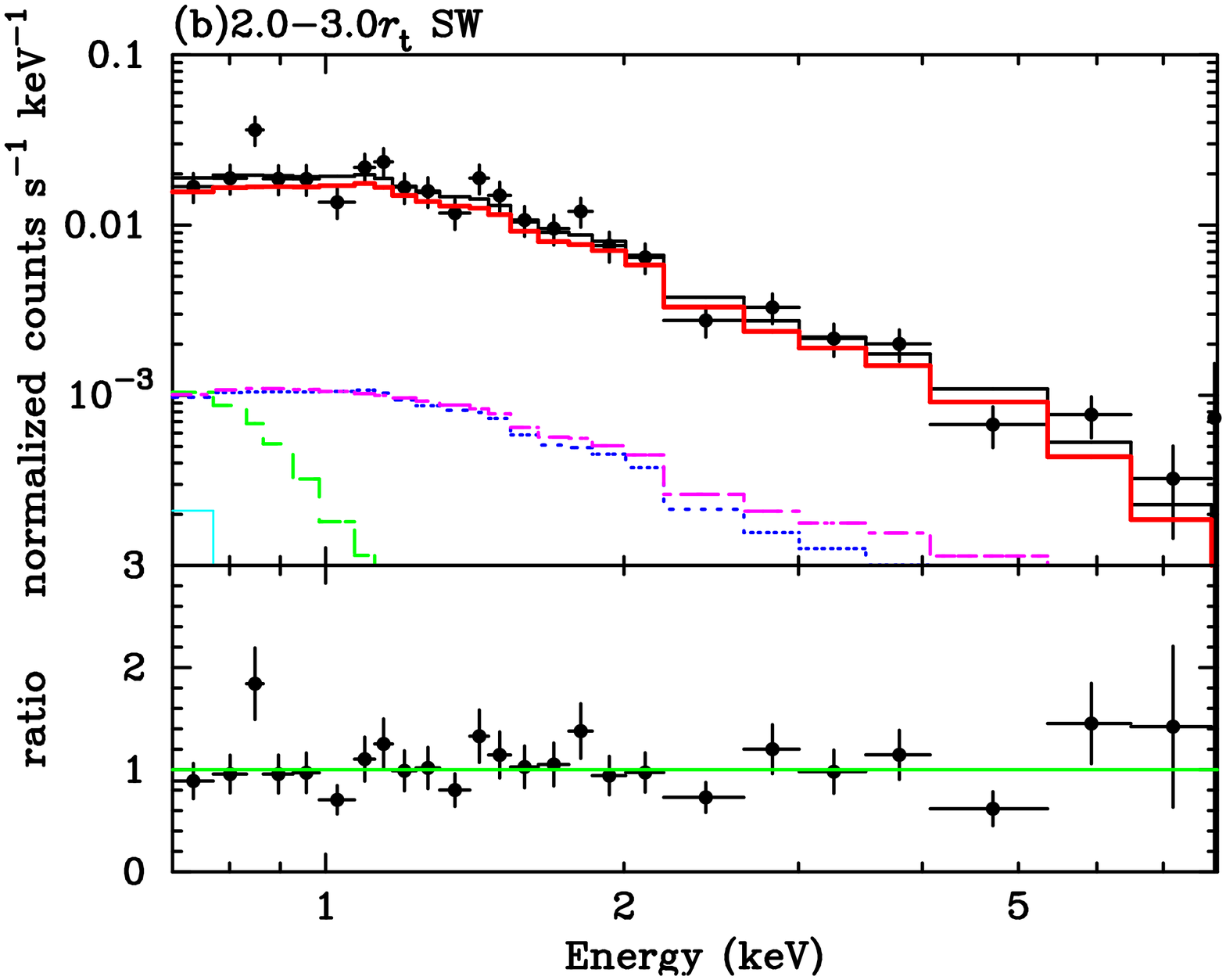}
\includegraphics[width=0.30\textwidth,angle=0,clip]{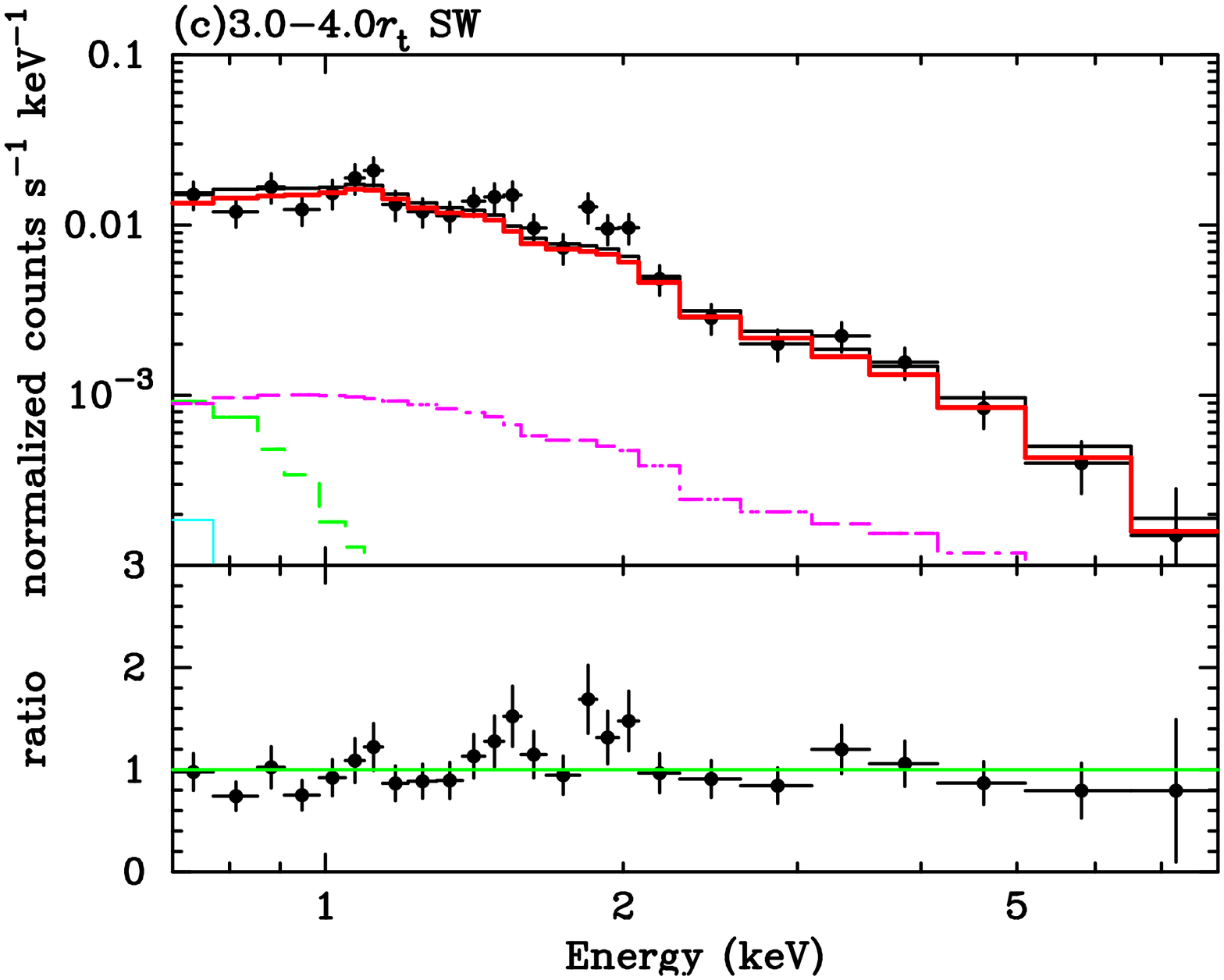}
\caption{
The same as figure \ref{fig:spec2TTail}, but for the ``SW" sector.  
}
\label{fig:spec2TSW}
\end{center}
\end{figure*}

\begin{figure*}
\begin{center}
\includegraphics[width=0.30\textwidth,angle=0,clip]{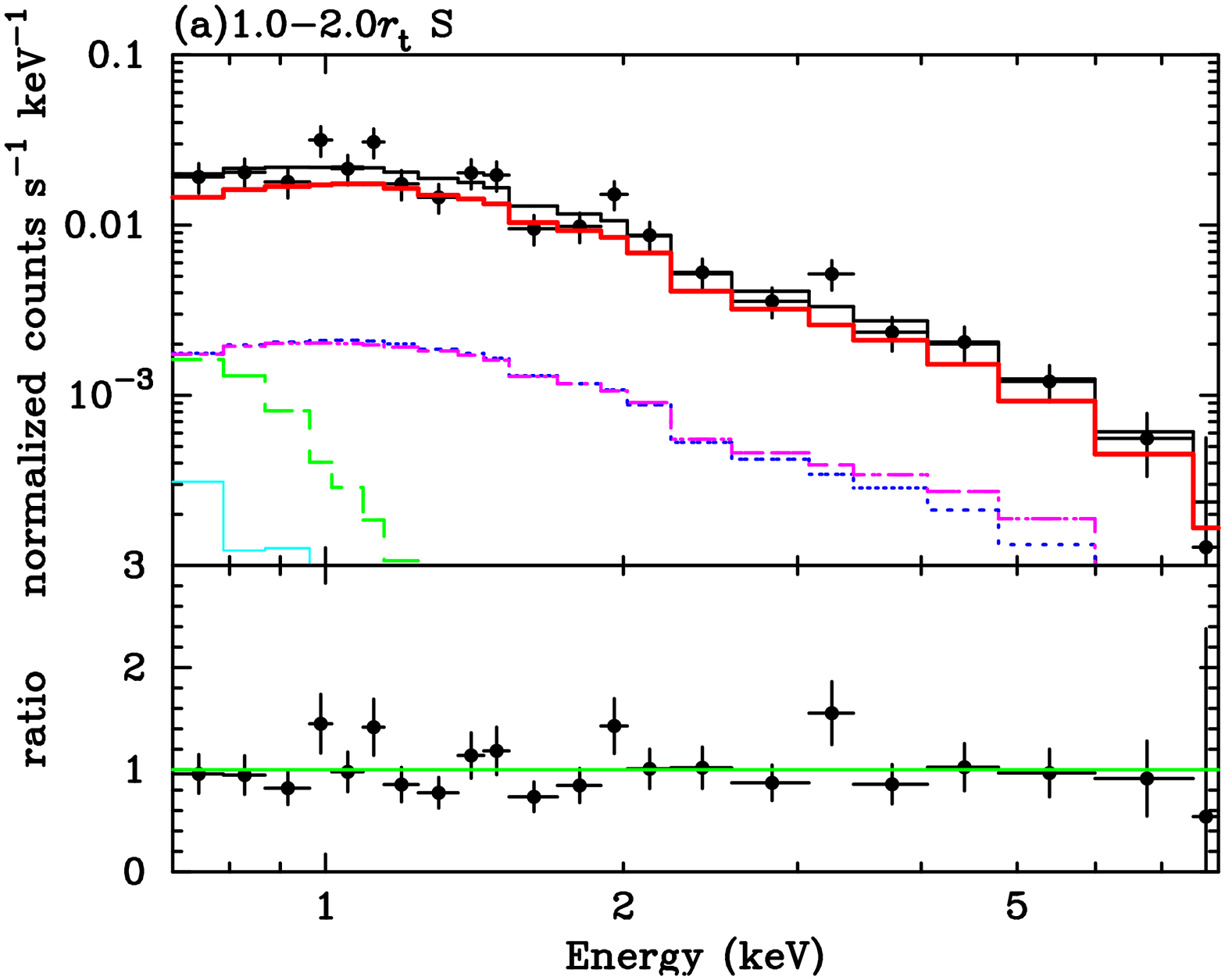}
\includegraphics[width=0.30\textwidth,angle=0,clip]{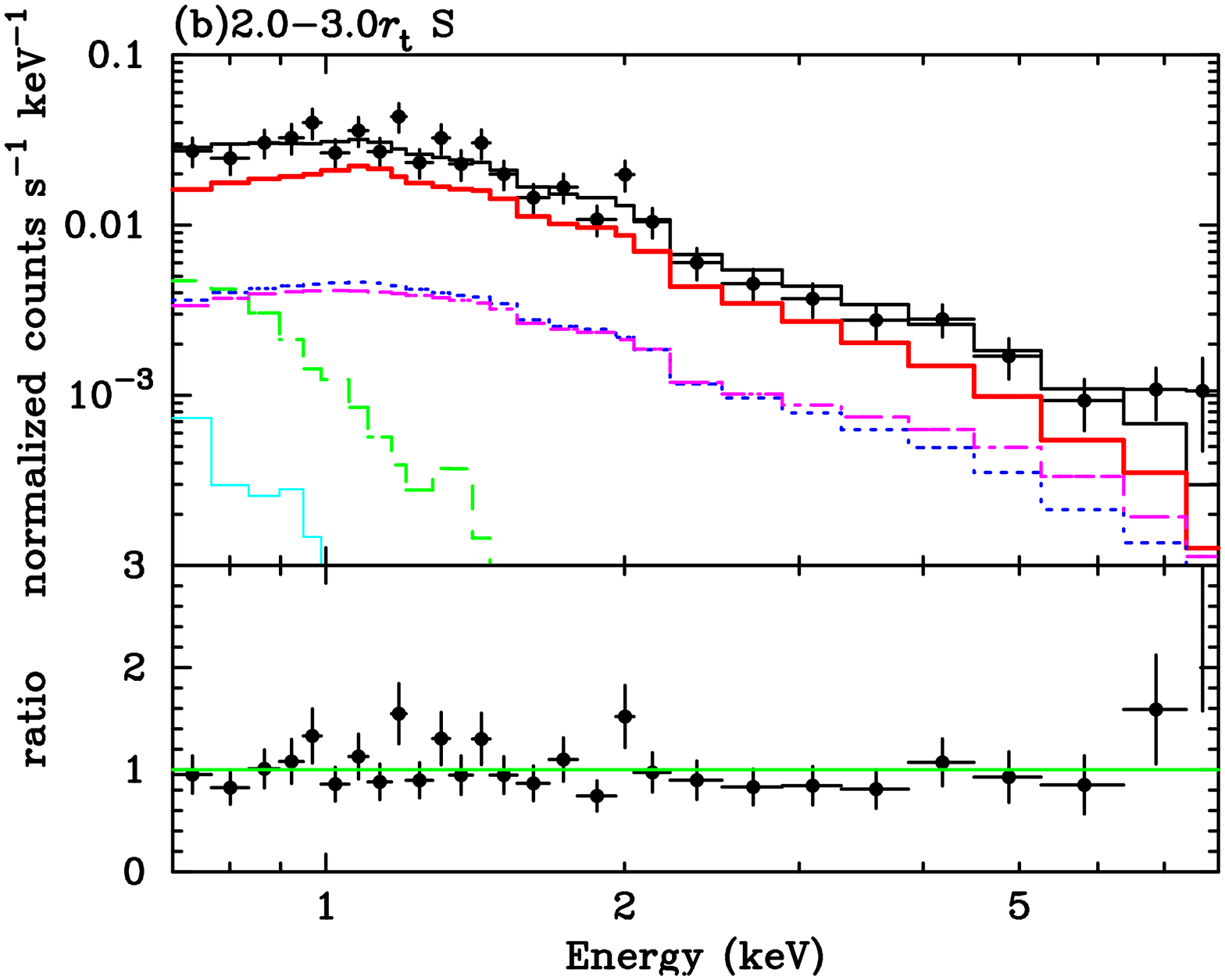}
\includegraphics[width=0.30\textwidth,angle=0,clip]{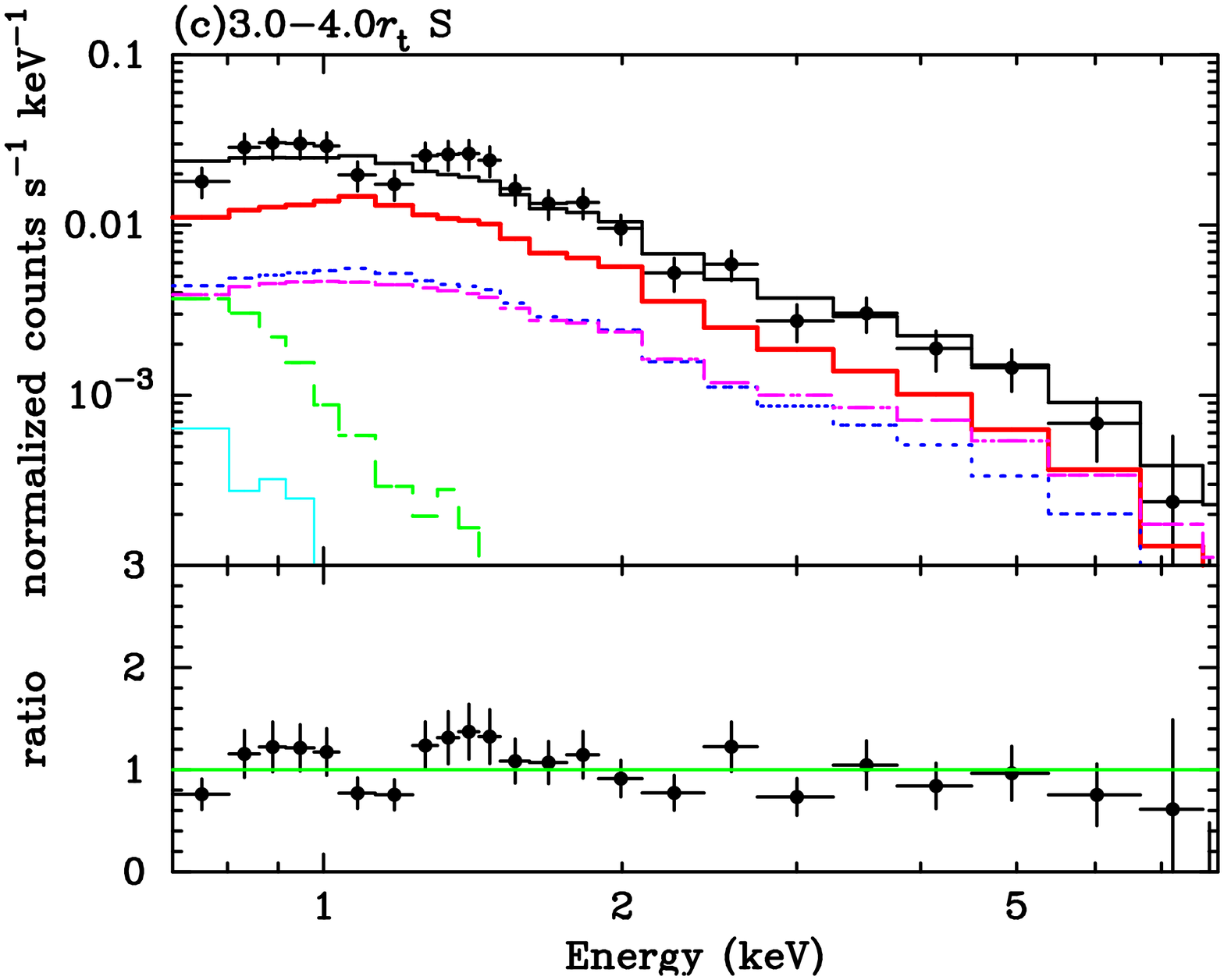}
\caption{
The same as figure \ref{fig:spec2TTail}, but for the ``S" sector.  
}
\label{fig:spec2TS}
\end{center}
\end{figure*}

\begin{figure*}
\begin{center}
\includegraphics[width=0.30\textwidth,angle=0,clip]{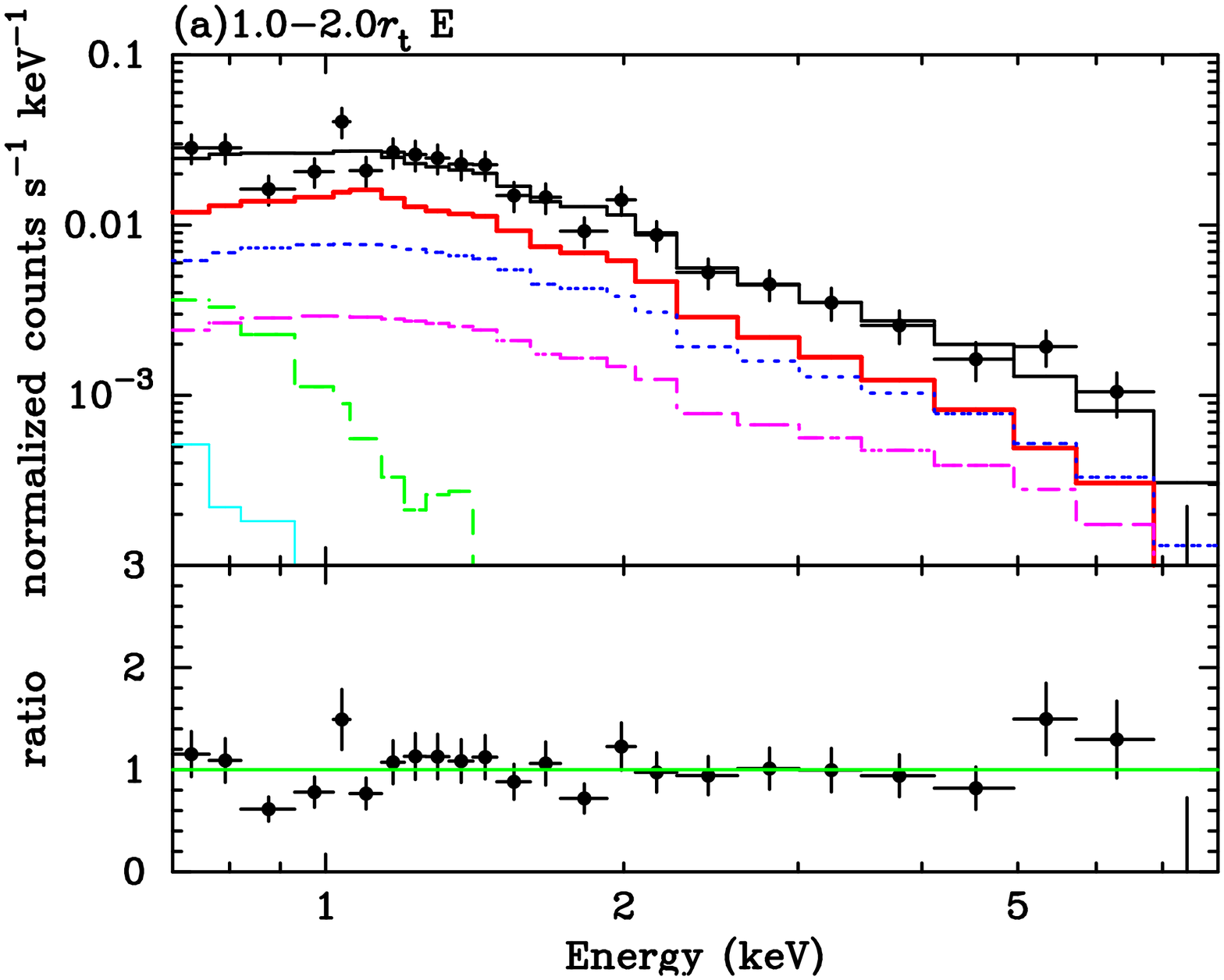}
\includegraphics[width=0.30\textwidth,angle=0,clip]{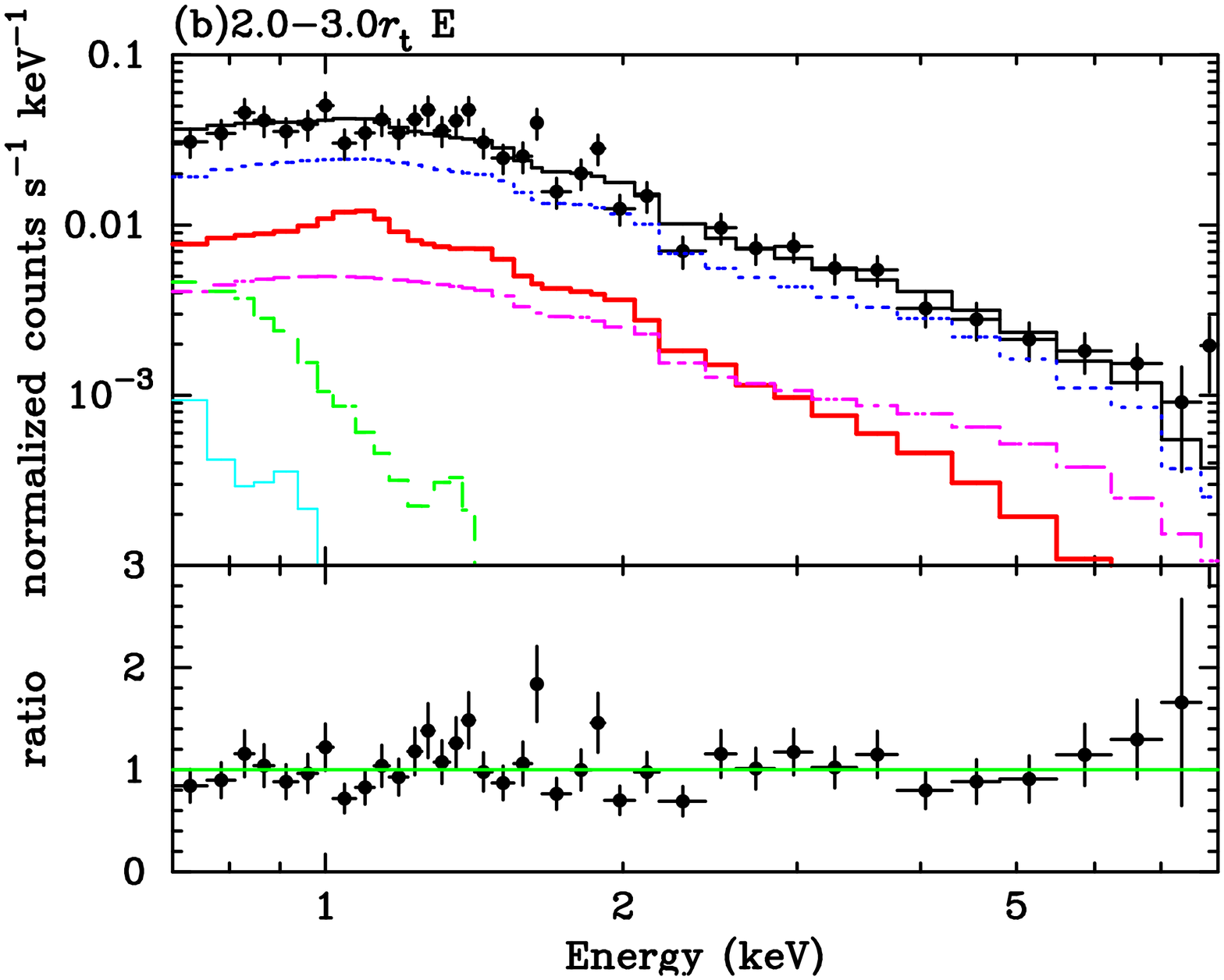}
\includegraphics[width=0.30\textwidth,angle=0,clip]{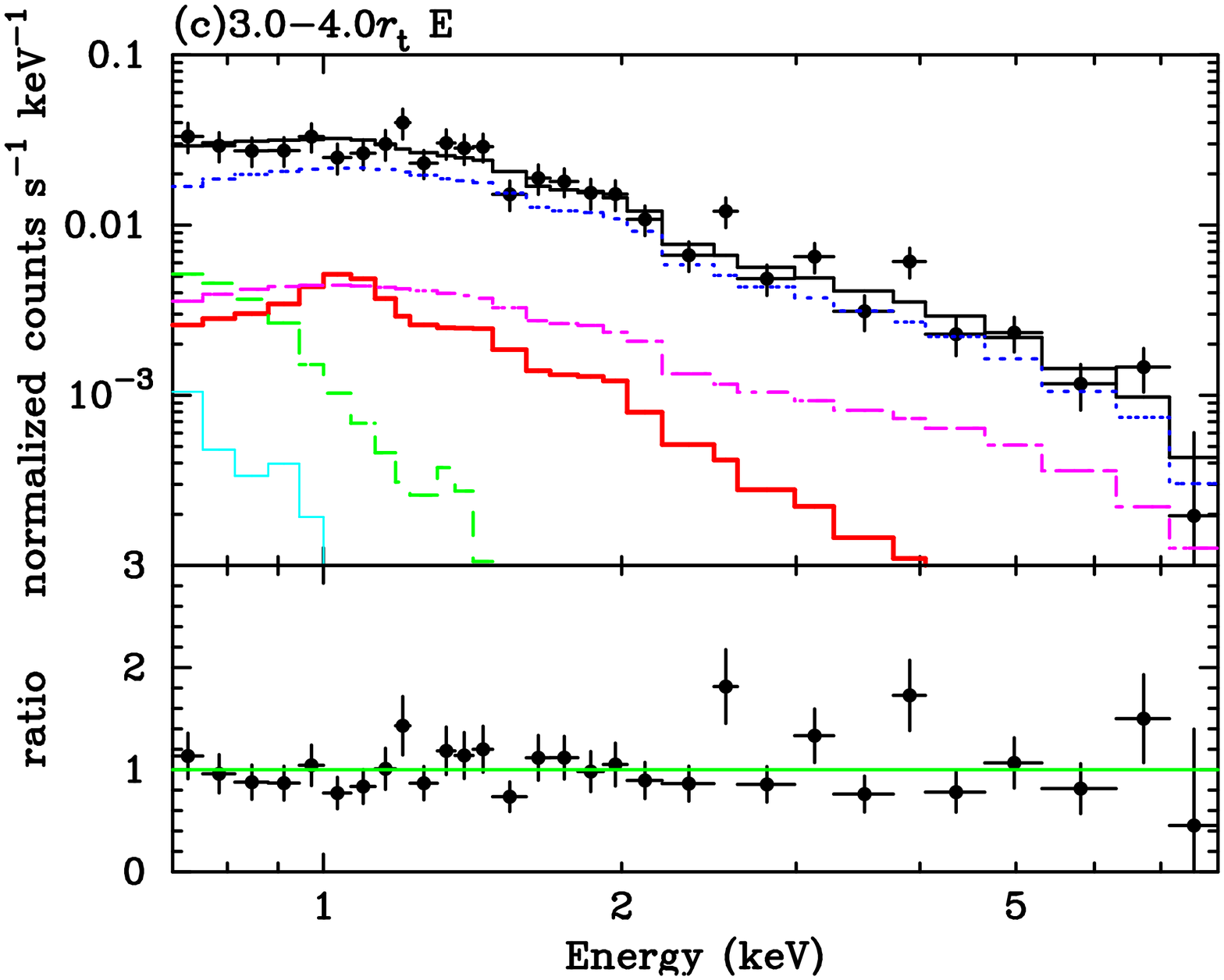}
\caption{
The same as figure \ref{fig:spec2TTail}, but for the ``E" sector.  
}
\label{fig:spec2TE}
\end{center}
\end{figure*}

Figure \ref{fig:spec2TTail}-\ref{fig:spec2TE} show the other spectra extracted around the subhalo `ID~9".

\end{document}